\documentclass[preprint,12pt]{elsarticle}
\usepackage{geometry}
\usepackage[hyphens]{url}
\usepackage{amsmath, amsfonts, amssymb, amsthm}

\newtheoremstyle{named}{}{}{\itshape}{}{\bfseries}{.}{.5em}{\thmnote{#1} #3}
\theoremstyle{named}
\usepackage[hyphens]{url}
\usepackage[hidelinks]{hyperref}
\hypersetup{colorlinks=false,
            linkcolor = black, 
            citecolor = black, 
            urlcolor = black}
\usepackage{algorithm}
\usepackage{algpseudocode}
\usepackage{bbm}
\usepackage{graphicx}
\usepackage{subcaption}
\usepackage{tabularx}
\usepackage{multicol}
\usepackage{multirow}
\usepackage{caption}
\usepackage[countmax]{subfloat}
\usepackage{float}
\usepackage{textcomp}
\usepackage[table]{xcolor}
\usepackage{xcolor}
\usepackage{textcomp}
\usepackage{array}
\usepackage{booktabs}
\usepackage{mathtools}
\usepackage{nicematrix}
\usepackage[official]{eurosym}
\usepackage{xcolor}

\usepackage{environ}
\NewEnviron{revised}{
    \textcolor{red}{\BODY}
}
\usepackage{nicefrac}
\usepackage[capitalize]{cleveref}
\usepackage{pifont}
\usepackage{nomencl}
\makenomenclature

\usepackage{etoolbox}
\renewcommand\nomgroup[1]{%
  \item[\bfseries
  \ifstrequal{#1}{S}{Indices and Sets}{%
  \ifstrequal{#1}{V}{Decision Variables}{%
  \ifstrequal{#1}{M}{Machine Learning Variables}{%
  \ifstrequal{#1}{P}{Parameters}{}}}}%
]}

\usepackage[acronym, nonumberlist]{glossaries}
\makeglossaries
\newacronym{mfrr}{mFRR}{manual frequency restoration reserve}
\newacronym{nn}{NN}{neural network}
\newacronym{qr}{QR}{quantile regression}
\newacronym{bnn}{BNN}{Bayesian neural network}
\newacronym{mcd}{MCD}{Monte Carlo dropout}
\newacronym{cp}{CP}{conformal prediction}
\newacronym{pv}{PV}{photovoltaic} 
\newacronym{hems}{HEMS}{home energy management system}
\newacronym{dam}{DAM}{day-ahead market}
\newacronym{uq}{UQ}{uncertainty quantification}
\newacronym{sota}{SOTA}{state-of-the-art}
\newacronym{afrr}{aFRR}{automatic frequency restoration reserve}
\newacronym{ml}{ML}{machine learning}
\newacronym{mse}{MSE}{mean square error}
\newacronym{rmse}{RMSE}{root mean square error}
\newacronym{mae}{MAE}{mean absolute error}
\newacronym{mcp}{MCP}{multivariate conformal prediction}
\newacronym{cqr}{CQR}{conformalised quantile regression}
\newacronym{mmcp}{MMCP}{max-Mahalonobis conformal prediction}
\newacronym{pcp}{PCP}{probabilistic conformal prediction}
\newacronym{ccp}{CCP}{copula conformal prediction}
\newacronym{picp}{PICP}{prediction interval coverage probability}
\newacronym{mpiw}{MPIW}{mean prediction interval width}
\newacronym{is}{IS}{interval score}
\newacronym{pi}{PI}{perfect-information}
\newacronym{idm}{IDM}{intraday market}
\newacronym{fcas}{FCAS}{frequency controlled ancillary services}

\makeatletter
\def\ps@pprintTitle{%
  \let\@oddhead\@empty
  \let\@evenhead\@empty
  \def\@oddfoot{\footnotesize\itshape Superseded by the published version in Applied Energy 427 (2027) 129059\hfill}%
  \let\@evenfoot\@oddfoot}
\makeatother

\begin{document}

\begin{frontmatter}



\title{Calibrated uncertainty quantification for prosumer flexibility aggregation in ancillary service markets\tnoteref{t1}}
\tnotetext[t1]{This preprint is superseded by the peer-reviewed version published in \textit{Applied Energy}, 427 (2027) 129059, \url{https://doi.org/10.1016/j.apenergy.2026.129059} (open access). Please refer to and cite the published version.}

\author[1]{Yogesh Pipada Sunil Kumar} %% Author name
\author[1]{S. Ali Pourmousavi}
\author[2]{Jon A.R. Liisberg}
\author[2]{Julian Lemos-Vinasco}

\affiliation[1]{organization={School of Electrical and Mechanical Engineering, The University of Adelaide},%Department and Organization
            addressline={Ingkarni Wardli Building}, 
            city={Adelaide},
            %postcode={5005}, 
            state={SA 5005},
            country={Australia}}
\affiliation[2]{organization={Watts A/S},%Department and Organization
            addressline={Brogade 19D}, 
            city={K\o ge 4600},
            country={Denmark}}

%% Abstract
\begin{abstract}

Reliable forecasting of prosumer flexibility is critical for demand response aggregators participating in frequency controlled ancillary service markets, where reliability requirements such as the P90 standard are enforced. Limited behavioural data and dependence on exogenous forecasts introduce uncertainty into flexibility estimates, which can lead to reserve overestimation and financial penalties. This study applies and systematically evaluates established uncertainty quantification approaches based on Monte Carlo dropout (MCD), conformal prediction (CP), and conformalised quantile regression (CQR) to construct prediction intervals for aggregated prosumer flexibility and support uncertainty-aware aggregator bidding.

The approaches are evaluated for a behind-the-meter aggregator participating in the Danish manual frequency restoration reserve capacity market. A large-scale synthetic dataset is generated using an industry-grade home energy management system, combined with publicly available load, solar, price, activation, and device-level data. Multiple multivariate CP strategies are evaluated using MCD predictions, alongside CQR using a separately trained quantile regression model. Under nominal synthetic test conditions, conformal methods substantially improve joint coverage compared with uncalibrated MCD baselines. More conservative methods satisfy the 90\% lower-tail reliability requirement, while CQR and MCP provide comparatively narrow intervals and higher retained economic value. CQR and MCP retain approximately 69\% and 68\% of perfect-information profit, respectively, under the stricter penalty interpretation. A controlled robustness analysis further shows that increasing perturbations of load, solar, price, and activation inputs increase overbidding and reduce empirical coverage and economic performance. The results highlight the trade-off between reliability, economic value, and robustness to input uncertainty when applying established uncertainty quantification approaches to aggregator bidding.

\end{abstract}


% \begin{highlights}
% \end{highlights}

%% Keywords

\end{frontmatter}



\printglossary[type=\acronymtype, title=List of Abbreviations]
\nomenclature[S]{$\mathcal{T}$}{Set of time indices $\{1, \dots, T\}$}
\nomenclature[S]{$\mathcal{I}$}{Set of prosumer indices}
\nomenclature[S]{$\mathcal{J}$}{Set of input scenarios}
\nomenclature[P]{$\boldsymbol{\lambda^{\text{U}}}, \boldsymbol{\lambda^{\text{D}}}$}{mFRR up/down regulation capacity market prices}
\nomenclature[P]{$\boldsymbol{\lambda^{\text{B}}}$}{mFRR imbalance price profile}
\nomenclature[P]{$\boldsymbol{a}^{\uparrow}, \boldsymbol{a}^{\downarrow}$}{mFRR up/down regulation capacity activation signal}
\nomenclature[P]{$\zeta$}{Contextual information for forecasting model}
\nomenclature[P]{$\beta$}{Prosumer revenue factor}
\nomenclature[P]{$T$}{Time horizon}
\nomenclature[V]{$\boldsymbol{x^{\text{u}}}, \boldsymbol{x^{\text{d}}}$}{Reserved up/down regulation flexibility capacity}
\nomenclature[V]{$\boldsymbol{y^{\text{u}}}, \boldsymbol{y^{\text{d}}}$}{Random vectors for the maximum available up/down flexibility capacity}
\nomenclature[V]{$\boldsymbol{\lambda^{\text{u}}}, \boldsymbol{\lambda^{\text{d}}}$}{Up/down regulation incentive profiles offered to prosumer group}
\nomenclature[M]{$\boldsymbol{\Theta^{i}}, \boldsymbol{\Theta}$}{Input set at individual and aggregated prosumer levels}
\nomenclature[M]{$\boldsymbol{y^{\text{u}, i}}, \boldsymbol{y^{\text{u,agg}}}$}{Available up-regulation capacity at individual and aggregated prosumer levels}
\nomenclature[M]{$\boldsymbol{x^{\text{s}, i}}, \boldsymbol{x^{\text{s,agg}}}$}{Solar generation profile at individual and aggregated prosumer levels}
\nomenclature[M]{$\boldsymbol{x^{\text{l}, i}}, \boldsymbol{x^{\text{l,agg}}}$}{Load profile at individual and aggregated prosumer levels}
\nomenclature[M]{$\boldsymbol{b^{\text{cap}, i}}, \boldsymbol{b^{\text{cap,agg}}}$}{Maximum battery capacity at individual and aggregated prosumer levels}
\nomenclature[M]{$\boldsymbol{b^{\text{dis}, i}}, \boldsymbol{b^{\text{dis,agg}}}$}{Maximum battery power at individual and aggregated prosumer levels}
\nomenclature[M]{$\xi$}{Calibrated prediction interval}
\nomenclature[M]{$\mathcal{B}_{\xi}$}{Bounding hyper-rectangle of the calibrated prediction interval}
\nomenclature[M]{$\boldsymbol{l}, \boldsymbol{u}$}{Lower/upper bounds of the calibrated prediction interval}
\nomenclature[M]{$\boldsymbol{Y}$}{MCD sample prediction set}
\nomenclature[M]{$s(\cdot)$}{Conformity score for calibration}
\nomenclature[M]{$\alpha$}{Desired miscoverage level}
\nomenclature[M]{$\hat{q}$}{Calibrated conformity score}
\nomenclature[M]{$S$}{Number of Monte Carlo samples}
\nomenclature[M]{$\mathcal{D}$}{Training, calibration and validation dataset}
\nomenclature[M]{$\mathcal{D}_{\beta}$}{Dataset to perform economic evaluation of bidding model}
\printnomenclature[2cm]
%------------------------------------
\section{Introduction}
\label{sec: intro}
\subsection{Motivation}
\label{ssec: motivation}
Independent demand response aggregators combine distributed energy resources (DERs) such as photovoltaic systems, residential batteries, electric vehicles, and flexible loads to provide demand side flexibility to energy and balance markets. They act as intermediaries between small scale prosumers, who individually offer limited flexibility, and system operators, who increasingly depend on fast and affordable flexibility as conventional generation retires. Effective aggregator models can reduce consumer costs, stimulate innovation, and accelerate DER adoption while enhancing system efficiency and supporting grid stability~\cite{Attarha2021}. Aggregators therefore play a pivotal role in enabling a cost effective and resilient energy transition. Hence, the European Union has introduced measures under the Electricity Directive (EU) 2019/944~\cite{EU_Directive_2019_944} and the Electricity Regulation (EU) 2024/1747~\cite{EU_Regulation_2024_1747} to ensure equitable participation of aggregators in electricity markets. These initiatives, complemented by ongoing consultations on a demand response network code~\cite{EC_2025_demand_response_consultation}, aim to remove regulatory barriers and strengthen the operational and market integration of aggregators across Europe.

One of the central challenges in aggregator business models is predicting prosumer price responsiveness, that is, the extent of flexibility offered at a given incentive level and a given time. This relationship is inherently complex, as prosumer behaviour depends on interdependent factors such as weather, load variability, solar generation, and behavioural inconsistencies~\cite{duck_aware,Nam_2}. Data-driven models have therefore emerged as an effective solution, enabling accurate and scalable estimation of prosumer behaviour from historical data without restrictive assumptions~\cite{dsf_quantification_review}. However, the limited availability of behavioural data and the reliance on forecasts of weather, load, solar generation, and price introduce significant epistemic uncertainty into these models. This uncertainty directly affects the procurement and delivery of critical balancing services, where reserved capacity may not be available when required. 

From an economic perspective, imposing 100\% availability requirements on aggregators relying on inherently uncertain flexibility sources would impose excessive financial risk, potentially rendering participation unprofitable and discouraging market entry. To balance the risk of capacity unavailability with the need to incentivise aggregator participation, the Danish system operator has adopted a P90 compliance standard for \gls{fcas} capacity market participants, allowing a capacity deviation with a maximum probability of 10\%~\cite{jalal_p90}. Non-compliance can result in financial penalties or suspension from market participation, highlighting the need for predictive models with integrated \gls{uq}. Consequently, there is a need for data-driven \gls{uq} frameworks that capture epistemic uncertainty while providing finite-sample marginal coverage guarantees under the assumptions required by conformal prediction. Such frameworks can improve the reliability of aggregator forecasts under data scarcity and market reliability constraints.

% To mitigate this risk, the Danish system operator has adopted a P90 compliance standard for \gls{fcas} market participants, allowing a capacity deviation with a maximum probability of 10\%~\cite{jalal_p90}. Non-compliance can result in financial penalties or suspension from market participation, highlighting the need for predictive models with integrated uncertainty quantification. Consequently, this study aims to develop a data-driven framework that captures epistemic uncertainty while providing formal coverage guarantees, thus improving the robustness of aggregator forecasts under data scarcity and market reliability constraints.
\subsection{Related work}
\label{ssec: related work}
\gls{uq} has become a central research focus in energy forecasting, particularly for aggregator business models that depend on reliable flexibility estimation under limited data availability and forecast dependence. Methods that capture both aleatoric and epistemic uncertainty can support more informed and risk-aware decision-making, particularly when flexibility estimates are subject to uncertainty and must satisfy regulatory requirements such as the P90 compliance standard in ancillary service markets.

\glspl{bnn} have been widely studied for demand and flexibility forecasting due to their ability to capture both epistemic and aleatoric uncertainty by learning distributions over model parameters~\cite{uq_load_forecasting_review}. Several studies report improved uncertainty quantification for residential demand forecasting using \glspl{bnn} compared to deterministic neural networks~\cite{bnn_1,bnn_2,bnn_3}. More recently, \cite{bnn_4} applied a \gls{bnn}-based transformer architecture to electric vehicle flexibility forecasting for frequency regulation markets, demonstrating strong predictive performance. However, the high computational cost of \glspl{bnn} limits their scalability, particularly for complex architectures, frequent, and longer forecast horizons. More generally, exact Bayesian inference for practical \glspl{bnn} remains analytically and computationally intractable, motivating the use of approximate Bayesian inference methods~\cite{bayesian_lit_survey}

Approximate Bayesian methods, such as \gls{mcd} provide a computationally efficient alternative~\cite{abcd_cp,dropout_Yarin}. By retaining dropout layers during inference, \gls{mcd} produces a predictive distribution that can be interpreted as an approximation of a deep Gaussian process~\cite{dropout_Yarin}. This approach has been successfully applied in medical imaging~\cite{mcd_medical}, object detection~\cite{mcd_object_detection}, and energy demand forecasting~\cite{mcd_energy}. However, \gls{mcd} does not produce a true Bayesian posterior and can result in miscalibrated uncertainty estimates, particularly in multi-output or long-horizon settings~\cite{dropout_folgoc, dropout_shridhar}.
 
To address calibration issues, \gls{cp} provides a lightweight and distribution-free framework for obtaining prediction intervals with finite-sample coverage guarantees under data exchangeability~\cite{cp_vovk, cp_angelopolous}. Applications in electricity price~\cite{qr_cp_electricity_price} and photovoltaic generation forecasting ~\cite{qr_cp_pv} commonly employ \gls{cqr}, which combines \gls{qr} with \gls{cp} to obtain calibrated prediction intervals. However, standard \gls{cqr} primarily addresses aleatoric uncertainty and may be less effective in settings where epistemic uncertainty is substantial~\cite{cqr_issues}. In contrast, \gls{mcd} is well suited to representing epistemic uncertainty but can produce miscalibrated uncertainty estimates. The established integration of \gls{mcd} with \gls{cp} provides a complementary approach, combining MCD-based epistemic uncertainty representation with conformal calibration and finite-sample coverage guarantees~\cite{mcd_cp}. Both \gls{mcd}--\gls{cp} and \gls{cqr} therefore provide established approaches for uncertainty quantification, but their application to prosumer flexibility aggregation and aggregator bidding remains limited.

Despite these developments, the application of conformal uncertainty quantification to prosumer flexibility forecasting and aggregator bidding remains limited. Existing applications of both \gls{mcd}--\gls{cp} and \gls{cqr} have predominantly focused on single-output forecasting tasks, whereas aggregator decision-making requires multivariate, multi-time-step flexibility predictions. Although multivariate \gls{cp} is an emerging area of research~\cite{cp_algo}, the application of conformal methods to aggregator bidding and their implications for market reliability and economic performance remain largely unexplored.

\subsection{Contributions}
\label{ssec: contributions}
In this study, we implement and evaluate \gls{uq} approaches, including \gls{mcd}, \gls{mcd}–\gls{cp}, and \gls{cqr}, to model price responsiveness of prosumers for an aggregator participating in the European \gls{fcas} markets. A decision-dependent chance-constrained optimisation program was developed to simulate the aggregator’s bidding problem under uncertainty. To enable a representative evaluation, a synthetic dataset was generated using a modified version of the industry-grade \gls{hems} algorithm from Watts A/S~\cite{watts_homepage}, which was the project industry partner, combined with publicly available datasets for load and solar profiles, market prices, activation data, and device-level size distributions of residential batteries and solar panels. The evaluated approaches weree trained and assessed for both P90 compliance and economic performance.

The main contributions of this work relative to the aggregator operation \gls{sota} are as follows:

\begin{enumerate}
\item In contrast to prior studies that rely on computationally expensive \glspl{bnn} and are largely restricted to short-term settings~\cite{bnn_1,bnn_2,bnn_3,bnn_4}, this work applies and evaluates uncertainty quantification approaches, including \gls{mcd}, \gls{mcd}--\gls{cp}, and \gls{cqr}, within a multi-time-step, multivariate aggregator bidding problem. The results demonstrate that conformal calibration can substantially improve flexibility-bidding reliability while retaining economically meaningful profitability under the nominal conditions considered in the Danish \gls{fcas} capacity market.
\item Unlike existing applications of \gls{mcd}~\cite{mcd_energy}, \gls{mcd}--\gls{cp}~\cite{mcd_cp}, and \gls{cqr}~\cite{qr_cp_pv,qr_cp_electricity_price}, which predominantly address single-output forecasting tasks, this study systematically compares these approaches within a multi-step aggregator optimisation problem. The analysis evaluates their predictive reliability and economic performance, including their robustness to perturbations in the input data. 
\item We develop a large-scale synthetic dataset using a modified industry-grade \gls{hems} algorithm from Watts A/S, combined with publicly available datasets for load and solar profiles, market prices, activation signals, and residential device size distributions. This enables evaluation of predictive reliability and economic performance using representative market and device characteristics, while retaining a controlled synthetic representation of aggregate prosumer behaviour. For reproducibility, the generated synthetic datasets are publicly available~\cite{Pipada2025SyntheticData}. 
\end{enumerate}

The remainder of this paper is structured as follows. Section~\ref{sec: problem} formulates the aggregator bidding optimisation problem. Section~\ref{sec: methodology} introduces the hybrid \gls{mcd}–\gls{cp} framework. Section~\ref{sec: case study} outlines the case study setup, data generation strategy, and the learning architecture. Section~\ref{sec: results} presents the results, including the nominal \gls{uq} performance, economic evaluation, and robustness analysis. Section~\ref{sec: discussion} discusses the practical implications, limitations, and future research directions, and Section~\ref{sec: conclusions} concludes the study.
%------------------------------------
%------------------------------------
\section{Aggregator business model}
\label{sec: problem}
This section presents the business model of the behind-the-meter flexibility aggregator and outlines the associated mathematical framework considered in this study.  
\subsection{Background and context}
\label{ssec: background}
\begin{figure}[!htpb]
    \centering 
\includegraphics[trim=250pt 150pt 250pt 150pt, clip, width=0.55\linewidth]{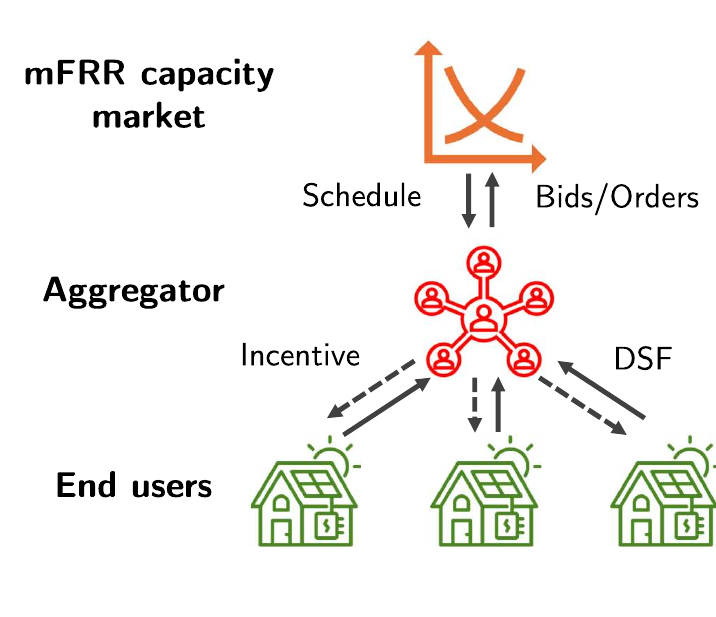}
    \caption{Aggregator business model}
    \label{fig: Aggregator business model}
\end{figure}

In this study, we consider the operation of a behind-the-meter flexibility aggregator participating in the Danish \gls{mfrr} capacity market. Figure~\ref{fig: Aggregator business model} illustrates the framework based on indirect load-control adopted in this case study. The aggregator sends incentive price signals to the \gls{hems} of multiple prosumers, who voluntarily adjust their consumption to provide flexibility. This aggregated flexibility is traded by the aggregator in relevant electricity markets and services. Although market structures vary between regions, the analysis focuses on the Danish \gls{mfrr} capacity market operated by Energinet. Nevertheless, the proposed mathematical framework is designed to be adaptable to similar market mechanisms, such as automatic frequency regulation reserves markets. However, this generalisability is not explicitly evaluated in this study.
% \begin{figure}[!htpb]
%     \centering
% \includegraphics[trim=90 70 50 152, clip, width=\linewidth]{mfrr_timeline_v2.pdf}
%     \caption{\gls{mfrr} capacity and activation market timeline}
%     \label{fig: mfrr timeline}
% \end{figure}

The Danish \gls{mfrr} capacity market procures tertiary reserves through a pay-as-clear, day-ahead auction, securing reserve commitments for the subsequent delivery day. The market is asymmetric, allowing participants to offer up-regulation, down-regulation, or both. Up-regulation corresponds to increased generation or reduced consumption, while down-regulation involves increased consumption or reduced generation. Accepted capacity bids are remunerated at the market-clearing price irrespective of activation, subject to reserve availability. Participation is governed by Energinet’s P90 reliability requirement, which permits reserve shortages or overbidding with a maximum probability of 10\%. Actual energy delivery is coordinated through the \gls{mfrr} activation market, operated 45 minutes before delivery using a pay-as-clear mechanism with one-price imbalance settlement. More details on Danish ancillary service markets and regulatory frameworks are available in~\cite{energinet_ancillary_2024, energinet_prequalification_2024}.

% The \gls{mfrr} capacity market uses a pay-as-clear market mechanism to procure tertiary reserves to balance demand and supply during emergencies. Up-regulation corresponds to exporting energy or reducing consumption, while down-regulation involves importing energy or increasing consumption. The market is asymmetric, allowing participants to offer up-regulation, down-regulation, or both, depending on their strategy.

% The market operates one day ahead, securing hourly reserves for the following day. Accepted participants are remunerated at the capacity price regardless of activation, but must ensure reserve availability or face financial penalties. Participants must ensure that the P90 requirement is satisfied when participating in Danish \gls{fcas} capacity markets~\cite{jalal_p90, energinet_prequalification_2024}. According to this requirement, Energinet allows for reserve shortage or capacity overbidding with a maximum probability of 10\%.

% Actual activations are procured through the \gls{mfrr} activation market, held 45 minutes before delivery, also following a pay-as-clear principle. Energinet applies a one-price imbalance settlement for energy activation, and participants with accepted capacity bids are required to offer at least the same volume in the activation market. Aggregators may offer additional capacity closer to real time if greater flexibility is anticipated. 
% The exact market timeline is illustrated in Figure~\ref{fig: mfrr timeline}. 
\subsection{Assumptions and mathematical framework}
\label{ssec: assumptions}
Based on the background and context presented in Section~\ref{ssec: background}, the aggregator’s bidding framework requires a flexibility prediction algorithm capable of estimating the flexibility available under the varying incentive profiles offered to prosumers. Furthermore, this prediction model must be embedded within the bidding optimisation problem of the aggregator, where the incentive profile offered to all prosumer’s \gls{hems} is treated as a decision variable. Finally, to satisfy the P90 reliability criterion, the bidding formulation must adopt a chance-constrained programming approach. Consequently, the following assumptions are made to formulate the aggregator’s profit maximisation problem.

\begin{itemize}
    \item The aggregator is a price-taking entity that participates in the \gls{mfrr} capacity market. It has access to point predictions of \gls{mfrr} up- and down-regulation market prices and \gls{mfrr} up- and down-regulation reserve activation forecasts, which are used within its decision-making algorithms. 
    % \item Each prosumer in the aggregator’s portfolio is equipped with a \gls{pv} system and a residential battery operated by a \gls{hems} algorithm (Watts' Homegrid \textregistered~\cite{watts_homegrid}), with all relevant nameplate parameters known to the aggregator. Additionally, the aggregator uses predictive models that generate point forecasts for the aggregated prosumer electric load and solar generation.
    % \item All prosumers are subject to an hourly electricity price in real time (Tariff ``C''). Under this tariff, prosumers receive the \gls{dam} prices when exporting energy to the grid, while energy imports are priced by adding the retailer’s premium, grid tariffs and VAT to the \gls{dam} price. The aggregator is assumed to have access to point forecasts of these prices.
    \item Distribution grid related constraints, such as congestion, are ignored within this framework. However, the model can be adapted to account for the same. 
    \item The available up- and down-flexibility of the prosumer cluster is uncertain and are characterised using probabilistic forecast models. These models are further described in Section~\ref{sec: methodology}.
\end{itemize}

Let $\mathcal{T} = \{1, \dots, T\}$ denote the set of time intervals considered in the optimisation horizon, where $T \in \mathbb{N}$. The price profiles of the \gls{mfrr} capacity market for up- and down-regulation are indicated by $\boldsymbol{\lambda}^{\text{U}}, \boldsymbol{\lambda}^{\text{D}} \in \mathbb{R}^{T}$ (DKK/MWh), while $\boldsymbol{\lambda}^{\text{B}} \in \mathbb{R}^{T}$ represents the imbalance price profile of the \gls{mfrr} activation market (DKK/MWh). The corresponding activation signals from Energinet are represented by $\boldsymbol{a}^{\uparrow}, \boldsymbol{a}^{\downarrow} \in [0,1]^{T}$. Let $\boldsymbol{x}^{\text{u}}, \boldsymbol{x}^{\text{d}} \in \mathbb{R}_{+}^{T}$ denote the reserved up- and down-regulation flexibility (MWh), and $\boldsymbol{\lambda}^{\text{I,u}}, \boldsymbol{\lambda}^{\text{I,d}} \in \mathbb{R}_{+}^{T}$ the incentive price profiles offered to the prosumer group (DKK/MWh) to provide up- and down-flexibility, respectively. The random vectors $\boldsymbol{y}^{\text{u}}, \boldsymbol{y}^{\text{d}} \in \mathbb{R}_{+}^{T}$ represent the maximum up- and down-flexibility available from the prosumer group (MWh). Finally, let $\mathcal{F}(\boldsymbol{y}^{\text{u}}, \boldsymbol{y}^{\text{d}})$ denote a statistical or probabilistic measure associated with these random variables, such as the cumulative distribution function or relevant quantiles. Thus, the aggregator’s bidding optimisation problem can be formulated as follows.
\begin{subequations}
    \begin{align}
    \label{eq: ch5 objective value}
    \max_{\mathcal{X}} \quad & \sum_{t \in \mathcal{T}} \Big[\left(\lambda_{t}^{\text{U}} + a^{\uparrow}_{t}\lambda^{\text{B}}_{t} - \lambda_{t}^{\text{I,u}}\right) x^{\text{u}}_{t} 
    + \left(\lambda_{t}^{\text{D}} - a^{\downarrow}_{t}\lambda^{\text{B}}_{t} -  \lambda_{t}^{\text{I,d}}\right) x^{\text{d}}_{t}\Big] \\
    % \max_{\mathcal{X}}& \sum_{t \in \mathcal{T}} \left[\left(\lambda_{t}^{\text{U}} - \lambda_{t}^{\text{I,u}}\right) x^{\text{u}}_{t} 
    % + \left(\lambda_{t}^{\text{D}} - \lambda_{t}^{\text{I,d}}\right) x^{\text{d}}_{t} 
    % + \left(a^{\uparrow}_{t} x^{\text{u}}_{t} - a^{\downarrow}_{t} x^{\text{d}}_{t} \right) \lambda^{\text{B}}_{t} \right] \\
    %
    \label{eq: ch5 joint ccp}
    \text{s.t.} \quad & \mathbb{P} \left( x^{i}_{t} \leq y^{i}_{t}, \ \forall t \in \mathcal{T} \right) \geq 1 - \alpha 
    \quad \forall i \in \{\text{u}, \text{d}\} \\
    \label{eq: ch5 estimated flexibility}
    & \mathcal{F}(\boldsymbol{y}^{\text{u}}, \boldsymbol{y}^{\text{d}}) = \Phi \left( 
        \boldsymbol{\lambda}^{\text{I,u}},\ 
        \boldsymbol{\lambda}^{\text{I,d}},\ 
        \boldsymbol{\zeta}
    \right)
    \end{align}
\end{subequations}
\allowdisplaybreaks

The set of decision variables is defined as $\mathcal{X} = \{\boldsymbol{x}^{\text{u}},\ \boldsymbol{\lambda}^{\text{I,u}},\ \boldsymbol{y}^{\text{u}},\ \boldsymbol{x}^{\text{d}},\ \boldsymbol{\lambda}^{\text{I,d}},\ \boldsymbol{y}^{\text{d}}\}$. Equation~\eqref{eq: ch5 objective value} represents the aggregator’s profit function, while Equation~\eqref{eq: ch5 joint ccp} introduces a joint chance constraint that enforces the availability requirement of Energinet's P90 on the offered bid capacity. The statistical characteristics of the random variables $\mathcal{F}(\boldsymbol{y}^{\text{u}}, \boldsymbol{y}^{\text{d}})$ are modelled using a data-driven algorithm denoted by $\Phi(\cdot)$, as shown in Eq.~\eqref{eq: ch5 estimated flexibility}. Thus, the optimisation model is a decision-dependent chance-constrained program, as it captures the influence of incentive prices offered to prosumers, $\boldsymbol{\lambda}^{\text{I,u}}$ and $\boldsymbol{\lambda}^{\text{I,d}}$ on the uncertain parameters. Moreover, it incorporates contextual information $\boldsymbol{\zeta}$, such as forecasts of load, solar generation, retailer prices, market activation signals, and aggregated data from prosumer devices. It is important to note that this optimisation problem is solved by the aggregator, not by individual prosumers’ \gls{hems}. Although the focus of this study is \gls{mfrr} markets, this optimisation framework can in principle be extended to other Danish \gls{fcas} markets. However, this generalisability is not evaluated here.

The proposed optimisation model simultaneously determines the bid submitted to the \gls{mfrr} market and the incentive offered to prosumers. Consequently, the objective function is bilinear. Owing to this bilinear structure and the potentially nonlinear nature of the probabilistic flexibility model, the resulting optimisation problem is non-convex, and subsequently computationally challenging to solve. To improve tractability, prosumer incentive prices are treated as fixed parameters rather than decision variables. This removes bilinear terms from the objective and allows flexibility estimates to be obtained externally via an estimator $\Phi_{1-\alpha}\!\left(\boldsymbol{\lambda}^{\text{I,u}},\boldsymbol{\lambda}^{\text{I,d}},\boldsymbol{\zeta}\right)=\left(\boldsymbol{\hat{y}}^{\text{u}},\boldsymbol{\hat{y}}^{\text{d}}\right),$ where $\boldsymbol{\hat{y}}^{\text{u}}, \boldsymbol{\hat{y}}^{\text{d}} \in \mathbb{R}^{T}$ are calibrated to satisfy the chance constraint in Eq.~\eqref{eq: ch5 joint ccp}. The optimisation is evaluated over multiple incentive-price scenarios, and the most profitable case is selected.

Logically, incentive prices must remain below market clearing prices for the aggregator to be profitable. A proportional revenue-sharing mechanism is therefore adopted, where $\beta \in [0,1]$ denotes the fraction of market revenue allocated to prosumers. Discretising $\beta$ yields a finite set of incentive scenarios $\mathcal{S}_{\beta}$. The resulting linearised optimisation model is presented in Eqs.~\eqref{eq: ch5 objective value 2}--\eqref{eq: ch5 joint ccp 2}.

\begin{subequations}
    \begin{align}
    \label{eq: ch5 objective value 2}
    \max_{\mathcal{X}} \quad & \sum_{t \in \mathcal{T}} \Big[\left(\lambda_{t}^{\text{U}} + a^{\uparrow}_{t}\lambda^{\text{B}}_{t} - \lambda_{t}^{\text{I,u}}\right) x^{\text{u}}_{t} 
    + \left(\lambda_{t}^{\text{D}} - a^{\downarrow}_{t}\lambda^{\text{B}}_{t} -  \lambda_{t}^{\text{I,d}}\right) x^{\text{d}}_{t}\Big] \\
    \label{eq: ch5 joint ccp 2}
    \text{s.t.} \quad & x^{i}_{t} \leq \hat{y}^{i}_{t}, \quad \forall t \in \mathcal{T},\, \forall i \in \{\text{u}, \text{d}\}\\
    \label{eq: up-regulation incentive}
    {} \quad & \lambda_{t}^{\text{I,u}} = \beta\left(\lambda_{t}^{\text{U}} + a^{\uparrow}_{t}\lambda^{\text{B}}_{t}\right) \quad \forall t \in \mathcal{T}\\
    \label{eq: down-regulation incentive}
    {} \quad & \lambda_{t}^{\text{I,d}} = \beta\left(\lambda_{t}^{\text{D}} - a^{\downarrow}_{t}\lambda^{\text{B}}_{t}\right) \quad \forall t \in \mathcal{T}
    \end{align}
\end{subequations}

The resulting simplified problem is straightforward to solve, allowing the aggregator to efficiently assess multiple incentive pricing scenarios and identify the one that maximises profit. However, by fixing incentive prices ex ante and decoupling them from the flexibility predictor $\Phi_{1-\alpha}(\cdot)$ within the optimisation, the identified maximum profit solution becomes inherently sub-optimal. The extent of this sub-optimality depends on the resolution and number of scenarios considered, reflecting a deliberate trade-off between computational tractability and solution optimality. While a revenue-sharing incentive mechanism is adopted in this study, the proposed optimisation framework is sufficiently general to support alternative incentive design strategies.
%------------------------------------
%------------------------------------
\section{Methodology}
\label{sec: methodology}
In this section, we describe the \gls{uq} approaches evaluated in this study to capture uncertainty in the behaviour of prosumers, that is, the flexibility prediction algorithm $\Phi_{1-\alpha}(\cdot)$. As noted earlier, the prediction made by the algorithm should guarantee satisfaction of the joint chance constraint. As introduced in Section~\ref{sec: intro}, a hybrid \gls{mcd} and \gls{cp} method is used to achieve this. The study also considers \gls{cqr}, whose multivariate formulation follows the approach described in~\cite{cp_algo}. Since the underlying \gls{cqr} methodology is well established, its mathematical formulation is not reproduced here; its implementation using a separately trained quantile regression model is detailed in Section~\ref{sec: case study}. The \gls{mcd}-\gls{cp} approach and its associated multivariate conformity scores are described in the following sections.
%----------------------------
\subsection{MCD--CP Preliminaries}
\label{ssec: uq framework}
Traditionally, dropout layers in \glspl{nn} are used to prevent overfitting by probabilistically ``switching off'' neurons during training using a user-defined dropout rate. These layers are typically disabled during inference. However, \cite{dropout_Yarin} showed that a \gls{nn} with dropout across all weight layers can be viewed as a probabilistic deep Gaussian process. By retaining dropout during inference and performing multiple stochastic forward passes, one can approximate samples from a deep Gaussian process. This approach, known as \gls{mcd}, offers a scalable and computationally efficient alternative to \glspl{bnn}, especially when the uncertainty structure is not highly complex. On the other hand, \gls{cp} is a distribution-free framework for uncertainty quantification that constructs statistically valid prediction sets under the assumption of data exchangeability~\cite{cp_vovk, cp_angelopolous}. It is simple, model-agnostic, and computationally efficient. Among its variants, this study adopts the split \gls{cp}.

Being a ``Bayesianesque'' approximation, \gls{nn} with \gls{mcd} effectively captures epistemic (model-related) uncertainty~\cite{dropout_Yarin}, though it may exhibit poor calibration for complex posteriors. Integrating it with \gls{cp} enhances uncertainty reliability while retaining tractability. Moreover, the marginal coverage guarantee of \gls{cp} allows the multivariate prediction region $\xi$ to be interpreted as element-wise upper and lower bounds that satisfy joint chance constraints, which makes this hybrid framework well-suited to our application. The Algorithm~\ref{alg: multivariate mcd-cp} defines the probabilistic prediction framework $\Phi_{1 - \alpha}(\cdot)$ used for uncertainty quantification. Various conformity scores can be adopted for multivariate predictions; see~\cite{cp_algo} for details. The scores used in this study are specified in Section~\ref{ssec: ch5 cp metrics}.

\allowdisplaybreaks
\begin{algorithm}[t]
\caption{Hybrid \gls{mcd} and split \gls{cp} for obtaining multivariate prediction intervals}
\label{alg: multivariate mcd-cp}
\begin{algorithmic}[1]
\Require \gls{nn} with dropout layers $f_{y}$ trained on $\mathcal{D}_{\text{train}}$, number of MC samples $S$, calibration set $\mathcal{D}_{\text{cal}}$, new input $\boldsymbol{\Theta} \not\subset \mathcal{D}_{\text{train}} \cup \mathcal{D}_{\text{cal}}$, coverage level $1 - \alpha$
\Ensure Prediction set $\boldsymbol{\xi}$
\Statex \textbf{CP calibration:}
\State Initialise calibration score vector $\mathbf{q} \in \mathbb{R}^{|\mathcal{D}_{\text{cal}}| + 1}$
\For{$j = 1$ to $|\mathcal{D}_{\text{cal}}|$}
\Statex \textbf{\gls{mcd} algorithm start}
\State Set model $f_{y}$ to \textbf{train mode} to activate dropout at inference
\State Initialise prediction matrix $\mathbf{Y}^{(j)} \in \mathbb{R}^{S \times d}$ where $d$ is the output dimension
    \For{$i = 1$ to $S$}
        \State $\boldsymbol{\hat{y}^{(i)}} \leftarrow f_{y}\left(\boldsymbol{\Theta}^{(j)}\right) \in \mathbb{R}^{d}$ \Comment{Vector output}
        \State Store $\boldsymbol{\hat{y}^{(i)}}$ as row $i$ in matrix $\mathbf{Y}^{(j)}$
    \EndFor  
\Statex \textbf{\gls{mcd} algorithm end}
    \State $q_j \leftarrow s\left(\mathbf{Y}^{(j)}, \boldsymbol{y}^{(j)}\right) \in \mathbb{R}$
\EndFor
\State Compute the adjusted quantile level:
\[
\hat{\alpha} = \frac{\lceil (|\mathcal{D}_{\text{cal}}| + 1)(1 - \alpha) \rceil}{|\mathcal{D}_{\text{cal}}| + 1}
\]
\State Compute the $\hat{\alpha}$ quantile of the calibration scores:
\[
\hat{q} = \text{Quantile}_{\hat{\alpha}}\left(\mathbf{q} \cup \{\infty\}\right) \in \mathbb{R}
\]
\Statex \textbf{\gls{cp} Inference:}
\State For a new random input $\boldsymbol{\Theta}$, compute \gls{mcd} sample predictions $\mathbf{Y} \in \mathbb{R}^{S \times d}$ using steps 3 -- 8 
\State Construct the prediction region of the output $\boldsymbol{y}$ as:
\[
\boldsymbol{\xi} = \left\{ \boldsymbol{y} \in \mathbb{R}^d : s\left(\mathbf{Y}, \boldsymbol{y} \right) \leq \hat{q} \right\}
\]
\\
\Return $\boldsymbol{\xi}$
\end{algorithmic}
\end{algorithm}

Marginal coverage ensures that, on average, the prediction region contains the true output with probability \(1 - \alpha\), though not necessarily for each individual input (conditional coverage). Thus, while some samples may be over- or under-covered, the expected coverage remains correct~\cite{cp_barber}. Although conditional coverage is generally unattainable without distributional assumptions, marginal coverage can be interpreted to yield conservative element-wise bounds on $\boldsymbol{y}$, as shown in Eqs.~\eqref{eq: ch5 cp region}–\eqref{eq: ch5 cp-ccp region}, where $\boldsymbol{u} = \sup \xi$ and $\boldsymbol{l} = \inf \xi$.
\begin{align}
    \label{eq: ch5 cp region}
    {} & \mathbb{P}\left[\boldsymbol{y} \in \xi \right] \geq 1-\alpha \\
    \label{eq: ch5 infinmum}
    {} & \mathcal{B}_{\xi} = \left\{ \boldsymbol{y} \in \mathbb{R}^d \;\middle|\; \boldsymbol{l} \leq \boldsymbol{y} \leq \boldsymbol{u}\right\}\\ 
    \label{eq: ch5 cp-ccp region}
    \therefore\quad & \mathbb{P}\left[\boldsymbol{y} \in \mathcal{B}_{\xi}\right] \geq \mathbb{P}\left[\boldsymbol{y} \in \xi \right] \geq 1- \alpha
\end{align}

\begin{figure}[htb]
    \centering
    \includegraphics[width=0.80\linewidth]{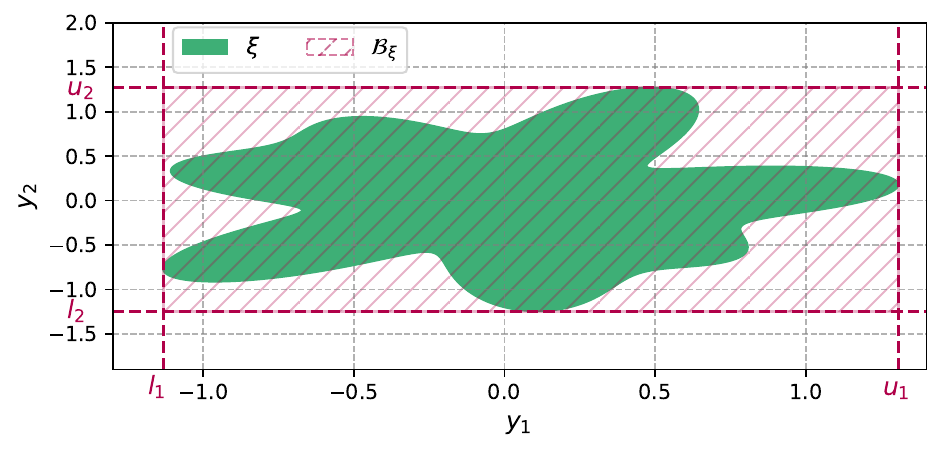}
    \caption{Example conformal prediction region $\xi$ for $\boldsymbol{y} \in \mathbb{R}^{2}$ with bounding rectangle $\mathcal{B}_{\xi}$}
    \label{fig: ch5 bounding box}
\end{figure}

Figure~\ref{fig: ch5 bounding box} illustrates this concept, where a bounding hyperrectangle $\mathcal{B}_{\xi}$ encloses the conformal prediction region $\xi$. By construction, $\mathbb{P}[\boldsymbol{y} \in \mathcal{B}_{\xi}] \geq 1 - \alpha$. These conservative bounds are particularly advantageous for downstream optimisation problems involving joint chance constraints, such as those defined in Eqs.~\eqref{eq: ch5 objective value}--\eqref{eq: ch5 estimated flexibility}. For example, in a chance-constrained optimisation problem with decision variables $\boldsymbol{x} \in \mathbb{R}^{d}$ and random vector $\boldsymbol{y}$ as defined above, the probabilistic constraint $\mathbb{P}\!\left[x_{i} \leq y_{i}, \ \forall i \in \{1, \dots, d\}\right] \geq 1 - \alpha$ can be conservatively reformulated as the deterministic constraint $\boldsymbol{x} \leq \boldsymbol{l}$. Similarly, the probabilistic constraint $\mathbb{P}\!\left[x_{i} \geq y_{i}, \ \forall i \in \{1, \dots, d\}\right] \geq 1 - \alpha$ can be conservatively reformulated as the deterministic constraint $\boldsymbol{x} \geq \boldsymbol{u}$. 
%----------------------------
%----------------------------
\subsection{Conformity scores}
\label{ssec: ch5 cp metrics}
As discussed in Section~\ref{ssec: uq framework}, the choice of the conformity score critically influences the quality and geometry of the multivariate prediction intervals generated by the evaluated \gls{uq} framework (Algorithm~\ref{alg: multivariate mcd-cp}). Assuming the dataset is exchangeable, any conformal prediction method guarantees marginal coverage; however, maintaining narrow prediction intervals may be essential to maximise downstream profitability.  

This section evaluates three conformity metrics within the split \gls{cp} framework (Algorithm~\ref{alg: multivariate mcd-cp}), along with a copula-based \gls{cp} approach commonly used in multivariate forecasting. Other conformity measures, such as those based on cumulative distributions or latent representations, are beyond the present scope but are discussed in~\cite{cp_algo}. The \gls{mcd} approach can be integrated with any of these methods, highlighting its flexibility in generating diverse statistical representations as precursors to \gls{cp}. The notation of Algorithm~\ref{alg: multivariate mcd-cp} is retained, with subscripts used to distinguish the prediction regions for each metric.

The first conformity metric, termed \gls{mcp}, follows~\cite{cp_algo, m_cp_ref}. Quantiles derived from \gls{mcd}-based predictions are calibrated using the conformity score defined in~\cite{cp_algo}, and the corresponding conformalised prediction interval $\xi_{MCP}$ is obtained using Eq.~\eqref{eq: ch5 mcp region}. 
\begin{align}
    \label{eq: ch5 mcp region}
    \xi_{\text{MCP}} = \text{\Large$\times\text{\normalsize}^{d}_{t=1}$}
    \left[
        \mathcal{Q}_{t}\!\left(\mathbf{Y}, \alpha\right) - \hat{q},\,
        \mathcal{Q}_{t}\!\left(\mathbf{Y}, 1-\alpha\right) + \hat{q}
    \right]
\end{align}

Here, $\mathcal{Q}_{t}(\cdot,\tau)$ denotes the $\tau$-quantile of the predicted samples at index $t$. The conformity score measures the maximum deviation of the true output from the predicted quantile band. Since the $\alpha$ to $1-\alpha$ quantile interval inherently provides coverage of $1-2\alpha$, Algorithm~\ref{alg: multivariate mcd-cp} calibrates $\hat{q}$ to achieve this target coverage, i.e., $\mathbb{P}\!\left[\boldsymbol{y} \in \xi_{\text{MCP}}\right] \geq 1 - 2\alpha$. Although the lower bound $\inf \xi_{\text{MCP}}$ is expected to satisfy $\mathbb{P}\!\left[\boldsymbol{y} \geq \inf \xi_{\text{MCP}}\right] \geq 1 - \alpha$, this may not always hold in practice, making \gls{mcp} less conservative than other approaches and potentially non-compliant with the P90 standard(see Section~\ref{ssec: ch5 uq performance}).

The second metric is based on the univariate problem specified in~\cite{cp_metric_mcdcp}, which we generalised to a multivariate setting. It uses the  univariate Mahalanobis distance to quantify the deviation of the true output from the predictive mean obtained from \gls{mcd} samples, expressed in units of the sampled standard deviation. The corresponding prediction interval is defined as the predictive mean plus or minus a scaled deviation, as shown in Eq.~\eqref{eq: ch5 mmcp region}.
\begin{subequations}
\begin{align} \label{eq: ch5 predictive mu} {} & \mu^{(j)}_{t} = \frac{1}{\mathcal{S}}\sum_{i=1}^{\mathcal{S}}Y_{i, t}^{(j)} \quad \forall t \in \{1, \dots, d\}\\ % 
\label{eq: ch5 predictive sigma} {} & \sigma^{(j)}_{t} = \sqrt{\frac{1}{\mathcal{S}}\sum_{i=1}^{\mathcal{S}}\left[Y_{i, t}^{(j)} - \mu^{(j)}_{t}\right]^{2}} \quad \forall t \in \{1, \dots, d\} \\
\label{eq: ch5 mmcp} {} & s_{MMCP}(\mathbf{Y}^{(j)},\boldsymbol{y^{(j)}})= \max_{t}\left\{\frac{y_{t}^{(j)}-\mu^{(j)}_{t}}{\sigma^{(j)}_{t}}\right\} \\ 
\label{eq: ch5 mmcp region} {} & \xi_{MMCP} = \text{\Large$\times\text{\normalsize}^{d}_{t=1}$} \left[\mu_{t} - \hat{q}\cdot\sigma_{t}, \mu_{t} + \hat{q}\cdot\sigma_{t} \right] 
\end{align}
\end{subequations}

We refer to this conformity score as \gls{mmcp}, which computes Mahalanobis distances independently for each output and retains the maximum distance across all outputs. This yields an axis-aligned hyperrectangular prediction region $\xi_{\text{MMCP}}$, equivalent to its bounding box $\mathcal{B}{\xi{\text{MMCP}}}$. The calibration threshold $\hat{q}$ is obtained using Algorithm~\ref{alg: multivariate mcd-cp} for coverage $1-\alpha$, using $\boldsymbol{\mu}$ and $\boldsymbol{\sigma}$ from~\eqref{eq: ch5 predictive mu}–\eqref{eq: ch5 predictive sigma} and the conformity score presented in Eq.~\eqref{eq: ch5 mmcp}.

The third metric, termed \gls{pcp} and proposed in~\cite{cp_metric_pcp}, determines the radius of the smallest hypersphere enclosing the true output, with each predicted sample from the \gls{mcd} sample serving as a centre, as shown in Eq.~\eqref{eq: ch5 pcp region}. 
\begin{align}
    \label{eq: ch5 pcp region}
    {} & \xi_{PCP} = \bigcup_{i \in \mathcal{S}} \left\{y : \left\|\boldsymbol{y} - \mathbf{Y}_{i}\right\|_{2} \leq \hat{q} \right\}
\end{align}

The prediction region comprises a union of hyperspheres of radius $\hat{q}$, where $\hat{q}$ is the calibrated conformity score for coverage $1-\alpha$. The corresponding axis-aligned bounding hyperrectangle is typically the most conservative among all methods.

The final method, \gls{ccp}, extends the copula-based approach of~\cite{cp_metric_copula}, originally developed for multi-step time series forecasting. Unlike the marginal calibration of the individual output dimensions used in the other split-\gls{cp} approaches considered here, \gls{ccp} explicitly models dependencies among the output residuals using copulas, thereby accounting for temporal and inter-dimensional correlations. The calibration yields a vector of thresholds $\mathbf{\hat{q}_{CCP}} \in \mathbb{R}^{T}$ that define the prediction region:
\begin{align}
    \label{eq: ch5 copula region}
    {} & \xi_{CCP} = \text{\Large$\times\text{\normalsize}^{d}_{t=1}$} \left[\mu_{t} - \hat{q}_{CCP, t}, \mu_{t} + \hat{q}_{CCP, t}\right]
\end{align}

Here, the conformity score is the absolute deviation between the true value and sample predictive mean from \gls{mcd}. Each dimension is independently calibrated using the corresponding element of $\mathbf{\hat{q}_{CCP}}$, which produces a hyperrectangular prediction region $\xi_{CCP}$ equivalent to its bounding box $\mathcal{B}_{\xi_{CCP}}$. The implementation follows the official repository in~\cite{copula_github}.

\begin{figure}[!htpb]
    \centering
    \begin{subfigure}[b]{0.45\linewidth}
        \centering
        \includegraphics[width=\linewidth]{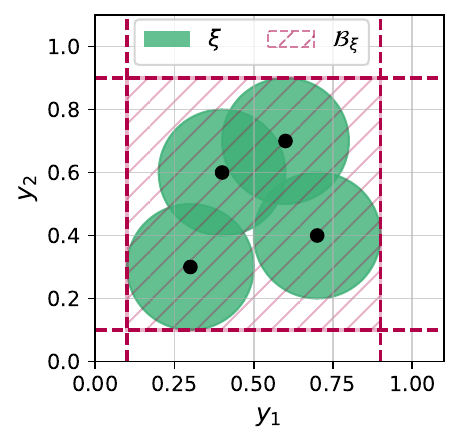}
        \subcaption{Example $\xi$ for PCP with $\mathcal{S}=4$}
        \label{fig: example A}
    \end{subfigure}
    \hfill
    \begin{subfigure}[b]{0.45\linewidth}
        \centering
        \includegraphics[width=\linewidth]{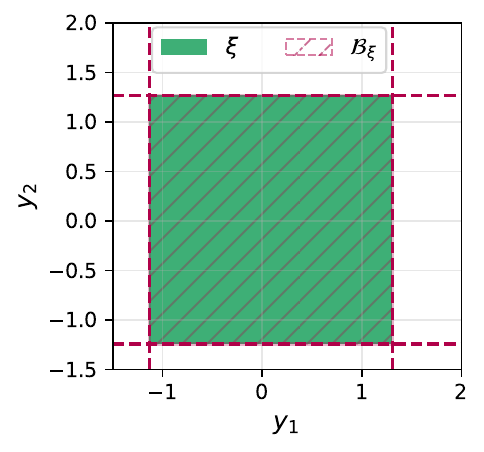}
        \subcaption{Example $\xi$ for CQR, MCP, MMCP and CCP}
        \label{fig: example B}
    \end{subfigure}
    \caption{Illustration of prediction regions $\xi$ and corresponding bounding hyperrectangles for $\boldsymbol{y} \in \mathbb{R}^{2}$, across the evaluated \gls{cp} methods.}
    \label{fig: Sample xi}
\end{figure}

As shown in Fig.~\ref{fig: Sample xi}, the bounding box of the \gls{pcp} method provides a conservative approximation, whereas for \gls{mcp}, \gls{mmcp}, and \gls{ccp}, the bounding box coincides with the prediction region.
%----------------------------
%----------------------------
%------------------------------------
%------------------------------------
\section{Case study}
\label{sec: case study}
This section outlines the simulation case study and the methodology adopted for training the probabilistic prediction algorithm $\Phi_{1-\alpha}$. 

In emerging flexibility markets, empirical data describing the behaviour of \gls{hems}-equipped prosumers may be limited or unavailable, particularly when an aggregator is evaluating a prospective business model before large-scale deployment. In such settings, modelling can provide a practical means of assessing the technical and economic feasibility of alternative business strategies before committing resources to field deployment and data collection. Accordingly, this study uses a modified version of the industry-grade \gls{hems} algorithm originally developed by the project’s industry partner, Watts~A/S, as detailed in~\cite{jullian_thesis}, to generate a controlled representation of aggregated prosumer flexibility. The baseline algorithm was subsequently modified to allow participation in the \gls{mfrr} up-regulation capacity market. As the algorithm is proprietary, its complete formulation cannot be disclosed; however, its essential characteristics and underlying assumptions, including those beyond~\cite{jullian_thesis}, are summarised as follows:
\begin{enumerate}
    % \item The algorithm minimises electricity costs while maximising self-consumption of locally generated solar energy.
    % \item Battery degradation due to cycling is modelled using a cycling cost.
    \item Each prosumer in the aggregator’s portfolio is equipped with a \gls{pv} system and a residential battery operated by a \gls{hems} algorithm (Watts' Homegrid \textregistered~\cite{watts_homegrid}), with all relevant nameplate parameters known to the aggregator. Additionally, the aggregator uses predictive models that generate point forecasts for the aggregated prosumer electric load and solar generation.
    \item All prosumers are subject to an hourly electricity price in real time (Tariff ``C''). Under this tariff, prosumers receive the day-ahead market prices when exporting energy to the grid, while energy imports are priced by adding the retailer’s premium, grid tariffs and VAT to the day-ahead market price. The aggregator is assumed to have access to point forecasts of these prices.
    \item Flexibility is provided exclusively by the prosumer’s battery. During periods reserved for capacity provision, the battery is scheduled to remain idle; thus, any deviation from this idle state indicates an external activation signal, which can be used to confirm the delivery of service. 
    \item The battery model does not explicitly account for degradation due to cycling or temperature-dependent variations in battery efficiency. The results should therefore be interpreted within the scope of the battery model implemented in the Watts \gls{hems}.
    \item The current \gls{hems} configuration supports participation only in the \gls{mfrr} up-regulation capacity market, and therefore this study is restricted accordingly. In addition, it uses deterministic forecasts for activation signals and electricity prices, without explicitly accounting for uncertainty.
    \item In case of reserve activation, prosumers are remunerated solely by the aggregator for their flexibility. The corresponding load deviation does not affect the electricity bill from the retailer, reflecting a regulatory assumption that dual compensation from both the aggregator and the retailer may be disallowed in the future.
\end{enumerate}

Since the \gls{hems} supports only \gls{mfrr} up-regulation capacity market participation, the aggregator’s participation is also limited to this market. This choice is motivated by the substantially higher average hourly procurement volume in the up-regulation market between August 2023 and April 2025, which is approximately 390~MW, compared to only 35~MW for down-regulation in Denmark, as obtained from~\cite{energidataservice_api_guide}. This market-specific constraint does not limit the applicability of the \gls{uq} framework introduced in Section~\ref{ssec: uq framework}, which remains generalisable to other \gls{fcas} market formulations. However, it directly affects the bidding optimisation model of the aggregator, as described in Section~\ref{sec: problem}, and is incorporated in Eqs.~\eqref{eq: ch5 objective value 3} and~\eqref{eq: ch5 joint ccp 3}. The prediction region $\xi$ is derived using the evaluated \gls{uq} framework described in Section~\ref{sec: methodology}, ensuring that $\mathbb{P}\left[\boldsymbol{y}^{u} \in \xi\right] \geq 1-\alpha$, with $\boldsymbol{l} = \inf \xi$. 
\begin{subequations}
    \begin{align}
    \label{eq: ch5 objective value 3}
    \max_{\mathcal{X}} \quad & \sum_{t \in \mathcal{T}} \left(1 -\beta \right)\lambda_{t}^{\text{U}} x^{\text{u}}_{t} \\
    \label{eq: ch5 joint ccp 3}
    \text{s.t.} \quad & x^{u}_{t} \leq l_{t}, \quad \forall t \in \mathcal{T}
    \end{align}
\end{subequations}

Finally, the modified Watts \gls{hems} algorithm is employed to generate synthetic data representing aggregated prosumer behaviour. This dataset forms the basis for training the probabilistic prediction model, $\Phi_{1-\alpha}(\cdot)$, using Algorithm~\ref{alg: multivariate mcd-cp}. It should be emphasised that the model aims to capture the aggregate flexibility of a population of prosumers, rather than individual household behaviour.
%-------------------------------------------------------------------
\subsection{Data generation for training $\Phi_{1-\alpha}(\cdot)$}
\label{ssec: ch5 data generation}
\allowdisplaybreaks
The input data for the \gls{hems} are categorised into common inputs at the cluster-level and individual prosumer-specific inputs. The common inputs comprise the predicted daily electricity purchase and sale price profiles, $\boldsymbol{\lambda}^{\text{P}}, \boldsymbol{\lambda}^{\text{S}} \in \mathbb{R}^{T}$ (DKK/kWh), the predicted \gls{mfrr} up-regulation activation signal, $\boldsymbol{a}^{\uparrow} \in \{0,1\}^{T}$, and a common up-regulation incentive price offered by the aggregator, $\boldsymbol{\lambda}^{\text{I,u}} \in \mathbb{R}^{T}$. Although fractional activations would more accurately represent partial reserve deployment, the activation signal is modelled as binary in this study due to data availability constraints. Let $\mathcal{I}$ denote the set of prosumers. For each prosumer $i \in \mathcal{I}$, the individual inputs include the predicted daily load and solar generation profiles, $\boldsymbol{x}^{\text{l},i}, \boldsymbol{x}^{\text{s},i} \in \mathbb{R}^{T}$ (kWh), and the battery specifications, characterised by the maximum discharge power $b^{\text{dis},i} \in \mathbb{R}^{+}$ (kW) and the energy capacity $b^{\text{cap},i} \in \mathbb{R}^{+}$ (kWh). Accordingly, the input set for the \gls{hems} of prosumer $i$ is defined as $\Theta^{i} = \{\boldsymbol{\lambda}^{\text{P}}, \boldsymbol{\lambda}^{\text{S}}, \boldsymbol{a}^{\uparrow}, \boldsymbol{\lambda}^{\text{I,u}}, \boldsymbol{x}^{\text{l},i}, \boldsymbol{x}^{\text{s},i}, b^{\text{dis},i}, b^{\text{cap},i}\}$. Based on this input set, the \gls{hems} determines the daily up-regulation capacity reserved by prosumer $i$, denoted by $\boldsymbol{y}^{\text{u},i} \in \mathbb{R}^{T}$ (kW).

The aggregator is primarily interested in the collective response of the prosumer cluster rather than individual \gls{hems} decisions. Consequently, the inputs to the \gls{ml} model mirror those of the \gls{hems}, with load profiles and battery specifications aggregated across all prosumers. Individual quantities are summed to form cluster-level inputs, and the same aggregation is applied to the outputs, producing a single daily \gls{mfrr} up-regulation capacity reservation profile. The input to the \gls{ml} model is therefore defined as $\Theta = \{ \boldsymbol{\lambda}^{\text{P}}, \boldsymbol{\lambda}^{\text{S}}, \boldsymbol{a}^{\uparrow}, \boldsymbol{\lambda}^{\text{I,u}},
\boldsymbol{x}^{\text{l,agg}}, \boldsymbol{x}^{\text{s,agg}},b^{\text{dis,agg}}, b^{\text{cap,agg}}\}$, where the superscript ``agg'' denotes quantities aggregated over all prosumers. The corresponding output is the aggregated daily \gls{mfrr} up-regulation capacity reservation profile, denoted by $\boldsymbol{y} = \boldsymbol{y}^{\text{u,agg}}$.

\begin{figure}[htpb!]
    \centering
    \includegraphics[
        trim=0pt 120pt 0pt 198pt,
        clip,
        width=\linewidth
    ]{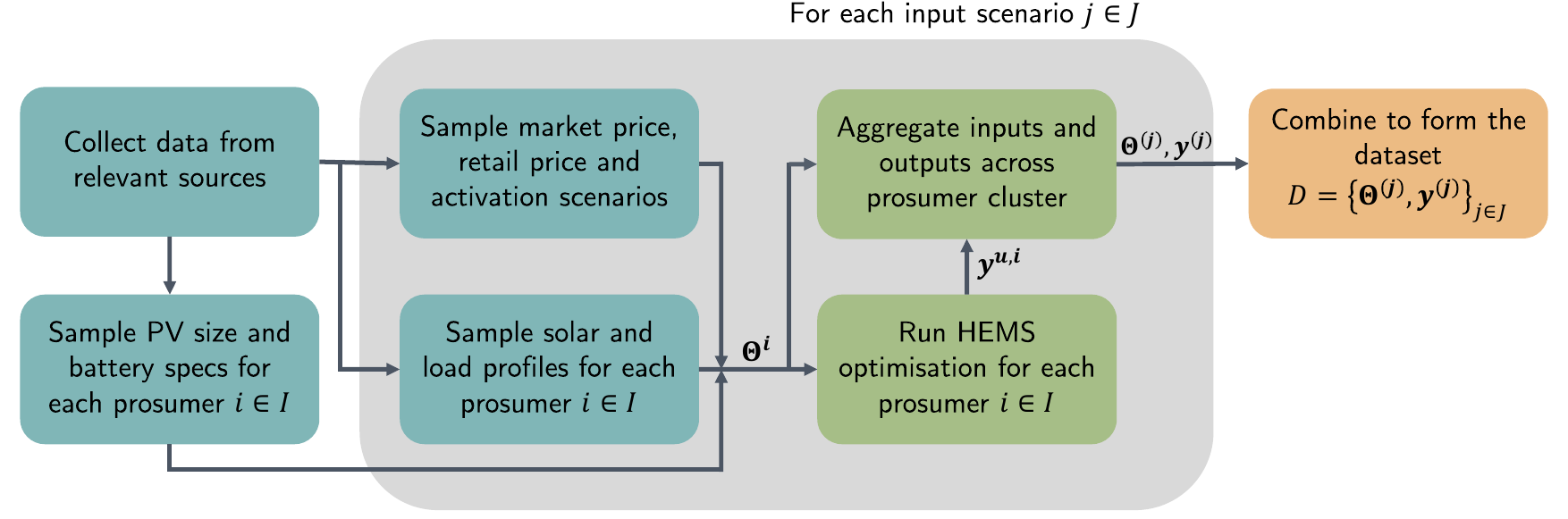}
    \caption{Synthetic data generation framework for aggregated HEMS flexibility}
    \label{fig: data gen block diagram}
\end{figure}
Figure~\ref{fig: data gen block diagram} illustrates the data generation workflow used to construct the training dataset to learn the aggregated behaviour of the \gls{hems}. The process begins with the collection of data from relevant sources, including electricity market prices, retail tariffs, and \gls{mfrr} activation signals. Prosumer-specific technology characteristics, namely photovoltaic system sizes and battery specifications, are sampled once for each prosumer $i \in \mathcal{I}$ and remain fixed across all scenarios, representing a static physical asset configuration. In contrast, market price trajectories, activation patterns, and daily load and solar generation profiles are sampled on a per-scenario basis to capture temporal and market-driven variability. For each scenario, these quantities are combined to form individual \gls{hems} input sets $\Theta^{i}$.

Given these inputs, the \gls{hems} optimisation problem is solved independently for each prosumer, resulting in the corresponding daily up-regulation capacity reservation $\boldsymbol{y}^{\text{u},i}$. Individual inputs and outputs are then aggregated across the prosumer cluster: load profiles, solar generation, and battery parameters are summed to form the input at the cluster-level $\boldsymbol{\Theta}$, while individual capacity reservations are aggregated to produce the corresponding output at the cluster-level $\boldsymbol{y} = \boldsymbol{y}^{\text{u,agg}}$. Repeating this procedure across multiple sampled scenarios indexed by $j \in \mathcal{J}$ results in the dataset $\mathcal{D} = \{(\boldsymbol{\Theta}^{(j)}, \boldsymbol{y}^{(j)})\}_{j \in \mathcal{J}}$. Using this data synthesis strategy, a dataset of size $|\mathcal{J}| = 22{,}000$ was generated for a prosumer cluster of size $|\mathcal{I}| = 100$, resulting in $2.2 \times 10^{6}$ individual \gls{hems} simulations. The dataset $\mathcal{D}$ is exchangeable, as each scenario depends only on its own inputs, and is used exclusively to train, calibrate and evaluate the \gls{ml} surrogate model $\Phi_{1-\alpha}(\cdot)$ following the training procedure described in Section~\ref{ssec: ch5 ml training}.

To evaluate the aggregator business model under the revenue-sharing incentive design defined in Eqs.~\eqref{eq: ch5 objective value 3} and~\eqref{eq: ch5 joint ccp 3}, an additional set of 100 input scenarios was generated using the same data collection and sampling strategy described in~\ref{appendix 1}, and the same set of prosumers with photovoltaic and battery parameters consistent with those used in dataset $\mathcal{D}$. Unlike $\mathcal{D}$, where incentive prices were generated using a randomly sampled time-varying incentive factor multiplied by the up-regulation market price, the incentive price in this evaluation was defined using a fixed and time-invariant incentive factor such that $\lambda_{t}^{\text{I,u}} = \beta \lambda_{t}^{\text{U}}$ for all $t \in \mathcal{T}$, with $\beta \in \mathcal{S}_{\beta} = \{0.1, 0.2, \dots, 0.9\}$. Each scenario was used to solve the \gls{hems} optimisation problem and subsequently aggregated in the same manner as before to construct a family of datasets $\mathcal{D}_{\beta}$ of size 100, for all $\beta \in \mathcal{S}_{\beta}$. These datasets are used exclusively to evaluate the economic performance of each \gls{uq} method to support bidding decisions under uncertainty. The complete datasets used in this study are available at~\cite{Pipada2025SyntheticData}. 

\subsection{$\Phi_{1-\alpha}(\cdot)$ training}
\label{ssec: ch5 ml training}
The \gls{ml} model $\Phi_{1-\alpha}(\cdot)$ is trained on dataset $\mathcal{D}$ following the procedure outlined in Algorithm~\ref{alg: multivariate mcd-cp}. The dataset is partitioned into a training set $\mathcal{D}_{\text{train}}$ (10,000 samples), a calibration set $\mathcal{D}_{\text{cal}}$ (2,000 samples), and a test set $\mathcal{D}_{\text{test}}$ (10,000 samples). A relatively large test set is employed to ensure a low-variance and statistically reliable evaluation of coverage performance. The training set is used to train the \gls{mcd}-based model and the separate \gls{qr} model, while the calibration set is used to calibrate the prediction intervals obtained from both the \gls{mcd}-\gls{cp} and \gls{cqr} approaches. For \gls{mcd}-\gls{cp}, each of the conformity metrics described in Section~\ref{ssec: ch5 cp metrics} is applied during calibration. For \gls{cqr}, the predicted lower and upper quantiles are calibrated using the \gls{mcp} conformity metric, following the multivariate formulation described in~\cite{cp_algo}. The test set is used for model and interval evaluation.

To accelerate training, min--max normalisation is applied to obtain \(\mathcal{D}^{'}\) and its subsets \(\mathcal{D}_{\text{train}}^{'}\), \(\mathcal{D}_{\text{cal}}^{'}\), and \(\mathcal{D}_{\text{test}}^{'}\). All inputs are min--max scaled except for the aggregated discharge power \(b^{\text{dis,agg}}\) and energy capacity \(b^{\text{cap,agg}}\), which are scaled by \(b^{\text{dis,agg}}\) to ensure physical consistency. Consequently, \(b^{\text{dis,agg}}\) is excluded from the final input features, as its effect is already embedded in the scaling. The normalised dataset therefore contains \(6T + 1\) input features and \(T\) output features. Although recent market regulations have reduced the market time unit from one hour to 15 minutes, the publicly available datasets used in this study are provided at hourly resolution. Accordingly, the market time unit is assumed to be one hour, and the scheduling horizon is set to $T = 24$.

The \gls{nn} $f_{\boldsymbol{y}}(\cdot)$ is trained on $\mathcal{D}_{\text{train}}^{'}$ and validated using an 80:20 split. Hyperparameter tuning is conducted with the Optuna framework~\cite{optuna} over 100 optimisation trials, identifying the optimal configuration of the network architecture, dropout rates, learning rate, activation function, and other training parameters based on validation loss. The final network is then trained using the selected parameters. Dropout is retained during inference to generate the stochastic prediction samples required by the \gls{mcd} technique. Concrete dropout~\cite{concretedropout}, which automatically learns dropout rates, was also tested but yielded lower predictive accuracy than the Optuna-tuned regular dropout model. Thus, the regular dropout-based model is used for the \gls{mcd}-based approach evaluated in this study.

For comparison, a separate \gls{qr} model is trained to estimate the lower and upper conditional quantiles of the aggregated flexibility response. To provide a controlled comparison with the \gls{mcd} model, the \gls{qr} model uses the same network architecture, comprising two hidden layers with 256 neurons and PReLU activation. Its training hyperparameters are independently tuned using the Optuna framework ~\cite{optuna}, with the pinball loss used as the validation objective. The resulting quantile estimates are conformalised using the multivariate \gls{cqr} formulation described in~\cite{cp_algo}. The lower and upper quantiles are set to $\tau=0.10$ and $\tau=0.90$, respectively. Unlike the \gls{mcd}-based model, dropout is disabled during inference and no stochastic prediction samples are generated. The training configurations of the \gls{mcd} and \gls{qr} models are summarised in Table~\ref{tab: ch5 nn hyperparams}.

\begin{table}[htpb!]
\centering
\caption{Neural network training hyperparameters}
\label{tab: ch5 nn hyperparams}
\begin{tabular}{lll}
\toprule
\textbf{Hyperparameter}        & \textbf{\gls{mcd} model} & \text{\gls{qr} model}                        \\ \midrule
Batch size                    & 32          &  16                       \\
Number of epochs              & 1000         &  1900                     \\
Optimiser                     & Adam          &    Adam                   \\
Learning rate                 & $3 \times 10^{-4}$  &  $4.3 \times 10^{-5}$               \\
Loss metric                   & mean square error & Pinball loss                          \\
Network architecture          & 2 hidden layers   & 2 hidden layers                  \\
Neurons per hidden layer      & 256   & 256                        \\
Hidden layer activation function & PReLU & PReLU\\
Output size & 24 & 48 \\
Output activation function    & Linear    &   Linear                        \\
Dropout probability (Layer 1) & 0.22       &   0.39                     \\
Dropout probability (Layer 2) & 0.16        &    0.11                    \\
\bottomrule
\end{tabular}
\end{table}

\begin{figure}[htpb!]
    \centering
    \begin{subfigure}{0.70\linewidth}
        \includegraphics[width=\linewidth]{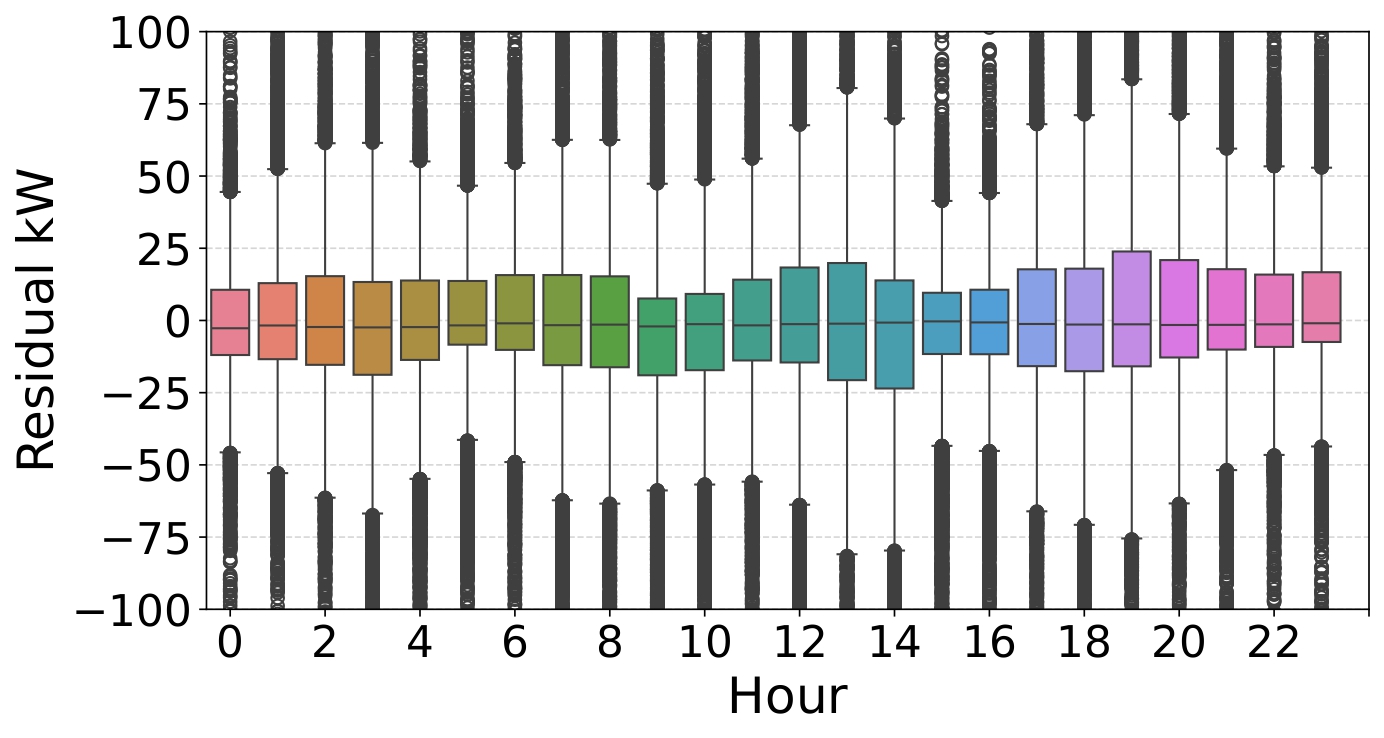}
        \subcaption{\gls{mcd} model}
        \label{fig: ch5 nn residual mcd}
    \end{subfigure}
    \hfill
    \begin{subfigure}{0.70\linewidth}
        \includegraphics[width=\linewidth]{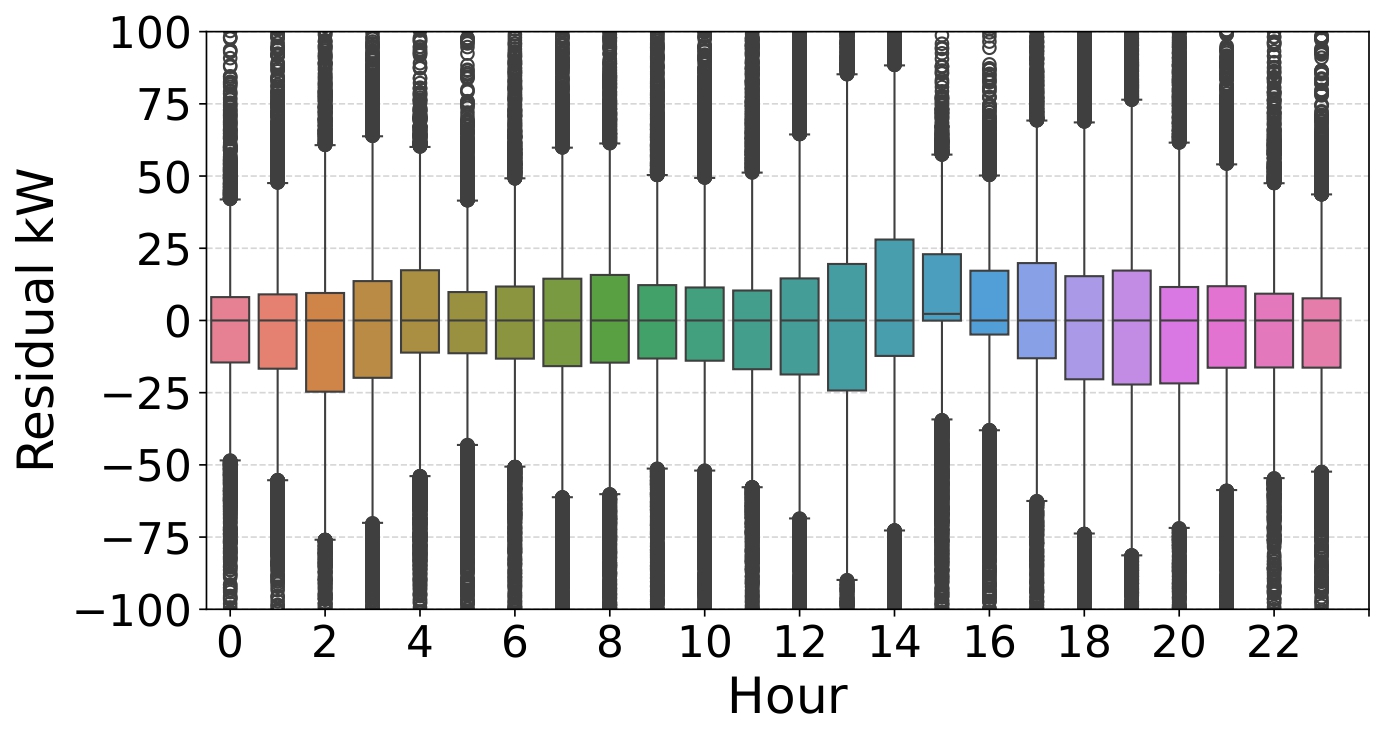}
        \subcaption{\gls{qr} model}
        \label{fig: ch5 nn residual qr}
    \end{subfigure}
    \caption{Hourly residuals between true and predicted aggregated up-regulation flexibility  for the MCD and QR models}
    \label{fig: ch5 nn residuals}
\end{figure}

\begin{figure}[htpb!]
    \centering
    \begin{subfigure}{0.70\linewidth}
        \includegraphics[width=\linewidth]{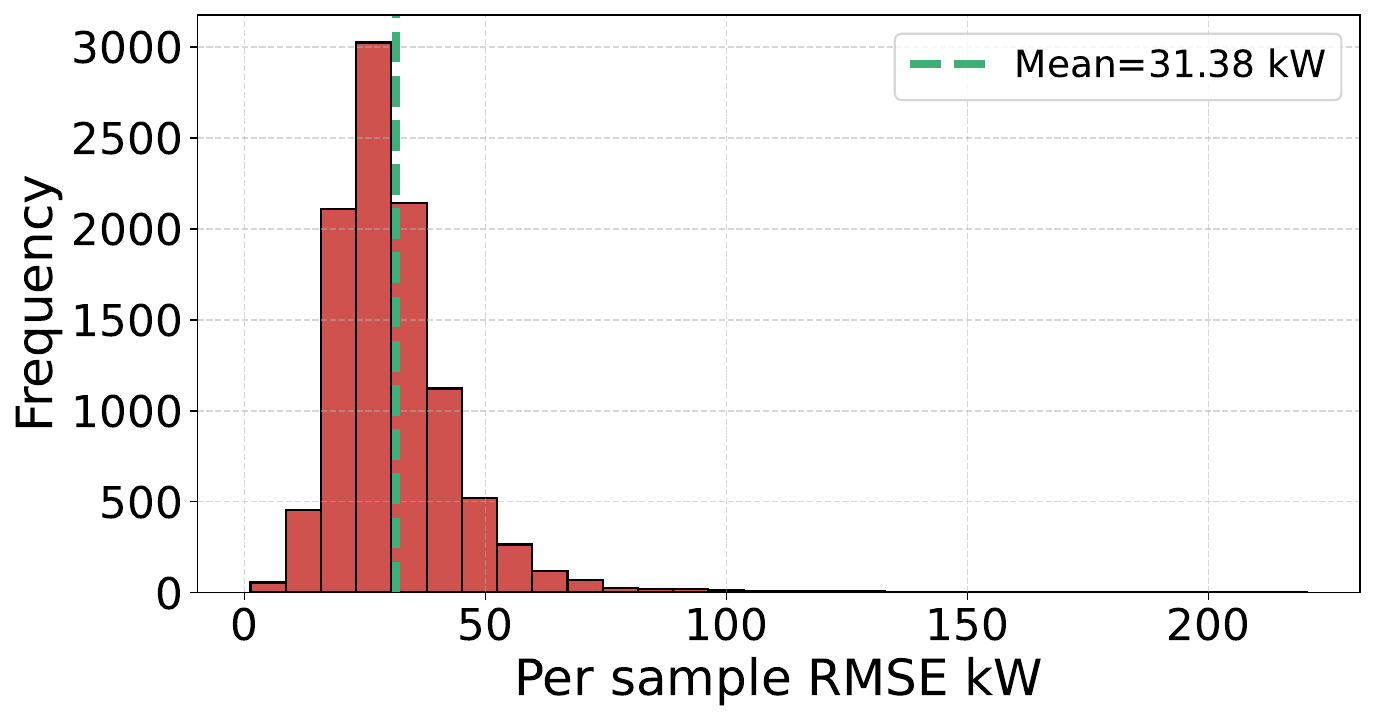}
        \subcaption{MCD model}
        \label{fig: ch5 nn rmse mcd}
    \end{subfigure}
    \hfill
    \begin{subfigure}{0.70\linewidth}
        \includegraphics[width=\linewidth]{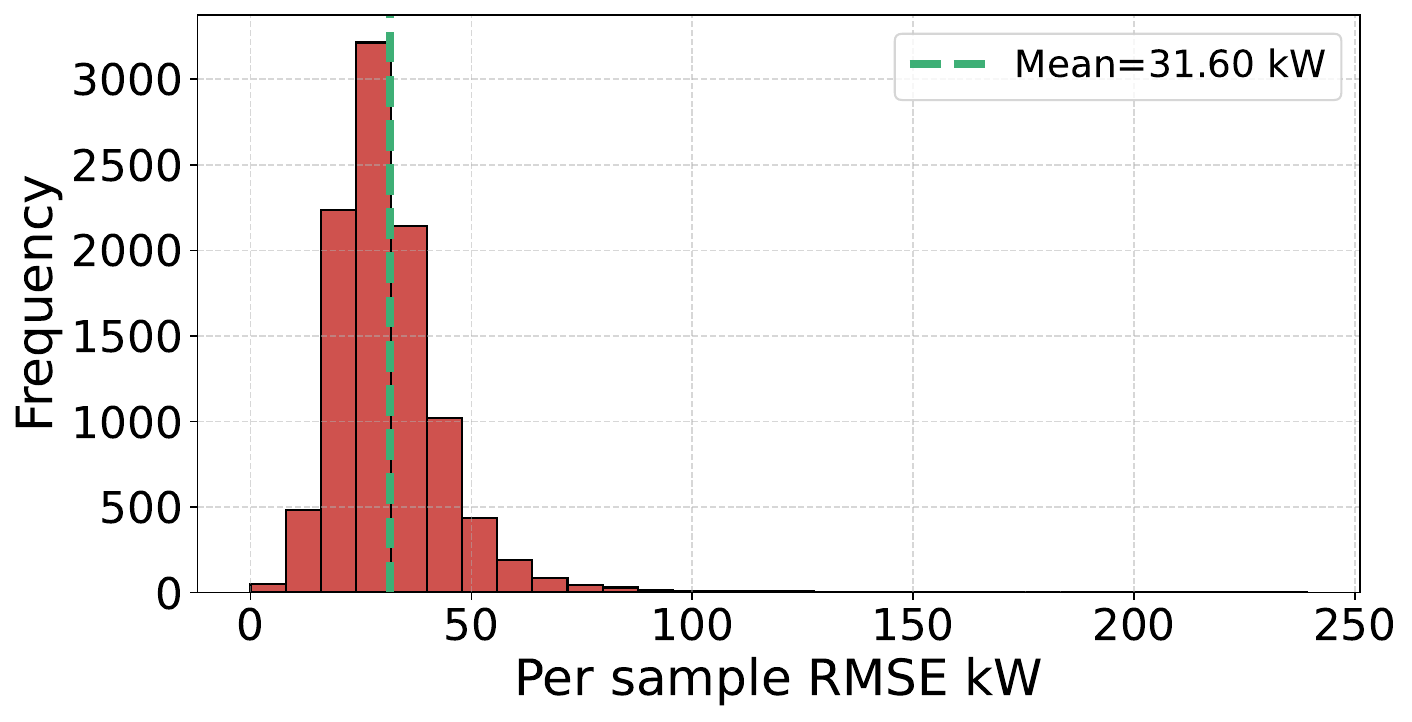}
        \subcaption{QR model}
        \label{fig: ch5 nn rmse qr}
    \end{subfigure}
    \caption{Distribution of per-sample RMSE between true and predicted aggregated up-regulation flexibility for MCD and QR models}
    \label{fig: ch5 nn histograms}
\end{figure}

Figures~\ref{fig: ch5 nn residuals} and~\ref{fig: ch5 nn histograms} compare the point-prediction errors of the \gls{mcd} and \gls{qr} models. For the \gls{qr} model, the point prediction used for this comparison is defined as the midpoint between the predicted lower and upper quantiles. The hourly residuals in Fig.~\ref{fig: ch5 nn residuals} are centred approximately around zero for both models, with comparable dispersion across the 24-hour horizon and no pronounced systematic hourly bias. The per-sample \gls{rmse} distributions in Fig.~\ref{fig: ch5 nn histograms} are also very similar, with mean \gls{rmse} values of 31.38~kW for the \gls{mcd} model and 31.60~kW for the \gls{qr} model. Thus, the two neural networks, which use the same underlying architecture, exhibit comparable point-prediction accuracy, providing a controlled basis for comparing their respective \gls{uq} approaches.

The error metrics are calculated without applying \gls{mcd} sampling at inference. For the \gls{mcd}-based framework, \gls{cp} operates on the min--max normalised dataset $\mathcal{D}_{\text{cal}}^{'}$, ensuring that differing output scales do not distort conformity scores. The resulting normalised up-regulation flexibility intervals are clipped to $[0,1]$, ensuring physical feasibility relative to the battery discharge limits. After calibration, two components are obtained: dropout-based \gls{nn} $f_{\boldsymbol{y}}(\boldsymbol{\Theta})$ and the conformal calibration distance(s) $\hat{q}$, used to construct the final prediction interval $\xi$ as in Steps 8--9 of Algorithm~\ref{alg: multivariate mcd-cp}.

Finally, the trained prediction algorithm \(\Phi_{1-\alpha}(\boldsymbol{\Theta})\) maps contextual inputs $\boldsymbol{\Theta}$, including load, solar generation, battery parameters, and market forecasts, to a prediction region $\xi$ for aggregated up-regulation flexibility $\boldsymbol{y}$. Under the
exchangeability assumption required by \gls{cp}, the calibrated region
satisfies the marginal coverage property $\Phi_{1-\alpha}\left(\boldsymbol{\Theta}\right) = \xi \text{ such that } \mathbb{P}\left[\boldsymbol{y} \in \xi\right] \geq 1 - \alpha.
$. This calibrated prediction region is subsequently incorporated into the aggregator’s bidding strategy.

\section{Results}
\label{sec: results}
\subsection{UQ performance}
\label{ssec: ch5 uq performance}
This section evaluates the trained probabilistic prediction model $\Phi_{1-\alpha}(\cdot)$ based on the approaches described in Sections~\ref{sec: methodology} and~\ref{sec: case study}. The test set $\left(\boldsymbol{\Theta}^{(j)}, \boldsymbol{y}^{(j)}\right) \in \mathcal{D}_{\mathrm{test}}$ of size $10{,}000$ is used to obtain prediction intervals $\xi^{(j)}$ with upper and lower bounds $\boldsymbol{u}^{(j)}$ and $\boldsymbol{l}^{(j)}$. For the \gls{mcd}-based model, three baseline prediction approaches are considered for comparison:

\begin{enumerate}
    \item \textbf{Sample mean} $\boldsymbol{\mu}^{(j)}$ from Eq.~\eqref{eq: ch5 predictive mu}, used as a point estimate. Although interval metrics do not apply, the overbid frequency (Eq.~\eqref{eq: ch5 frequency of overbid}) can still be evaluated.
    \item \textbf{Individual quantiles} $\left[\mathcal{Q}_{t}\left(\mathbf{Y}^{(j)}, \alpha\right), \mathcal{Q}_{t}\left(\mathbf{Y}^{(j)}, 1-\alpha\right)\right]$, computed independently per hour $t \in \mathcal{T}$, ignoring temporal dependence.
    \item \textbf{Na\"ive joint quantiles} \(\left[\mathcal{Q}_{t}\left(\mathbf{Y}^{(j)}, \frac{\alpha}{T}\right), \mathcal{Q}_{t}\left(\mathbf{Y}^{(j)}, 1-\frac{\alpha}{T}\right)\right]\), applying a Bonferroni correction as a simplified joint constraint~\cite{Kyri}.
\end{enumerate}

All methods adopt the P90 reliability target, corresponding to $\alpha=0.10$, such that the lower bound of the predicted flexibility should not exceed the realised flexibility in more than 10\% of the cases. For percentile-based two-sided intervals using the $\alpha$ and $1-\alpha$ quantiles, the corresponding nominal joint coverage is 80\%. The \gls{mcd}-based methods use $\mathcal{S}=1000$ stochastic forward passes from $\mathcal{D}_{\mathrm{test}}$ to generate prediction intervals, whereas \gls{cqr} produces a single pair of conditional quantile estimates for each input. The methods are evaluated using standard prediction interval metrics such as \gls{picp}, which measures empirical coverage; \gls{mpiw}, which reflects the average width of the prediction intervals; and \gls{is}, which penalises both wide intervals and coverage violations. In addition, the frequency of overbid metric from~\cite{jalal_p90} quantifies the frequency of reserve overestimation, which must remain below 10\% to satisfy the P90 compliance criterion. The mathematical definitions of these prediction interval evaluation metrics are presented in~\ref{appendix 2}.

\begin{table}[htpb!]
\centering
\caption{Joint prediction interval quality comparison across various methods}
\label{tab: ch5 joint_picp_comparison}
\begin{tabular}{lcccc}
\toprule
\textbf{\gls{uq} Method} & \textbf{Expected} & \textbf{Joint} & \textbf{Joint \gls{mpiw}} & \textbf{Joint \gls{is}} \\
{} & \textbf{Coverage} & \textbf{\gls{picp}} & \textbf{kW} & \textbf{kW} \\ 
\midrule
Individual quantiles & 0.80 & 0.00 & 33.20 & 128.26 \\
Na\"ive joint quantiles & 0.80 & 0.00 & 86.56 & 122.58 \\
\gls{cqr} & 0.80 & 0.80 & 144.18 & 147.99\\
\gls{mcp} & 0.80 & 0.80 & 163.90 & 167.86 \\
\gls{mmcp} & 0.90 & 0.90 & 211.37 & 213.54 \\
\gls{pcp} & 0.90 & 1.00 & 385.87 & 386.02 \\
\gls{ccp} & 0.90 & 0.91 & 311.73 & 316.56 \\
\bottomrule
\end{tabular}
\end{table}

Table~\ref{tab: ch5 joint_picp_comparison} shows that the conformal methods achieve empirical joint coverage at or above their respective target levels, whereas the uncalibrated baselines have near-zero joint \gls{picp}. The individual and na\"ive joint quantile methods produce narrow intervals but fail to provide adequate joint coverage, illustrating the difficulty of obtaining reliable multivariate intervals without calibration. Among the conformal methods, \gls{cqr} and \gls{mcp} achieve the expected 80\% joint coverage with relatively narrow intervals, with \gls{cqr} producing a narrower interval than \gls{mcp}. The more conservative \gls{mmcp}, \gls{ccp}, and \gls{pcp} methods provide wider intervals and higher empirical coverage. The relatively small difference between joint \gls{is} and \gls{mpiw} for the conformal methods indicates that penalty contributions from coverage violations are limited under the nominal test conditions.
 
\begin{figure}[htbp!]
    \centering
    % Row 1
    \begin{subfigure}[b]{0.48\linewidth}
        \centering
        \includegraphics[width=\linewidth]{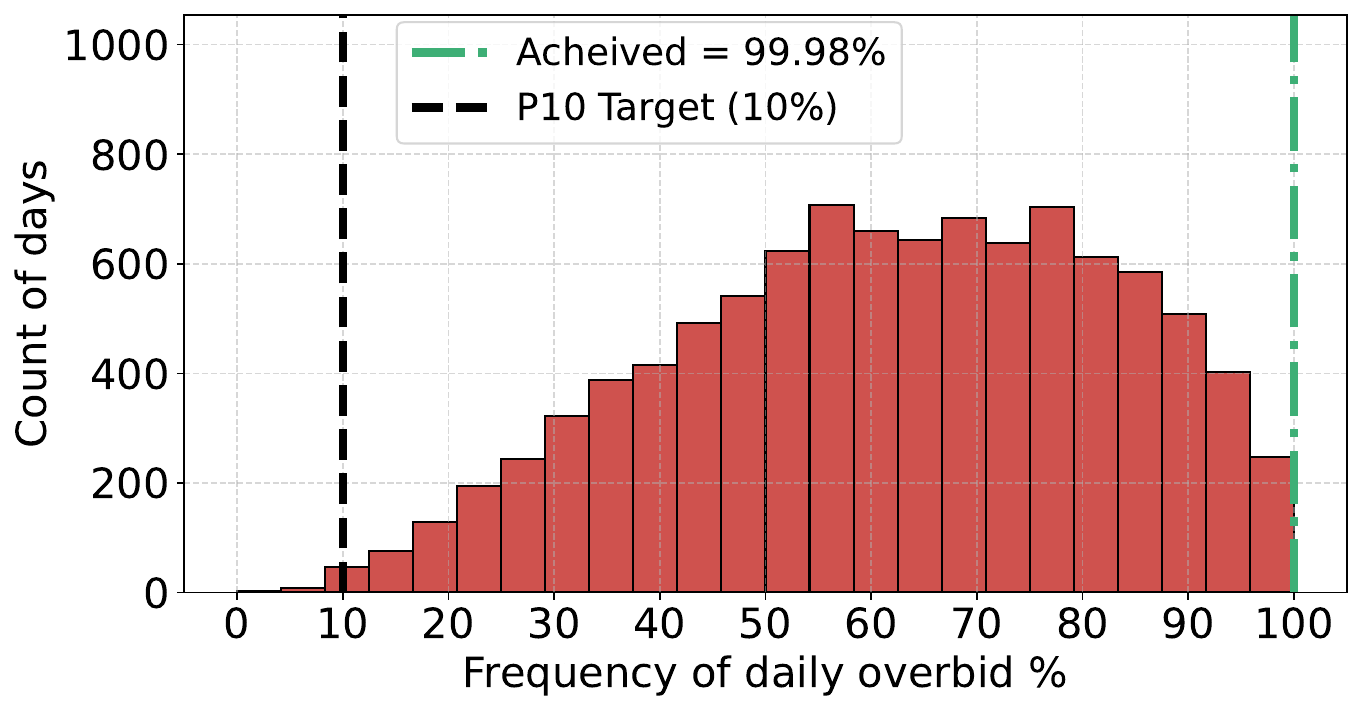}
        \subcaption{MCD sample mean $\boldsymbol{\mu}$}
        \label{fig: ch5 mean overbid}
    \end{subfigure}
    \hfill
    \begin{subfigure}[b]{0.48\linewidth}
        \centering
        \includegraphics[width=\linewidth]{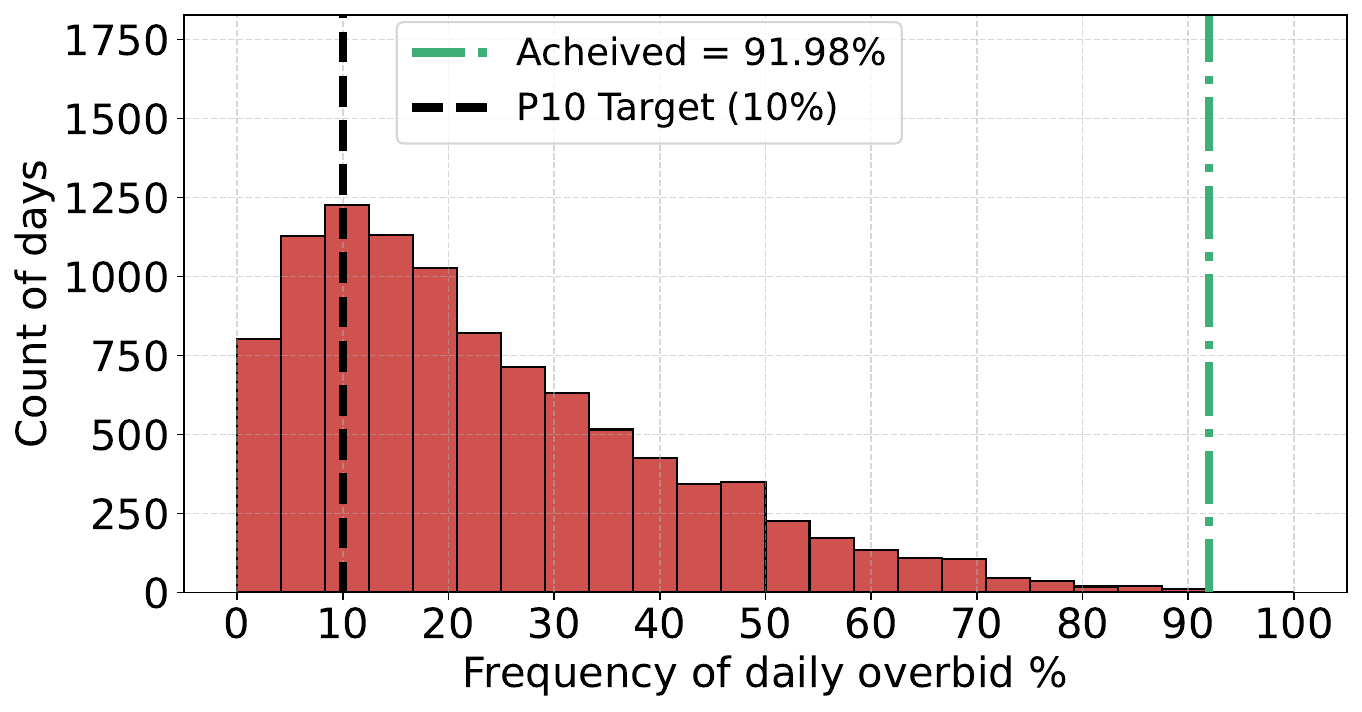}
        \subcaption{Individual percentiles}
        \label{fig: ch5 individual overbid}
    \end{subfigure}
    % Row 2
    \begin{subfigure}[b]{0.48\linewidth}
        \centering
        \includegraphics[width=\linewidth]{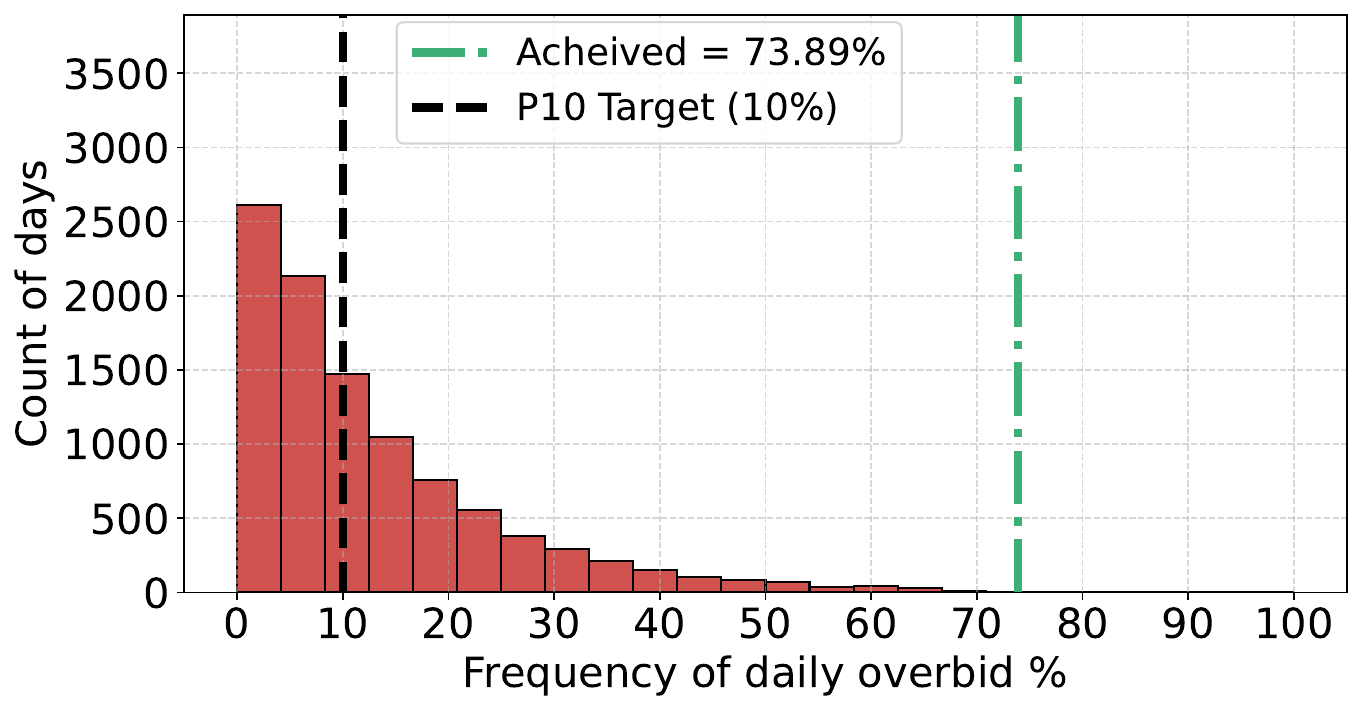}
        \subcaption{Na\"ive joint percentiles}
        \label{fig: ch5 naive overbid}
    \end{subfigure}
    \hfill
    \begin{subfigure}[b]{0.48\linewidth}
        \centering
        \includegraphics[width=\linewidth]{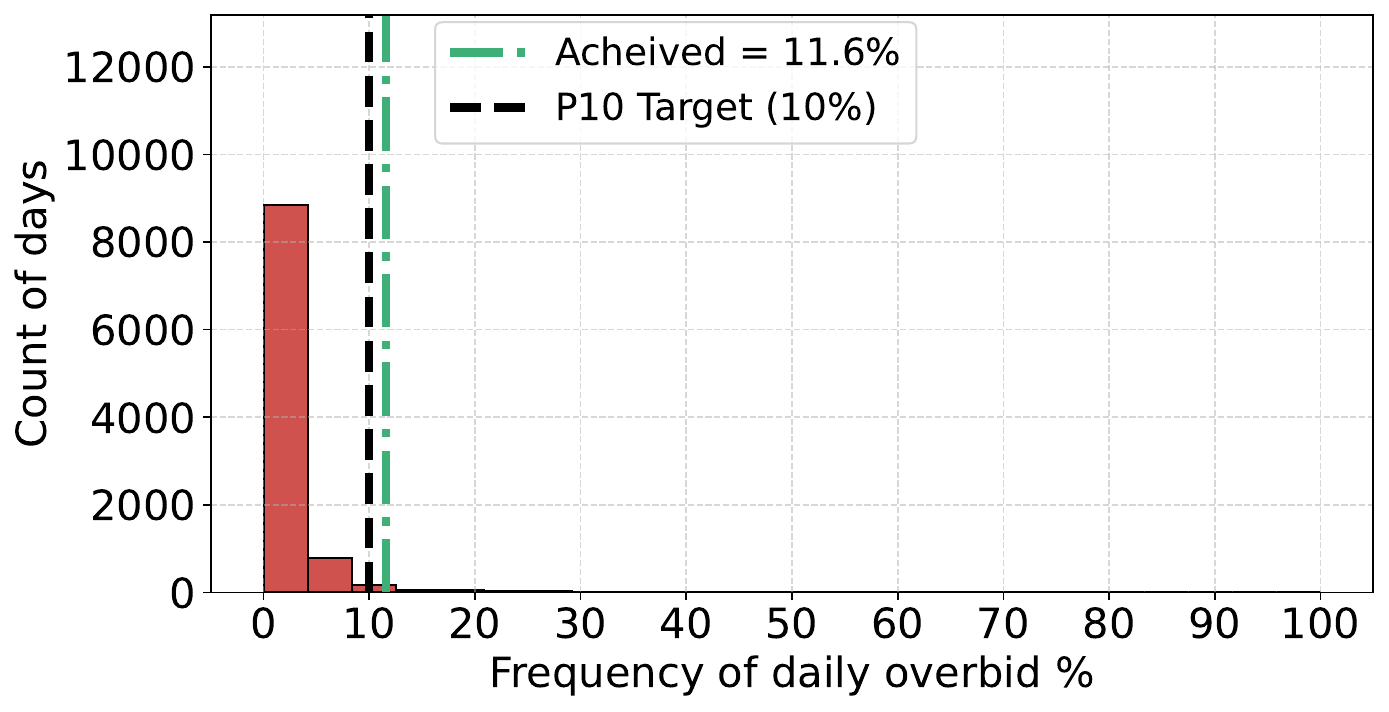}
        \subcaption{CQR}
        \label{fig: ch5 cqr overbid}
    \end{subfigure}
    % Row 3
        \begin{subfigure}[b]{0.48\linewidth}
        \centering
        \includegraphics[width=\linewidth]{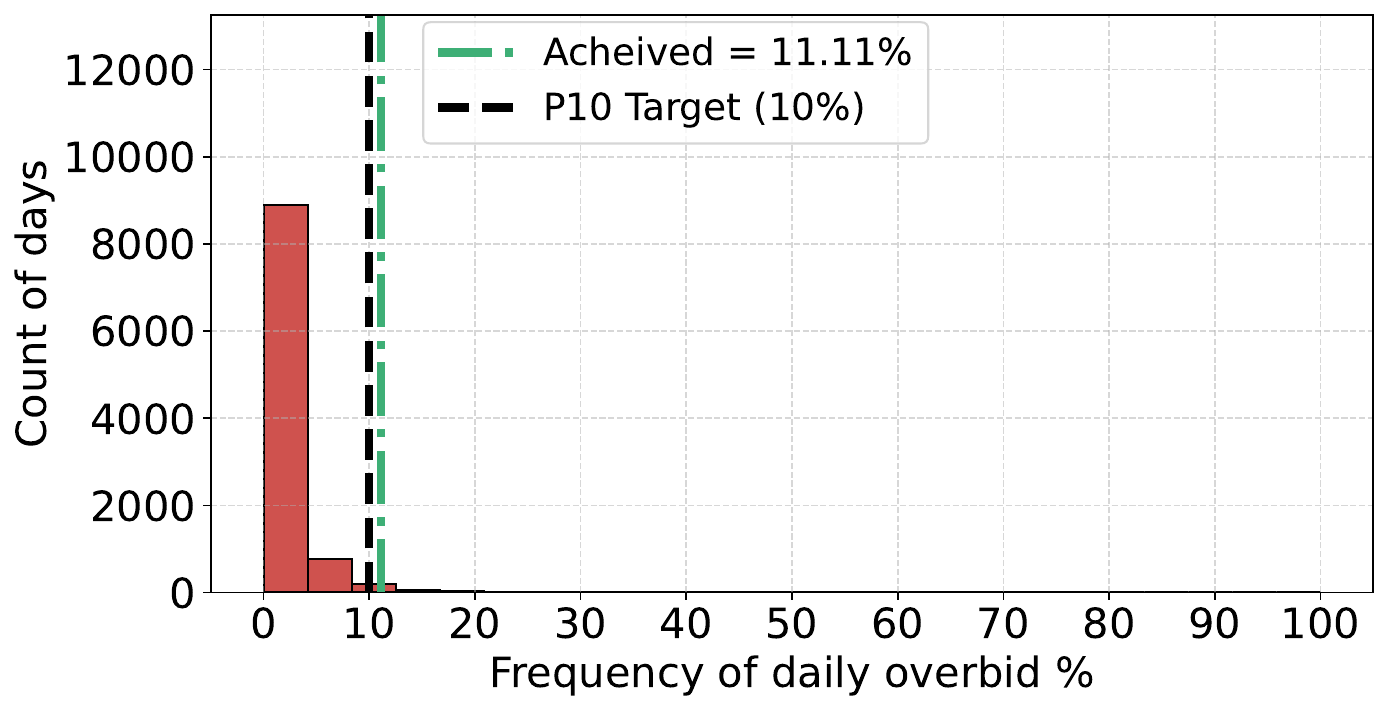}
        \subcaption{MCD using MCP}
        \label{fig: ch5 mcp overbid}
    \end{subfigure}
    \hfill
    \begin{subfigure}[b]{0.48\linewidth}
        \centering
        \includegraphics[width=\linewidth]{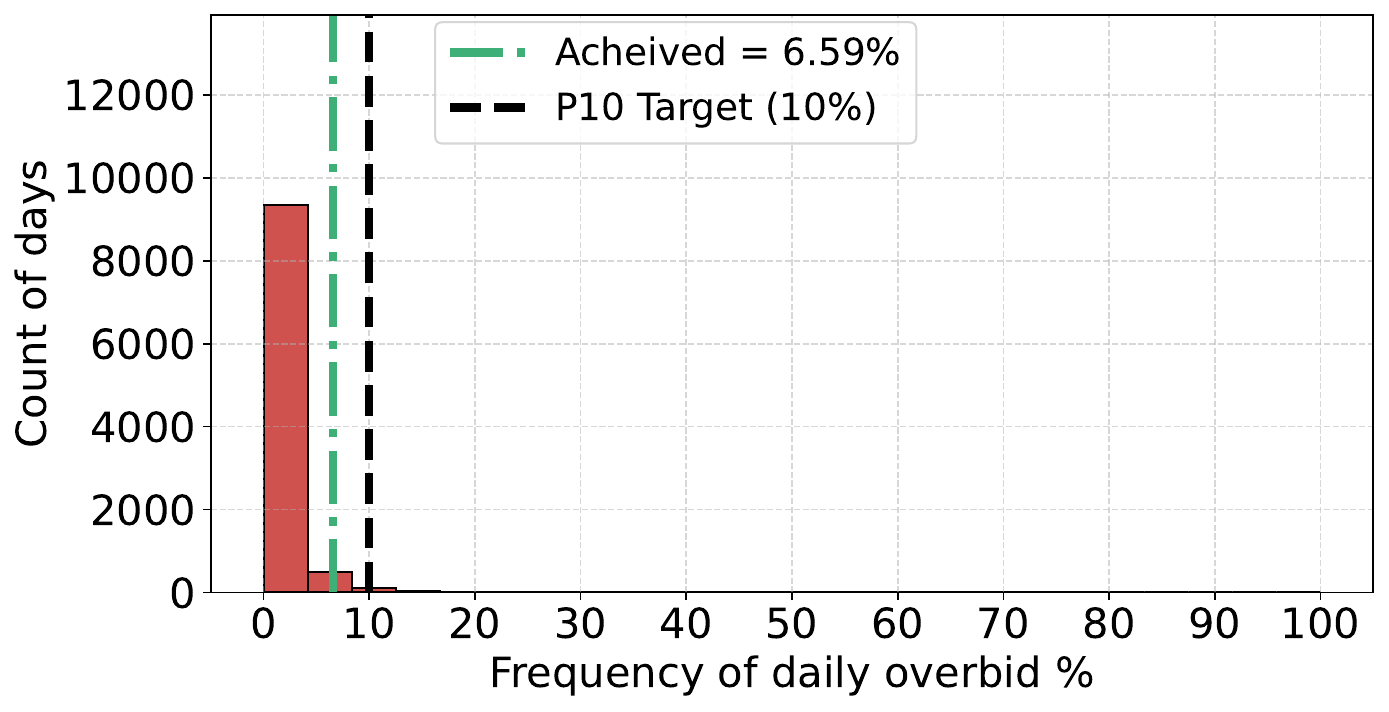}
        \subcaption{MCD using MMCP}
        \label{fig: ch5 mmcp overbid}
    \end{subfigure}

    % Row 4 - Centered
        \begin{subfigure}[b]{0.48\linewidth}
        \centering
        \includegraphics[width=\linewidth]{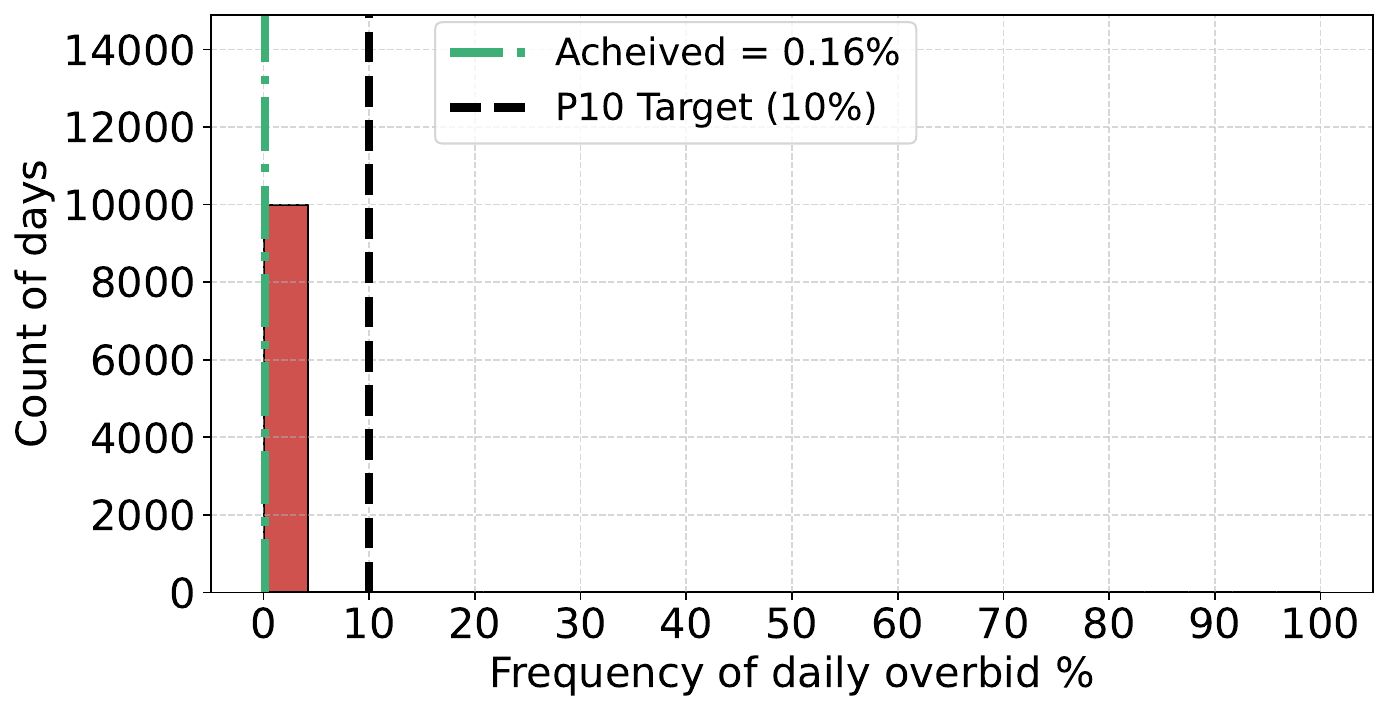}
        \subcaption{MCD using PCP}
        \label{fig: ch5 pcp overbid}
    \end{subfigure}
    \hfill
    \begin{subfigure}[b]{0.48\linewidth}
        \centering
        \includegraphics[width=\linewidth]{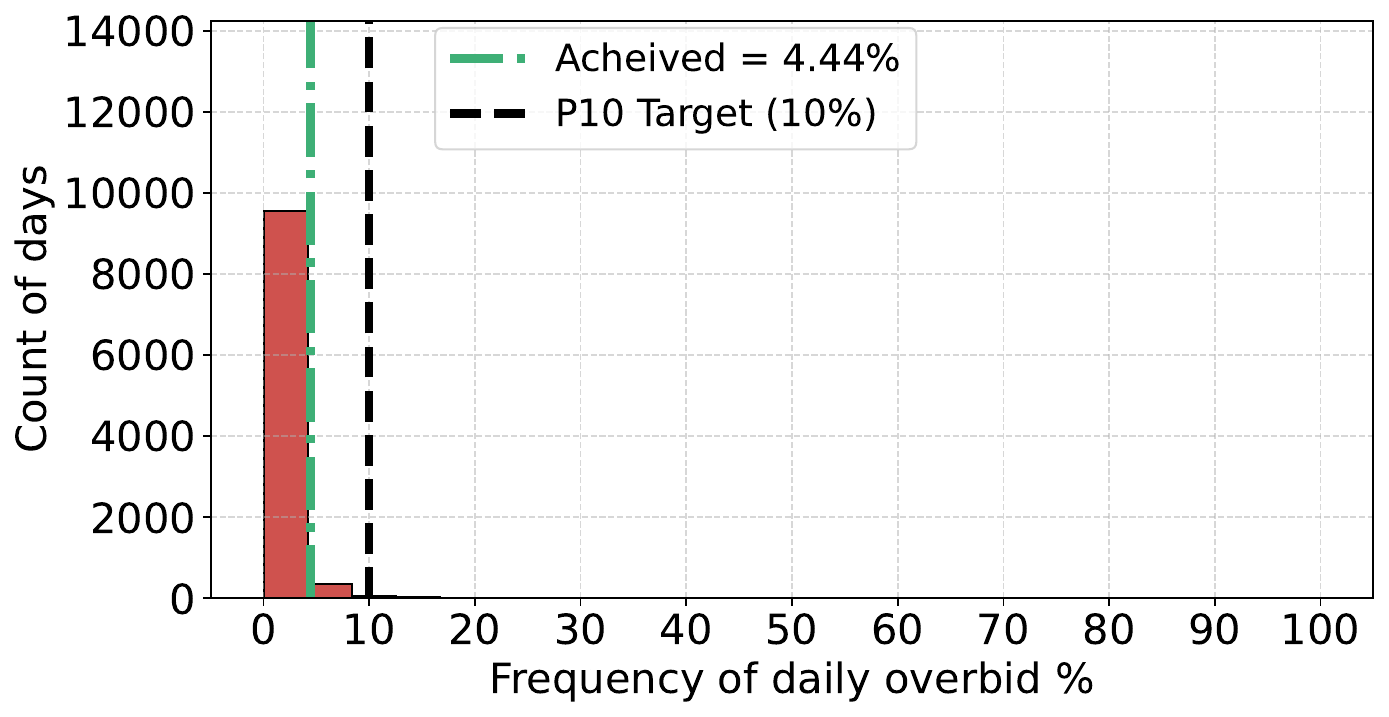}
        \subcaption{MCD using CCP}
        \label{fig: ch5 copula overbid}
    \end{subfigure}
    \caption{Distribution of daily overbid frequency for the evaluated prediction methods}
    \label{fig: ch5 overbid frequency}
\end{figure}

The lower bound must additionally satisfy the P90 requirement for market bidding. The frequency of daily overbids for each method is shown in Fig.~\ref{fig: ch5 overbid frequency}. The uncalibrated baselines fail to satisfy the 10\% overbid criterion, with the \gls{mcd}-based sample mean exhibiting an overbid frequency of 99.98\%. In contrast, \gls{mmcp}, \gls{pcp}, and \gls{ccp} satisfy the criterion, with overbid frequencies of 6.59\%, 0.16\%, and 4.44\%, respectively. The \gls{cqr} and \gls{mcp} methods are close to the target but slightly exceed it, with overbid frequencies of 11.60\% and 11.11\%, respectively. Thus, while \gls{cqr} and \gls{mcp} achieve their expected 80\% joint coverage, they do not quite satisfy the separate 90\% lower-tail reliability criterion in this case study.

\begin{figure}[htpb!]
    \centering

    % Row 1
    \begin{subfigure}[b]{0.48\linewidth}
        \centering
        \includegraphics[width=\linewidth]{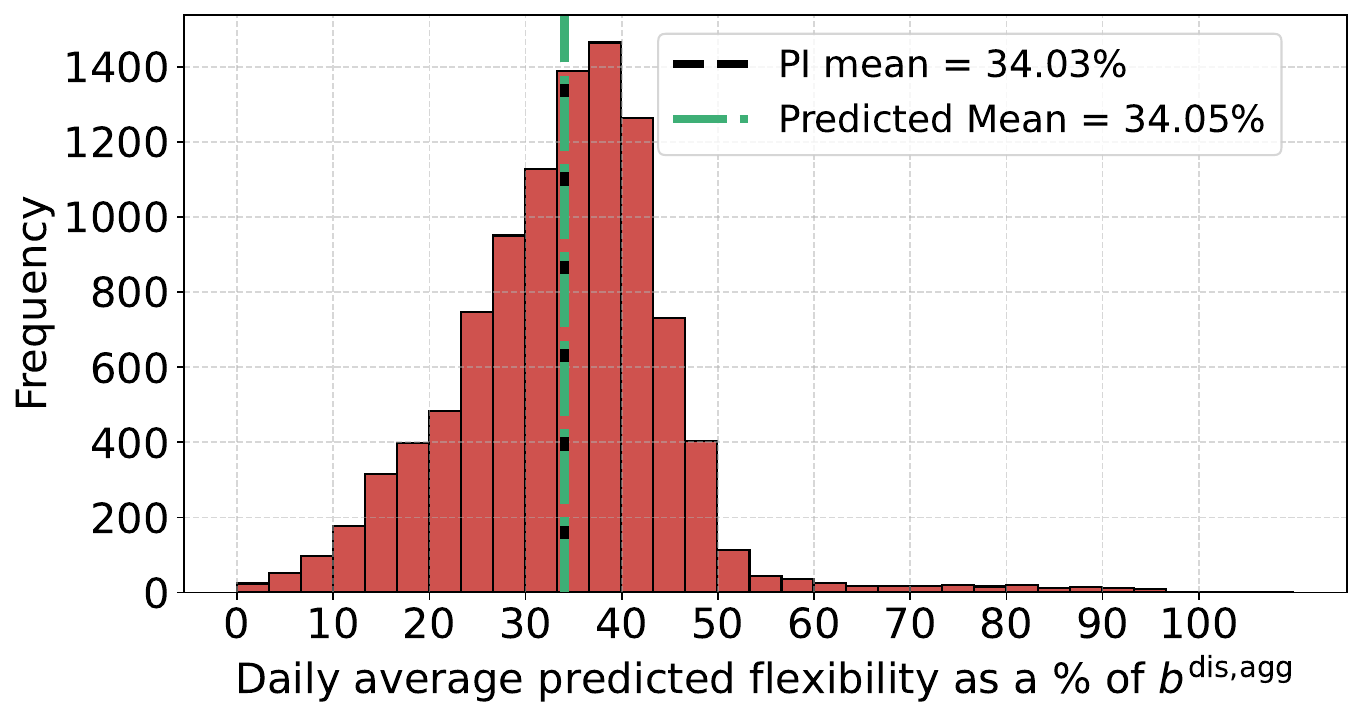}
        \subcaption{MCD sample mean $\boldsymbol{\mu}$}
        \label{fig: ch5 mean bid_volume}
    \end{subfigure}
    \hfill
    \begin{subfigure}[b]{0.48\linewidth}
        \centering
        \includegraphics[width=\linewidth]{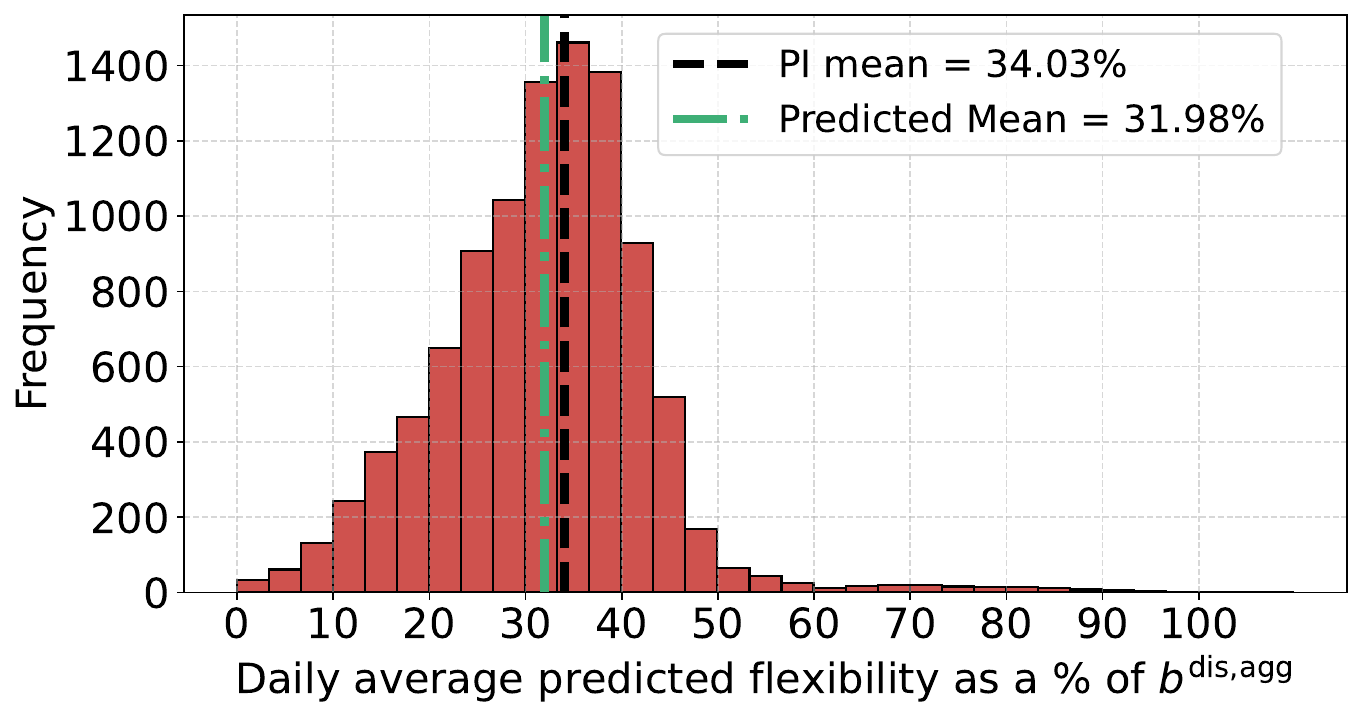}
        \subcaption{Individual percentiles}
        \label{fig: ch5 individual bid_volume}
    \end{subfigure}

    % Row 2
    \begin{subfigure}[b]{0.48\linewidth}
        \centering
        \includegraphics[width=\linewidth]{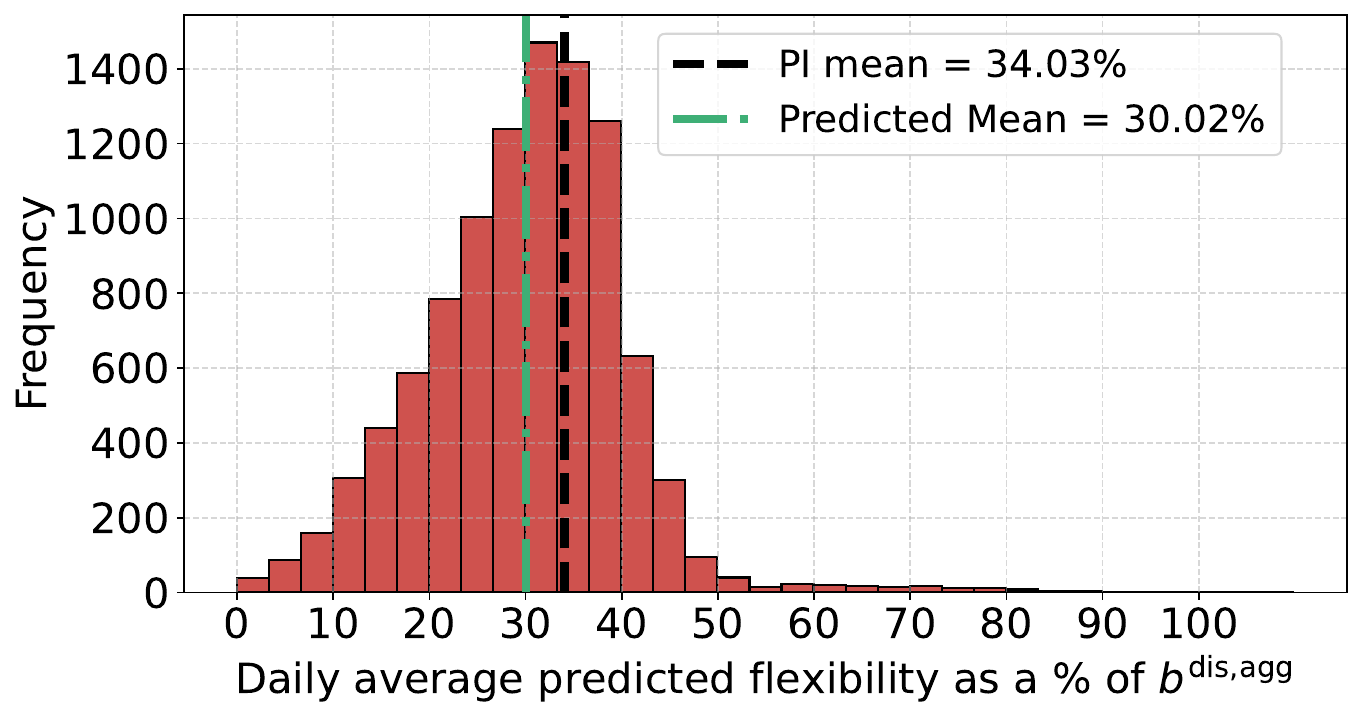}
        \subcaption{Na\"ive joint percentiles}
        \label{fig: ch5 naive bid volume}
    \end{subfigure}
    \hfill
    \begin{subfigure}[b]{0.48\linewidth}
        \centering
        \includegraphics[width=\linewidth]{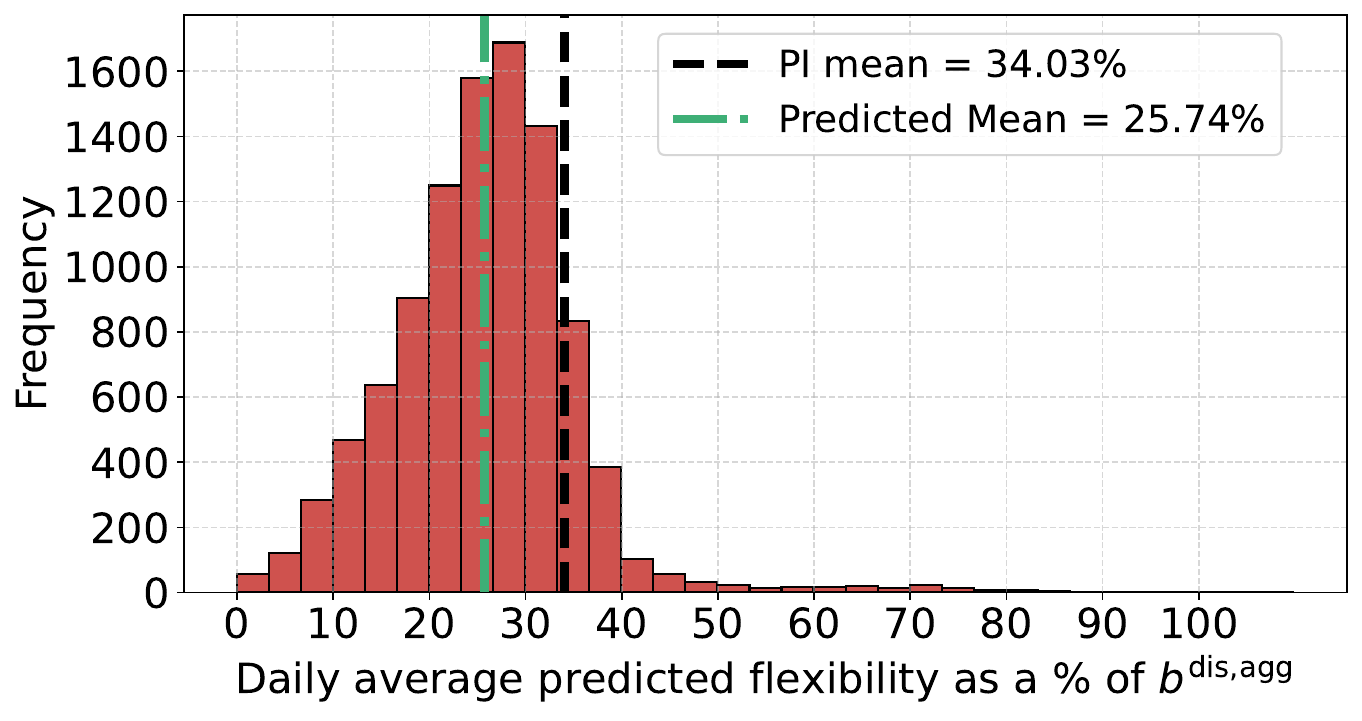}
        \subcaption{CQR}
        \label{fig: ch5 cqr bid volume}
    \end{subfigure}

    % Row 3
    \begin{subfigure}[b]{0.48\linewidth}
        \centering
        \includegraphics[width=\linewidth]{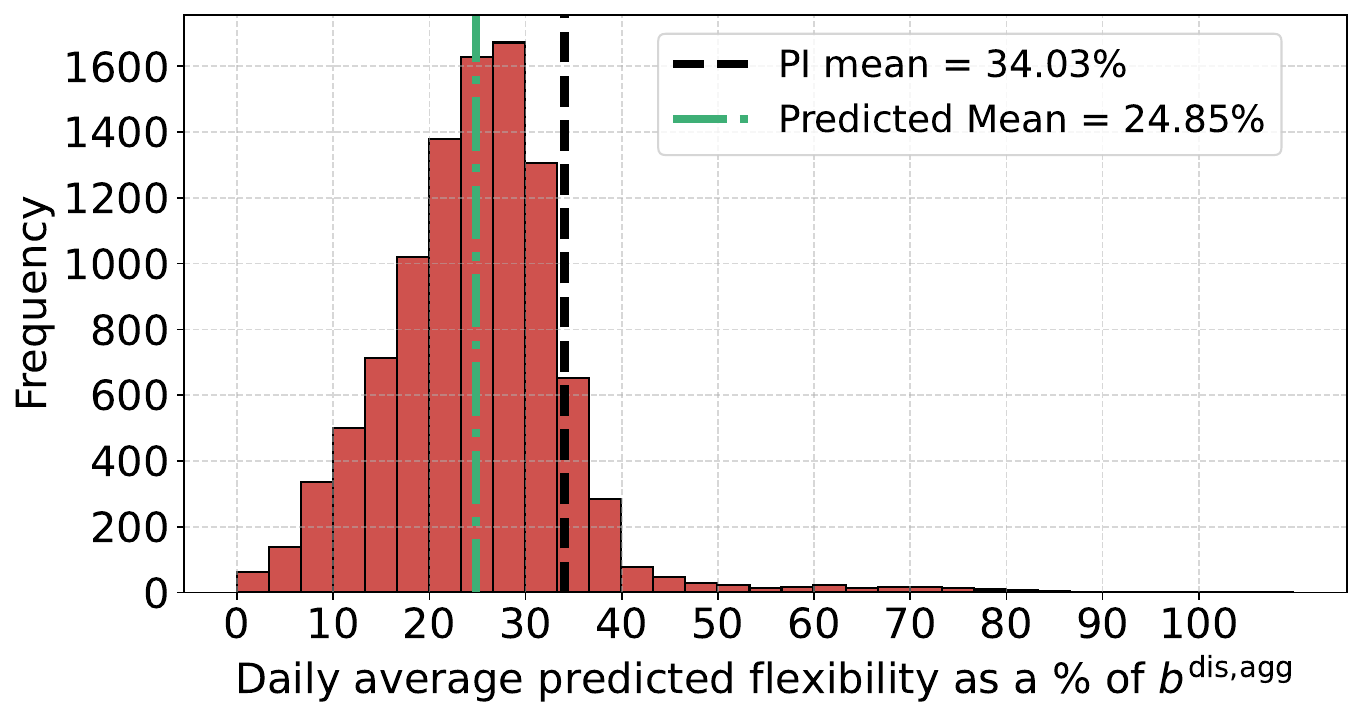}
        \subcaption{MCD using MCP}
        \label{fig: ch5 mcp bid volume}
    \end{subfigure}
    \hfill
    \begin{subfigure}[b]{0.48\linewidth}
        \centering
        \includegraphics[width=\linewidth]{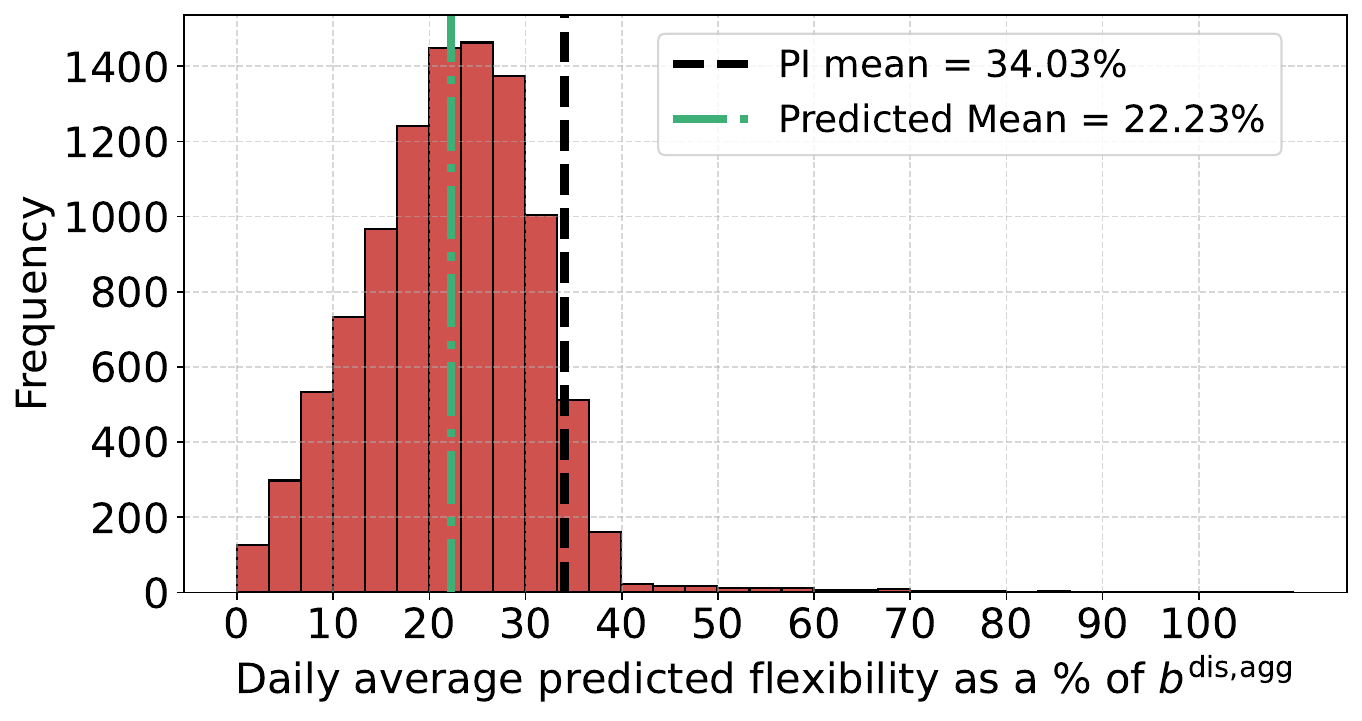}
        \subcaption{MCD using MMCP}
        \label{fig: ch5 mmcp bid volume}
    \end{subfigure}
    % Row 4 - Centered
    \begin{subfigure}[b]{0.48\linewidth}
        \centering
        \includegraphics[width=\linewidth]{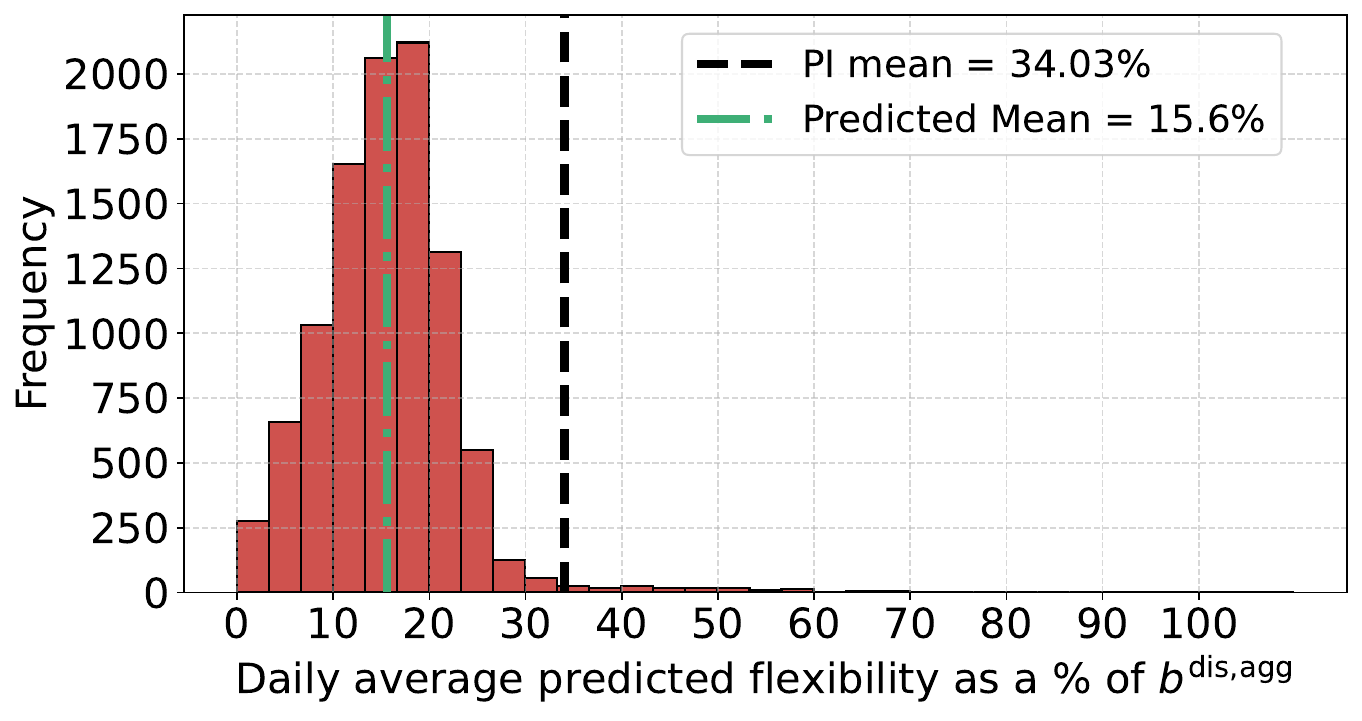}
        \subcaption{MCD using PCP}
        \label{fig: ch5 pcp bid volume}
    \end{subfigure}
    \hfill
    \begin{subfigure}[b]{0.48\linewidth}
        \centering
        \includegraphics[width=\linewidth]{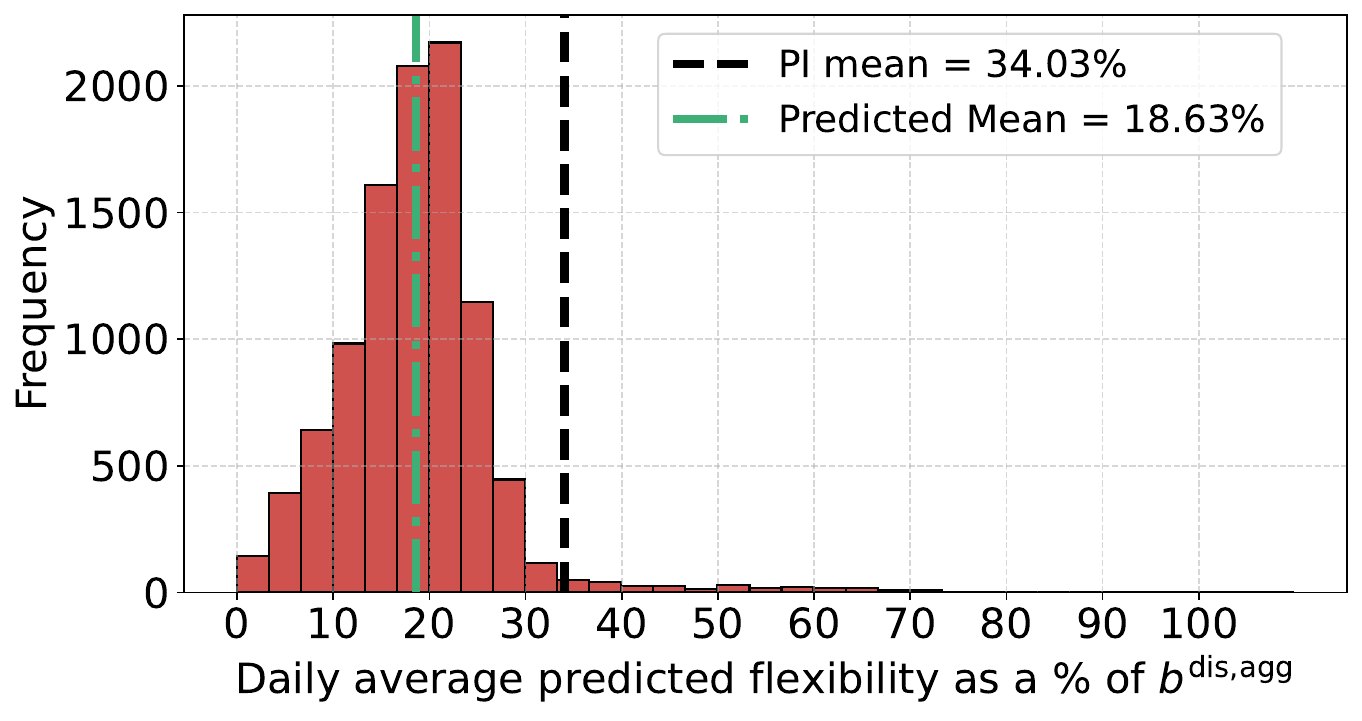}
        \subcaption{MCD using CCP}
        \label{fig: ch5 copula bid volume}
    \end{subfigure}
    \caption{Distribution of daily average predicted up-regulation flexibility lower bounds as a percentage of aggregated maximum battery discharge power}
    \label{fig: ch5 bid volume frequency}
\end{figure}

To satisfy the reliability requirement, the conformal methods generally reduce the lower-bound flexibility used for bidding. Figure~\ref{fig: ch5 bid volume frequency} presents the distribution of daily average lower bounds as a fraction of the aggregated discharge capacity $b^{\text{dis,agg}}$. The \gls{mcd} sample mean remains closest to the \gls{pi} benchmark, with mean lower-bound flexibility of 34.05\% compared with 34.03\% under \gls{pi}, but it fails to control overbidding. Among the calibrated methods, \gls{cqr} and \gls{mcp} provide relatively less conservative lower bounds, with mean values of 25.74\% and 24.85\%, respectively. \gls{mmcp}, \gls{ccp}, and \gls{pcp} provide progressively more conservative bounds, with mean values of 22.23\%, 18.63\%, and 15.60\%, respectively. These results illustrate the trade-off between reducing overbid risk and retaining flexibility for market bidding.

\begin{figure}[htpb!]
    \centering
    % Row 1: Two subfigures side by side
    \begin{subfigure}[b]{0.45\linewidth}
        \centering
        \includegraphics[trim=0 0 0 0, clip, width=\linewidth]{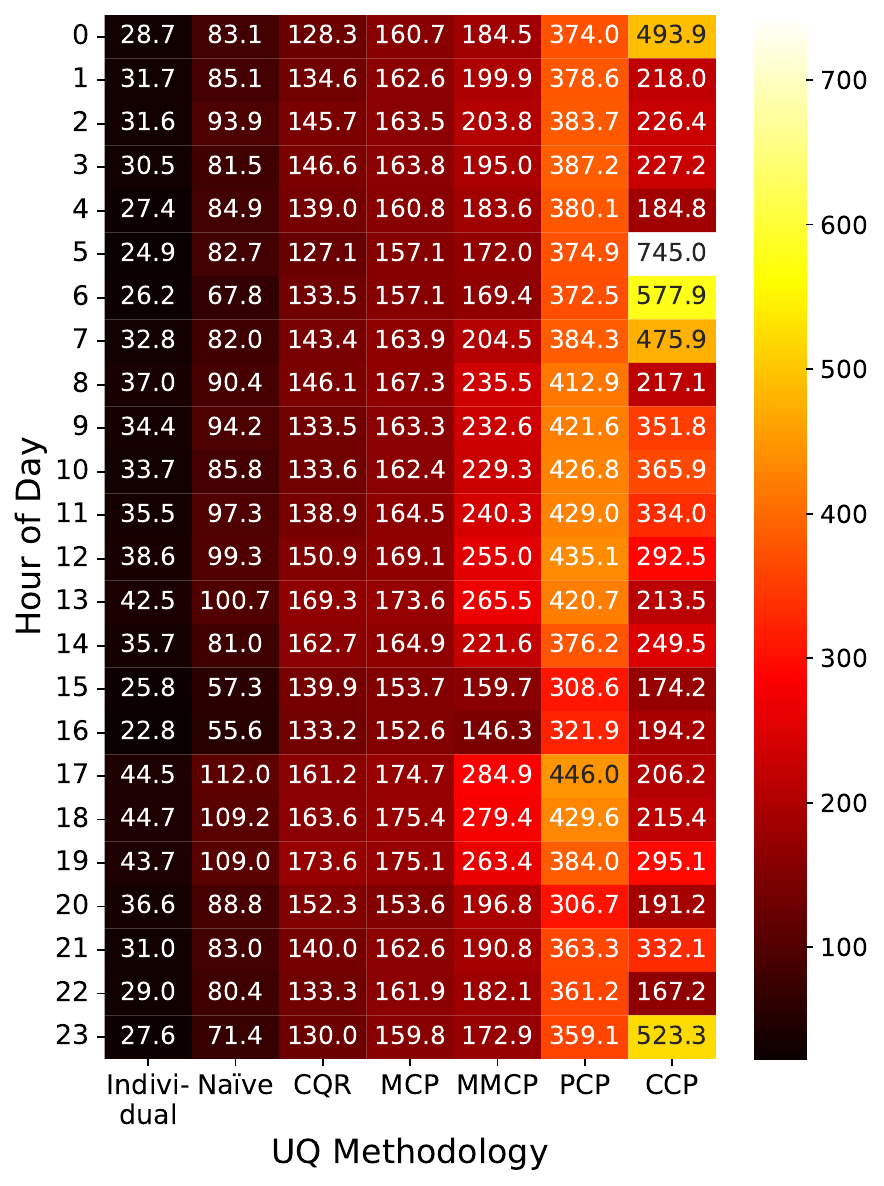}
        \subcaption{Hourly MPIW kW}
        \label{fig: ch5 mpiw heatmap}
    \end{subfigure}
    \begin{subfigure}[b]{0.45\linewidth}
        \centering
        \includegraphics[trim=0 0 0 0, clip, width=\linewidth]{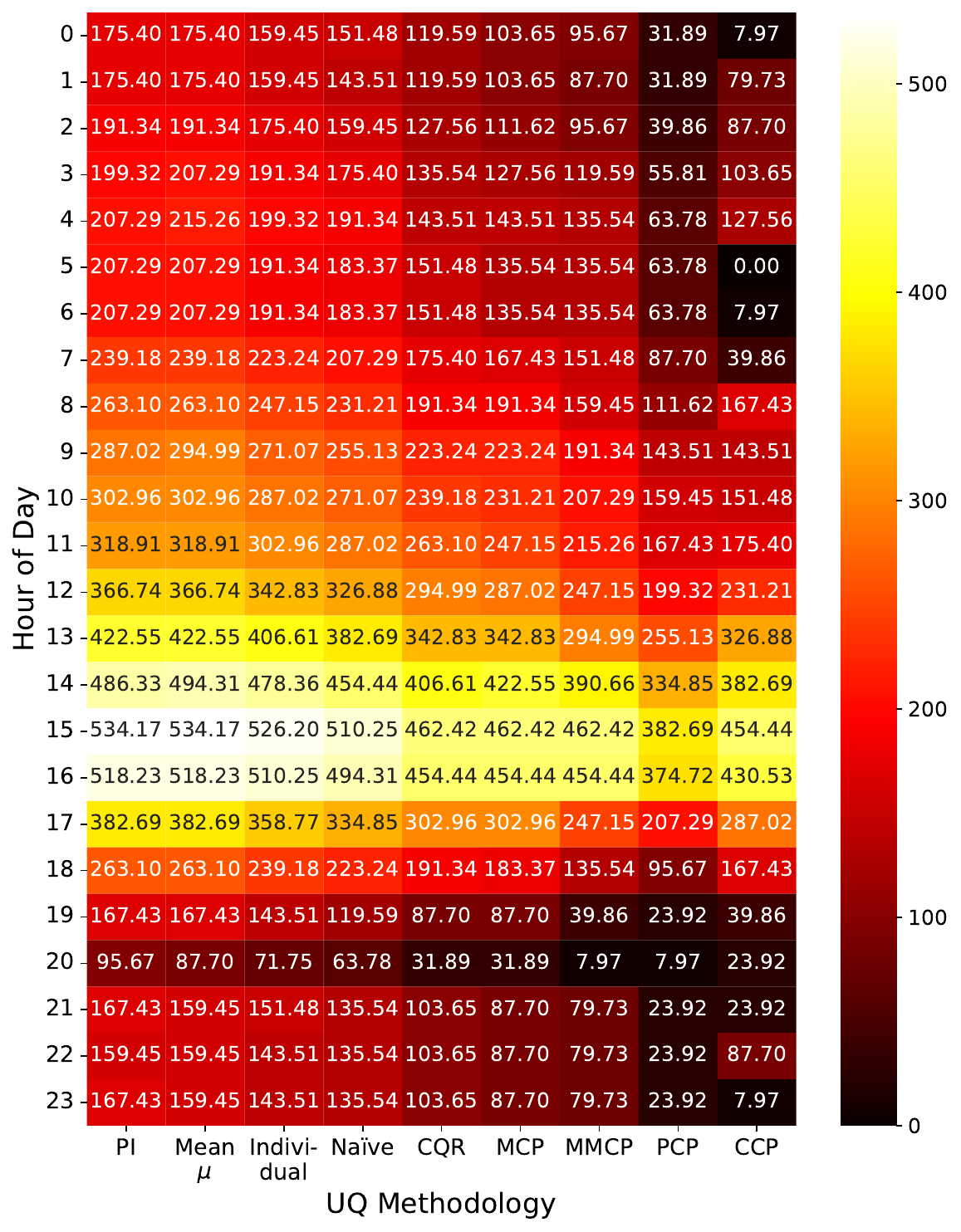}
        \subcaption{Hourly $\boldsymbol{l}^{(j)}$/  $b^{\text{dis,agg}}$}
        \label{fig: ch5 lower-bound heatmap}
    \end{subfigure}
    \caption{Hourly heatmap for different parameters across evaluated prediction methods}
    \label{fig: ch5 heatmap}
\end{figure}

Figure~\ref{fig: ch5 heatmap} further analyses the hourly variability in the prediction interval widths and lower-bound estimates. Due to Denmark's low solar capacity factor ($\approx12\%$)~\cite{DEA_LCoE_Calculator}, residential batteries play an important role in providing flexible capacity in Watts' \gls{hems} algorithm. The \gls{mpiw} generally increases during 08:00--13:00 and 17:00--19:00, indicating greater uncertainty in the predicted flexibility during these periods. The different \gls{uq} methods exhibit similar temporal patterns but different interval widths, with \gls{cqr} and \gls{mcp} producing narrower intervals than the more conservative \gls{mmcp}, \gls{ccp}, and \gls{pcp} methods. The hourly lower-bound heatmaps in Fig.~\ref{fig: ch5 heatmap}b indicate that the highest predicted flexibility generally occurs between 11:00--17:00, when the batteries in the simulated \gls{hems} fleet are more frequently available for up-regulation. These temporal patterns are consistent with the operational behaviour of the simulated residential \gls{hems} systems and further illustrate the reliability--sharpness trade-off across the evaluated \gls{uq} methods.

%--------------------------------------------------------------------
%--------------------------------------------------------------------
\subsection{Aggregator business model performance}
\label{ssec: ch5 aggregator performance}
This section evaluates the aggregator’s bidding problem formulated in Eqs.~\eqref{eq: ch5 objective value 3} and~\eqref{eq: ch5 joint ccp 3}, using the previously trained, calibrated, and validated \gls{uq} methods presented in Section~\ref{ssec: ch5 uq performance}. While Section~\ref{ssec: ch5 uq performance} focused on predictive reliability, this section examines economic performance—specifically, how \gls{uq} methods influence bid robustness and profit margins under uncertainty. These analyses are performed using the dataset $\mathcal{D}_{\beta}$.
\begin{align}
    \gamma = & \frac{\text{No. of hourly overbids in a day}}{\text{Total number of bids in a day}} \nonumber \\
    \label{eq: ch5 conditional overbid frequency}
    = &\frac{\sum_{t\in\mathcal{T}}\mathbbm{1}\left(y^{(j)}_{t} < x^{\text{u}}_{t} \,|\, x^{\text{u}}_{t} > 0\right)}{\sum_{t\in\mathcal{T}}\mathbbm{1}\left(x^{\text{u}}_{t} > 0\right)}
\end{align}
Energinet imposes penalties for overbidding as per Section~2.2.1 of its \gls{fcas} tender conditions~\cite{energinet_ancillary_2024}, which state that payments are reduced proportionally for hours of non-delivery or non-availability. These penalties apply separately to capacity and activation markets and directly impact realised profit. Interpreting this clause, if 6 of 24 contracted hours are unavailable, 25\% of the payment is forfeited. However, “contract period” may refer to individual hours or to the entire daily auction period. Hence, adjusted profits are reported under both interpretations. Equation~\eqref{eq: ch5 conditional overbid frequency} defines the conditional daily overbid frequency, quantifying the share of hours where overbidding occurs among all hours with bids ($x^{\text{u}}_{t}>0$). This term acts as a penalty factor when the contract period is interpreted as all hourly bids placed within a day.
\begin{subequations}
    \begin{align}
        \label{eq: unadjusted profit} 
        {} & R\left(\boldsymbol{x^{\text{u}}}, \beta\right) = \left[1 - \beta\right]\sum_{t \in \mathcal{T}}\lambda^{\text{U}}_{t} x^{\text{u}}_{t}\\
        \label{eq: ch5 interpretation 1 profit}
        {} & R_{1}\left(\boldsymbol{x^{\text{u}}}, \beta\right) = \sum_{t \in \mathcal{T}}\mathbbm{1}\left(x^{\text{u}}_{t} \leq y^{(j)}_{t}\right)\lambda^{\text{U}}_{t} x^{\text{u}}_{t} - \beta\sum_{t \in \mathcal{T}}\lambda^{\text{U}}_{t} x^{\text{u}}_{t} \\
        \label{eq: ch5 interpretation 2 profit}
        {} & R_{2}\left(\boldsymbol{x^{\text{u}}}, \beta\right) = \left[1 - \gamma - \beta\right]\sum_{t \in \mathcal{T}}\lambda^{\text{U}}_{t} x^{\text{u}}_{t}
    \end{align}
\end{subequations}

The unadjusted (baseline) profit is defined in Eq.~\eqref{eq: unadjusted profit}. Equation~\eqref{eq: ch5 interpretation 1 profit} defines the hourly profit, where each hour is settled independently; non-delivery forfeits that hour’s revenue. Equation~\eqref{eq: ch5 interpretation 2 profit} models daily settlement, penalising full-day revenue by $\gamma$, the fraction of overbid hours. In both, incentives are paid fully to prosumers, proportional to their capacity share, regardless of aggregator-level unavailability. 

\begin{figure}[htpb!]
    \centering
    \includegraphics[width=0.90\linewidth]{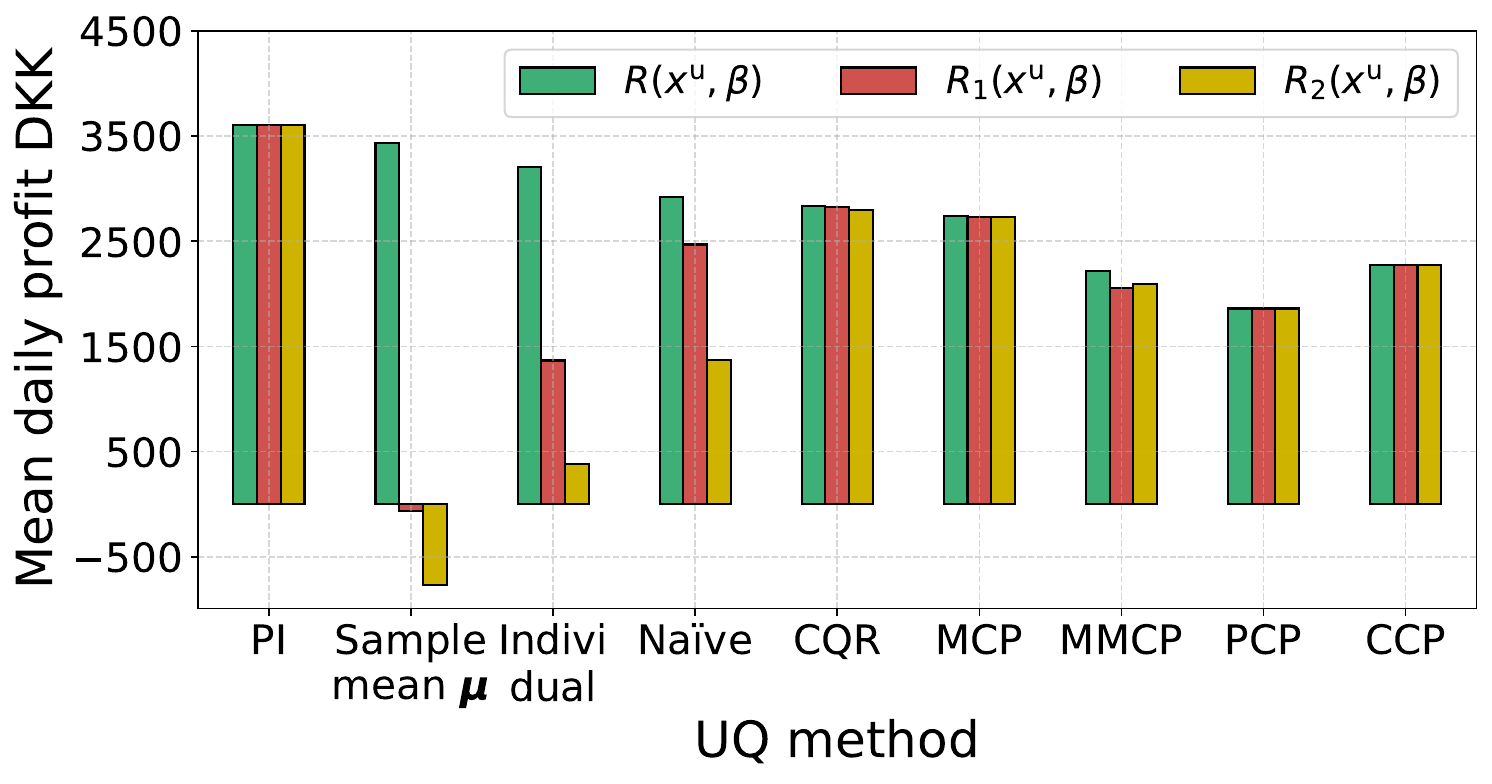}
    \caption{Mean daily unadjusted and adjusted profit for all \gls{uq} methods across all $\beta \in \mathcal{S}_{\beta}$}
    \label{fig: ch5 profit}
\end{figure}

Figure~\ref{fig: ch5 profit} compares unadjusted and adjusted daily profits for all \gls{uq} methods across the evaluated $\beta$ values. The \gls{pi} benchmark represents the maximum daily profit achievable under perfect information. The \gls{mcd}-based sample mean and individual quantile methods produce high unadjusted profits, but incur substantial reductions under the penalty schemes due to persistent overbidding. In particular, the sample mean results in a negative adjusted profit under both penalty interpretations. In contrast, conformal methods substantially reduce the economic impact of overbidding. \gls{cqr} and \gls{mcp} achieve the highest adjusted profits among the conformal methods, with \gls{cqr} providing a similar economic outcome to \gls{mcp} while using a separate \gls{qr}-based predictive model. The more conservative \gls{mmcp}, \gls{pcp}, and \gls{ccp} methods retain lower adjusted profits but reduce exposure to overbids through more conservative flexibility estimates.

\begin{table}[htpb!]
\centering
\caption{Mean percentage of PI profit for each UQ method}
\renewcommand{\arraystretch}{0.9} % tighter row spacing
\setlength{\tabcolsep}{3pt} % reduce column padding
\footnotesize % smaller font size
\begin{tabular*}{\textwidth}{@{\extracolsep{\fill}}lccc}
\toprule
\textbf{\gls{uq} Method} & \multicolumn{3}{c}{\textbf{Mean percentage of \gls{pi} profit (\%)}} \\
\cmidrule(lr){2-4}
& $R(\boldsymbol{x^{\text{u}}}, \beta)$ & $R_1(\boldsymbol{x^{\text{u}}}, \beta)$ & $R_2(\boldsymbol{x^{\text{u}}}, \beta)$ \\
\midrule
Sample mean $\boldsymbol{\mu}$ & 101 & -56 & -97 \\
Individual quantiles & 94 & 10 & -43 \\
Na\"ive joint quantiles & 87 & 46 & -10 \\
\gls{cqr} & 73 & 70 & 69\\
\gls{mcp} & 71 & 69 & 68 \\
\gls{mmcp} & 61 & 61 & 61 \\
\gls{pcp} & 43 & 43 & 43 \\
\gls{ccp} & 52 & 52 & 52 \\
\bottomrule
\end{tabular*}
\label{tab: ch5 profit comparison}
\end{table}

Table~\ref{tab: ch5 profit comparison} quantifies the mean profit of each
method relative to the \gls{pi} benchmark. The \gls{mcd}-based sample mean
and individual quantile methods initially match or exceed the \gls{pi}
profit because of overbidding, but experience substantial reductions once
penalties are applied. The conformal methods retain substantially higher
shares of the \gls{pi} profit after adjustment. Among these methods,
\gls{cqr} achieves the highest adjusted profit, retaining 70\% and 69\% of
the \gls{pi} profit under $R_1$ and $R_2$, respectively, closely followed 
by \gls{mcp} at 69\% and 68\%. The more conservative \gls{mmcp}, \gls{pcp}, and \gls{ccp} methods retain lower shares of the \gls{pi} profit while reducing overbid risk. These results highlight the practical value of \gls{cp} in balancing economic performance and delivery reliability under uncertainty. While more conservative \gls{cp} methods reduce the share of available flexibility committed to the capacity market, the resulting uncommitted flexibility could potentially remain available for subsequent market opportunities, such as intraday trading or activation markets.

\begin{figure}[htpb]
    \centering
    \begin{subfigure}{0.70\linewidth}
        \centering
        \includegraphics[width=\linewidth]{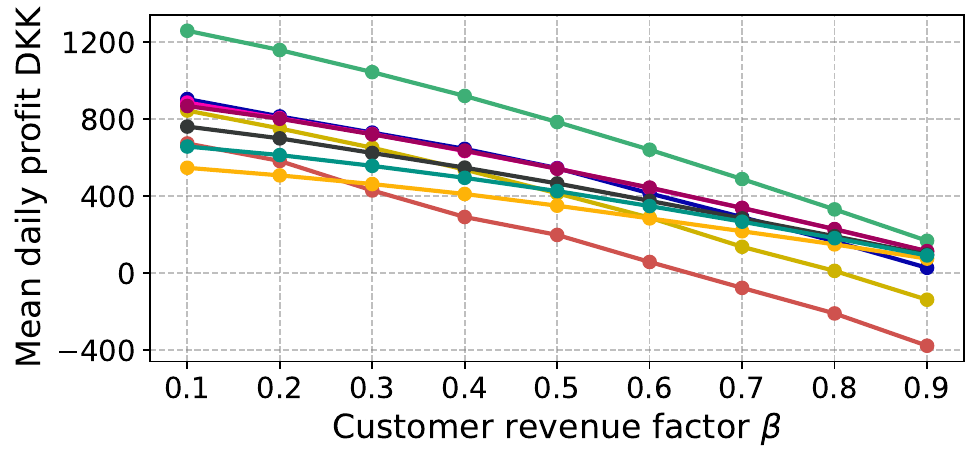}
        \subcaption{Mean daily adjusted profit sensitivity calculated using Eq.~\eqref{eq: ch5 interpretation 1 profit}}
    \end{subfigure}
    \hfill
    \begin{subfigure}{0.70\linewidth}
        \centering
        \includegraphics[width=\linewidth]{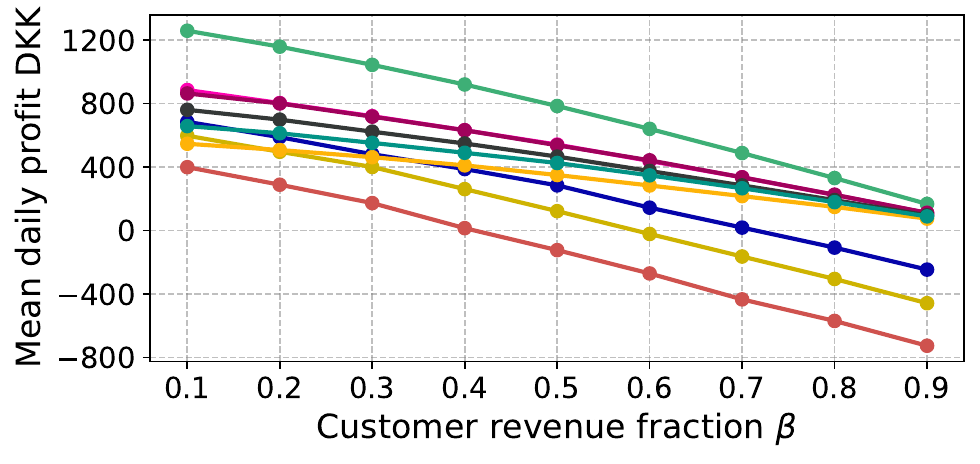}
        \subcaption{Mean daily adjusted profit sensitivity calculated using Eq.~\eqref{eq: ch5 interpretation 2 profit}}
    \end{subfigure}
    \hfill
    \begin{subfigure}{0.70\linewidth}
        \centering
        \includegraphics[width=\linewidth]{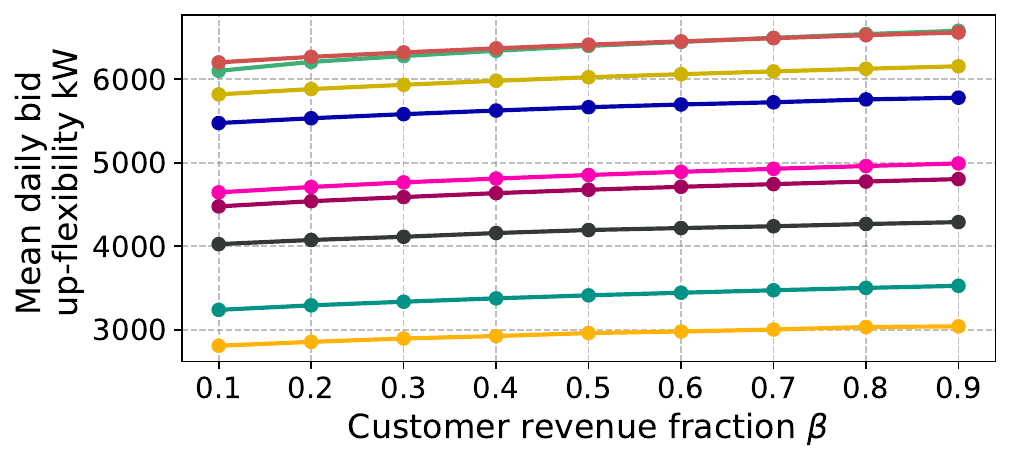}
        \subcaption{Mean daily bid up-regulation flexibility value sensitivity}
    \end{subfigure}
    \hfill
    \begin{subfigure}{\linewidth}
        \centering
        \includegraphics[width=\linewidth]{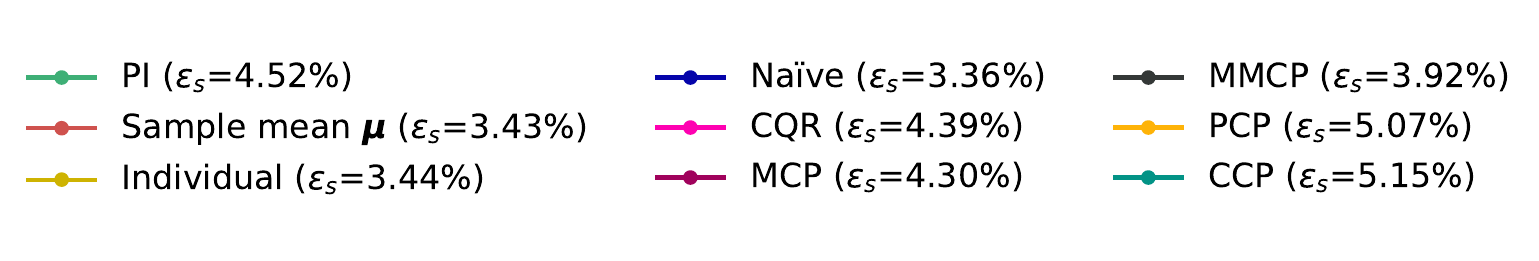}
    \end{subfigure}
    \caption{Sensitivity of aggregator performance to the revenue-sharing factor $\beta$}
    \label{fig: ch5 beta sensitivity}
\end{figure}

Figure~\ref{fig: ch5 beta sensitivity} presents the sensitivity of the aggregator performance to the revenue-sharing parameter $\beta$. Figures~\ref{fig: ch5 beta sensitivity}(a) and (b) show that the conformal methods generally achieve higher adjusted profits than the uncalibrated baselines across the evaluated values of $\beta$. As expected, the adjusted profit decreases monotonically as the prosumer incentive increases. This behaviour arises because the bid volume (Fig.~\ref{fig: ch5 beta sensitivity}(c)) exhibits only a weak response to increasing $\beta$, with an estimated supply elasticity $\epsilon_s$ in the range of approximately 3--5\%. Consequently, the simulated \gls{hems} exhibits only a modest increase in available flexibility in response to additional incentives under the techno-economic conditions considered in this study. This observation is further supported by the fact that, across 100 scenarios, the optimal incentive factor was $\beta^{\star} = 0.1$ in 97 cases and $\beta^{\star} = 0.2$ in the remaining three cases, indicating that allocating only 10--20\% of market revenue to prosumers typically maximises aggregator profit.
 
The relatively weak response to increasing incentives may be influenced by the relatively high retail electricity purchase prices faced by prosumers compared to aggregation incentives, which makes it economically rational within the simulated \gls{hems} to offer nearly all of the available flexibility even at relatively small incentive levels. Importantly, the estimated elasticity is a property of the simulated \gls{hems} decision-making model under the assumed techno-economic conditions, rather than an empirically estimated elasticity of human prosumers. Therefore, the results should not be interpreted as direct evidence of a barrier to prosumer engagement or as evidence that human prosumers would exhibit similarly low marginal gains from participation in flexibility. Nevertheless, the findings highlight the potential importance of both \gls{hems} algorithm design and market or incentive mechanisms in shaping the attractiveness of flexibility provision. Improvements to the \gls{hems} scheduling algorithm could be investigated along with measures such as electricity price or tax adjustments, grid tariff subsidies, or modest fixed participation incentives to assess their effects on the economic attractiveness and flexibility response of prosumers.

%------------------------------------------------------
%------------------------------------------------------
%------------------------------------
\subsection{Robustness to input uncertainty}
\label{ssec: ch5 robustness analysis}
To assess the sensitivity of the calibrated prediction intervals to uncertainty in the exogenous inputs, controlled perturbations were applied to the test inputs while the trained prediction models and conformal calibration thresholds were kept fixed. Thus, the experiment evaluates the sensitivity of the evaluated \gls{uq} methods to input perturbations and departure from the conditions represented in the calibration data, rather than the effect of retraining or recalibration. The detailed input perturbation generation procedure is specified in~\ref{app: noise generation}. 

Six perturbation levels of 1\%, 5\%, 10\%, 20\%, 30\%, and 50\% were considered. Repeated independent random realisations were used at each perturbation level to reduce sensitivity to individual random draws and provide more robust estimates of the reported performance metrics. The perturbations were applied to the load, solar generation, electricity prices and \gls{mfrr} activation signals, while the incentive prices and battery parameters were kept unchanged.

\begin{figure}[htpb]
    \centering
    \begin{subfigure}{0.70\linewidth}
        \centering
        \includegraphics[width=\linewidth]{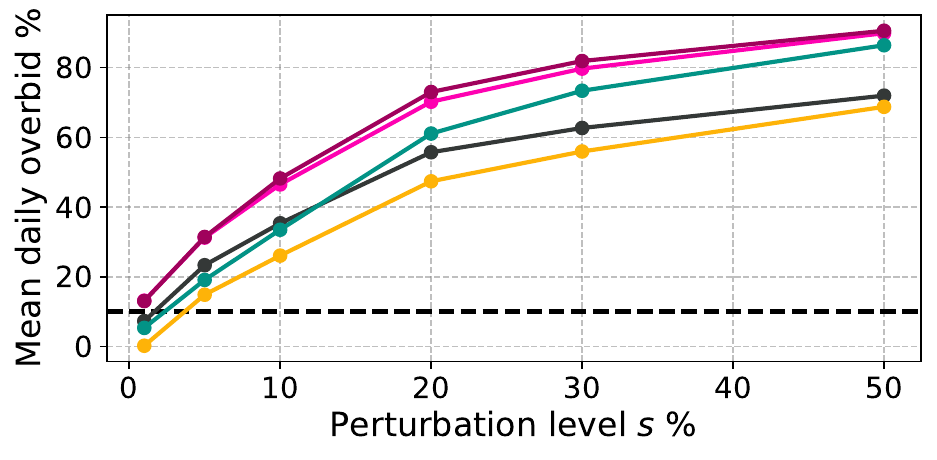}
        \subcaption{Mean daily overbid sensitivity to input noise}
    \end{subfigure}
    \hfill
    \begin{subfigure}{0.70\linewidth}
        \centering
        \includegraphics[width=\linewidth]{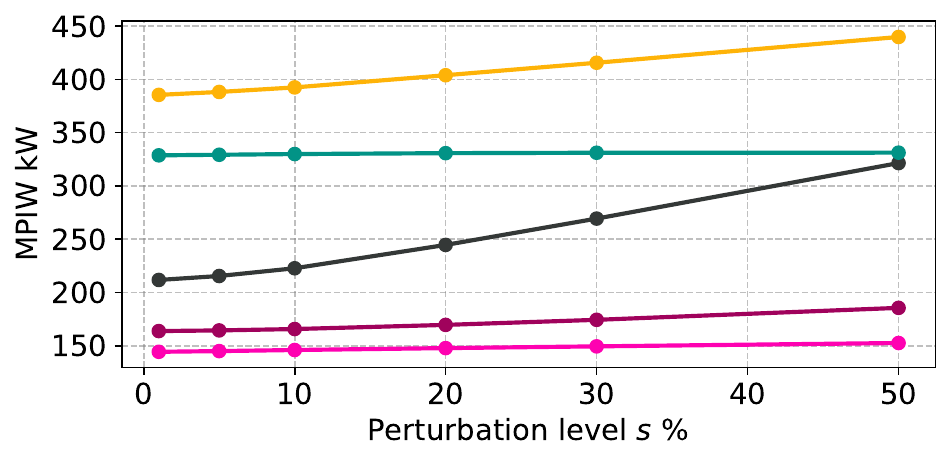}
        \subcaption{\gls{mpiw} sensitivity to input noise}
    \end{subfigure}
    \hfill
    \begin{subfigure}{0.70\linewidth}
        \centering
        \includegraphics[width=0.70\linewidth]{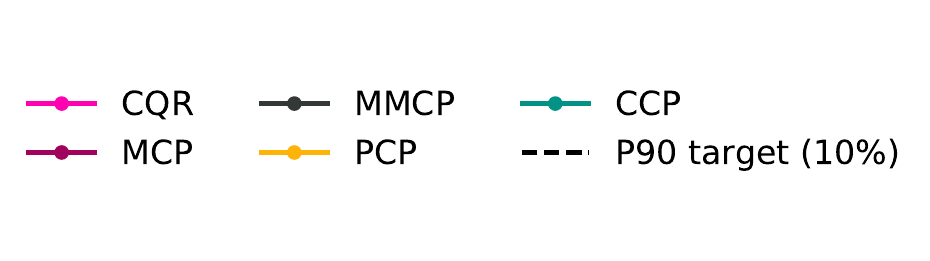}
    \end{subfigure}
    \caption{Sensitivity of interval evaluation metrics to simultaneous perturbations of all input sources.}
    \label{fig: ch5 aggregate noise results}
\end{figure}

Figure~\ref{fig: ch5 aggregate noise results} shows the effect of increasing
input perturbations on the mean daily overbid frequency and \gls{mpiw}. As
the perturbation level increases, all methods exhibit a progressive increase
in overbidding and a corresponding reduction in empirical joint coverage.
At the lowest perturbation level, the results remain close to the nominal
case reported in Section~\ref{ssec: ch5 uq performance}, whereas the degradation
becomes increasingly pronounced at higher perturbation levels. \gls{cqr} and
\gls{mcp} exhibit similar degradation trajectories, consistent with their
comparable nominal performance. More conservative \gls{cp} methods generally
maintain lower overbid frequencies under input perturbations, although the
extent of this robustness advantage varies across methods and diminishes at
high perturbation levels.

The prediction intervals do not widen sufficiently to compensate for the
increasing input uncertainty. For example, \gls{cqr} \gls{mpiw}
increases only from approximately 144~kW at the lowest perturbation level to
153~kW at the highest perturbation level, while its empirical joint coverage
decreases substantially. A similar behaviour is observed for \gls{mcp}, for
which \gls{mpiw} increases from approximately 164~kW to 185~kW while
joint coverage also decreases substantially. These results indicate that
fixed conformal calibration does not automatically adapt the prediction
intervals to increasingly perturbed input uncertainty conditions.

\begin{figure}[htpb]
    \centering
    \begin{subfigure}{0.70\linewidth}
        \centering
        \includegraphics[width=\linewidth]{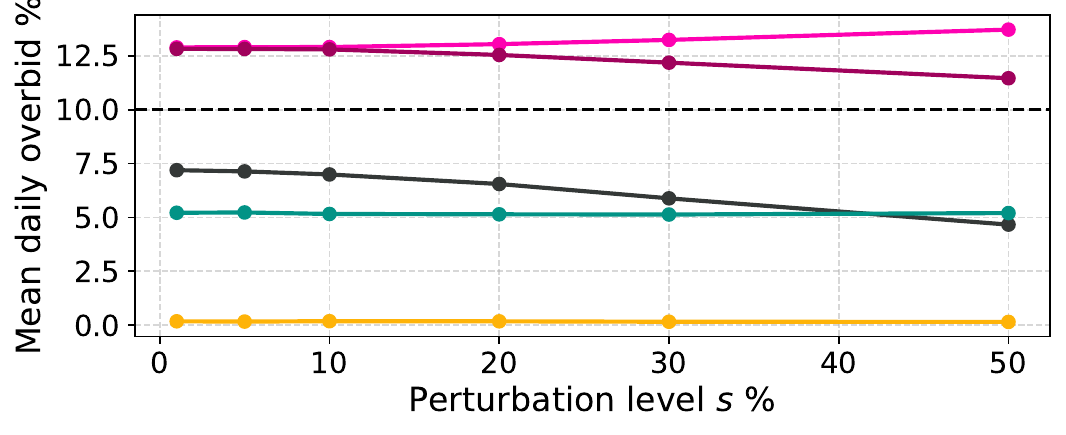}
        \subcaption{Load}
    \end{subfigure}
    \hfill
    \begin{subfigure}{0.70\linewidth}
        \centering
        \includegraphics[width=\linewidth]{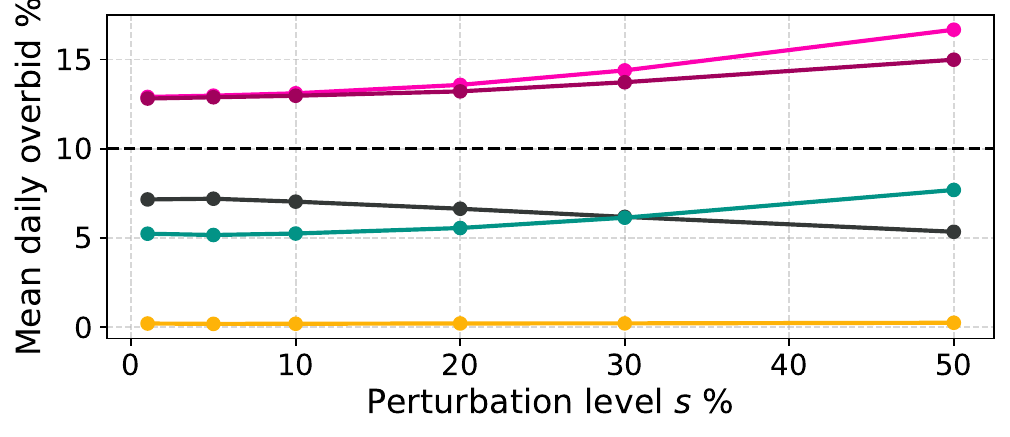}
        \subcaption{Solar}
    \end{subfigure}
    \hfill
    \begin{subfigure}{0.70\linewidth}
        \centering
        \includegraphics[width=\linewidth]{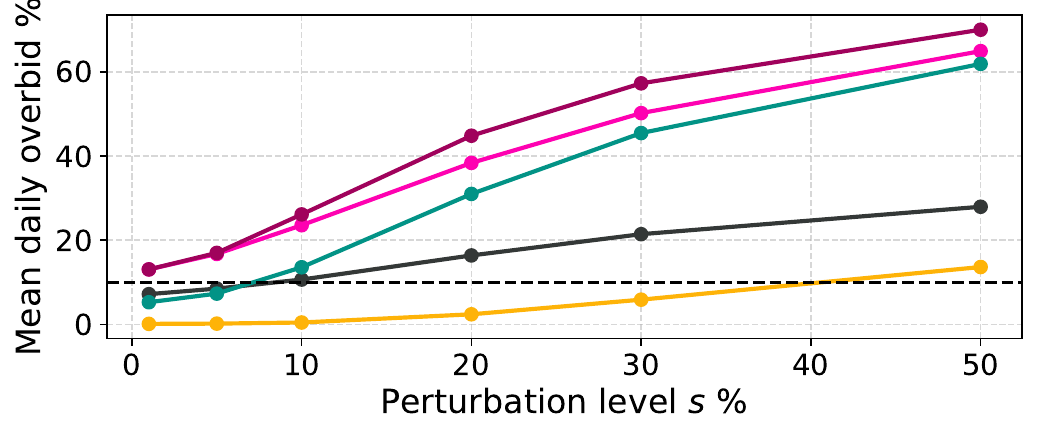}
        \subcaption{Retail electricity prices}
    \end{subfigure}
    \hfill
    \begin{subfigure}{0.70\linewidth}
        \centering
        \includegraphics[width=\linewidth]{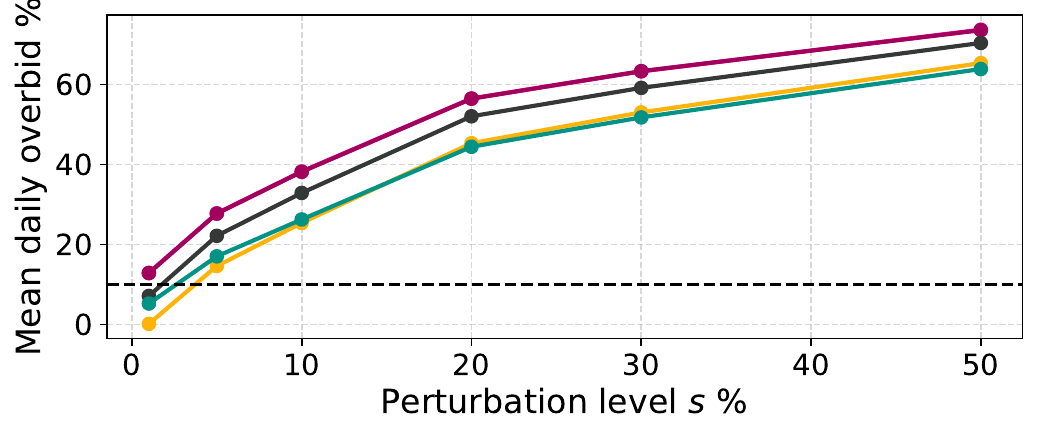}
        \subcaption{mFRR activation signal}
    \end{subfigure}
    \hfill
    \begin{subfigure}{0.70\linewidth}
        \centering
        \includegraphics[width=0.70\linewidth]{noise_analysis_legend.pdf}
    \end{subfigure}
    \caption{Sensitivity of mean daily overbid frequency to perturbations in individual input sources.}
    \label{fig: ch5 sourcewise noise results}
\end{figure}

Figure~\ref{fig: ch5 sourcewise noise results} further examines the sensitivity of P90 compliance to individual input sources. The magnitude of the effect depends on the perturbed input source, with activation and price inputs having a substantially greater influence on overbidding than load and solar inputs under the perturbation scheme used in this study. Notably, the more conservative \gls{pcp} method shows greater robustness to price perturbations, maintaining lower overbidding as the noise level increases. This suggests that greater conservatism in the prediction intervals can provide additional robustness to uncertainty in economically influential inputs. Consequently, reliable market and activation forecasts may be particularly important for flexibility bidding under the conditions considered, although the relative importance of individual input sources may depend on the characteristics of the underlying system and perturbation process.

\begin{figure}[htpb]
    \centering
    \begin{subfigure}{0.70\linewidth}
        \centering
        \includegraphics[width=\linewidth]{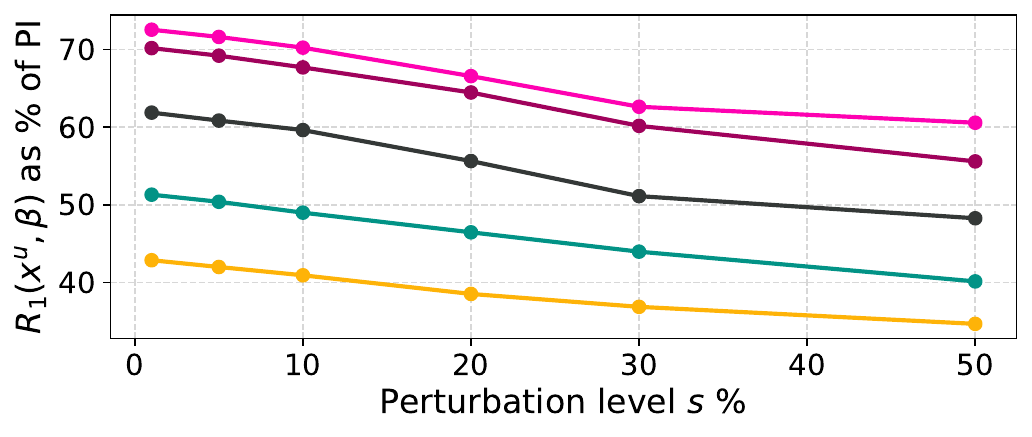}
        \subcaption{$R_{1}(\cdot)$ sensitivity to input perturbation}
    \end{subfigure}
    \hfill
    \begin{subfigure}{0.70\linewidth}
        \centering
        \includegraphics[width=\linewidth]{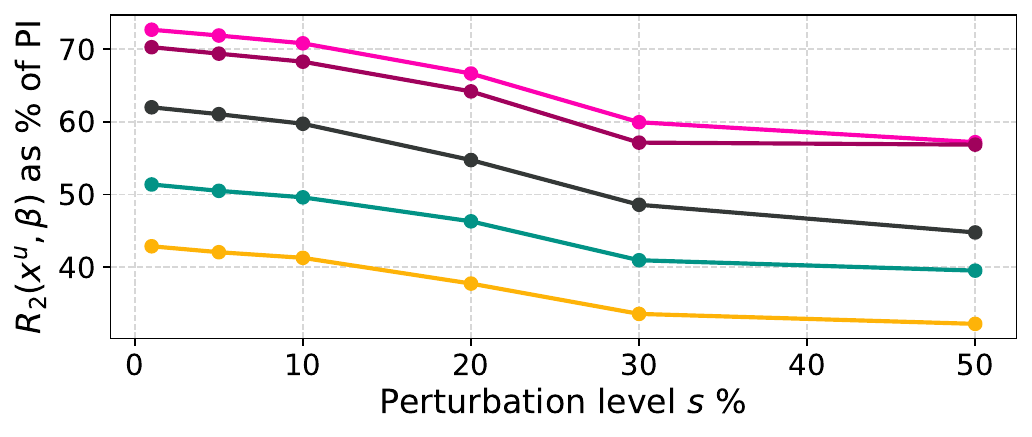}
        \subcaption{$R_{2}(\cdot)$ sensitivity to input perturbations}
    \end{subfigure}
    \hfill
    \begin{subfigure}{0.70\linewidth}
        \centering
        \includegraphics[width=0.70\linewidth]{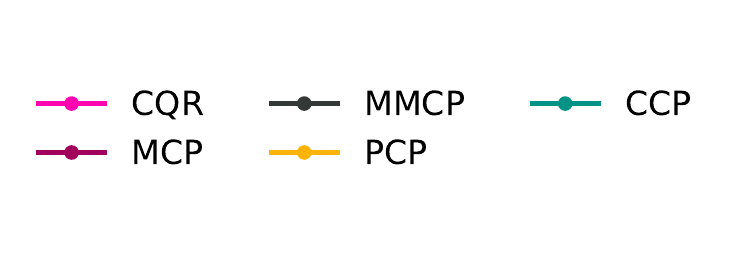}
    \end{subfigure}
    \caption{Sensitivity of adjusted profit to input perturbations for  $\beta=0.1$.}
    \label{fig: ch5 noisy profit}
\end{figure}

The corresponding economic effects are shown in
Fig.~\ref{fig: ch5 noisy profit} for the economically relevant case of
$\beta=0.1$, which is the optimal incentive factor in 97 of the 100 nominal
scenarios. Adjusted profit generally decreases as the perturbation level
increases, reflecting the increasing frequency of reserve shortfalls and
associated penalties. \gls{cqr} and \gls{mcp} retain relatively high adjusted
profits at low and moderate perturbation levels, while the relative economic
performance of the more conservative methods varies as the perturbation level
increases. Overall, the results show that increasing input uncertainty can
substantially reduce the economic benefit of flexibility bidding, with the
magnitude of the impact depending on the uncertainty quantification method and
penalty interpretation.

Overall, these results demonstrate that conformal calibration provides the
intended nominal reliability under the calibration conditions but does not
guarantee robustness to arbitrary distributional mismatch between the
calibration and test inputs. Increasing input perturbations progressively
deteriorate both P90 compliance and economic performance as the test inputs
depart further from the calibration conditions. These results therefore
highlight the importance of accounting for input forecast uncertainty when
applying conformal prediction for flexibility bidding. Forecast- or
input-uncertainty-aware conformal methods could potentially mitigate this
degradation by adapting the prediction intervals to the prevailing
uncertainty conditions. In particular, context-aware conformal approaches
and adaptive methods designed to account for distributional shifts in time
series provide promising directions for future work
\cite{moradi2026enhancedrenewableenergyforecasting, pmlr-v162-zaffran22a}.
%------------------------------------
%------------------------------------
\section{Discussion}
\label{sec: discussion}
\subsection{Practical implications and interpretation}
\label{ssec: discussion implications}
The results indicate that conformal calibration can substantially improve the reliability of flexibility bids while preserving a meaningful share of the aggregator's economic value under nominal conditions. The economic results also demonstrate that the different conformal approaches exhibit distinct trade-offs between reliability, conservatism, and retained flexibility. More conservative prediction intervals reduce the volume of flexibility committed to the day-ahead capacity market, while the resulting residual flexibility could potentially remain available for subsequent market opportunities. Thus, the choice of a conformal approach can be guided by the aggregator's reliability requirements and risk tolerance, with more conservative methods offering greater protection against overbidding at the expense of reduced day-ahead flexibility and profit.

The similar performance observed for \gls{cqr} and \gls{mcp} may be partly explained by similarities in their underlying predictive and calibration procedures. In this study, the separately trained \gls{qr} and \gls{mcd} models exhibit comparable predictive performance, while both approaches use the same calibration dataset and the \gls{mcp} conformity metric for calibration. Although the two approaches generate their initial prediction intervals differently, these similarities may contribute to comparable calibrated intervals and, consequently, similar coverage and economic performance. This observation should not be interpreted as equivalence between the two approaches, but rather as an empirical finding for the dataset and model architecture considered in this study.

The incentive sensitivity analysis shows a relatively weak flexibility supply elasticity of approximately 3--5\% within the simulated \gls{hems} under the techno-economic conditions considered. In the evaluated scenarios, the resulting optimal incentive factor is $\beta^\star=0.1$ in 97 of the 100 nominal scenarios and $\beta^\star=0.2$ in the remaining three. However, this elasticity reflects the decision-making behaviour of the simulated \gls{hems} rather than an empirically estimated response of human prosumers. Consequently, the results should not be interpreted as evidence of a general prosumer engagement barrier. Instead, they suggest that both \gls{hems} scheduling strategies and market incentive mechanisms can influence the economic attractiveness of flexibility provision and warrant further investigation.

The robustness analysis provides an additional qualification to the nominal reliability results. Increasing perturbations to exogenous inputs progressively reduce empirical joint coverage and increase overbidding, with price and activation inputs showing a particularly pronounced effect under the perturbation scheme considered. These results indicate that nominal conformal calibration does not automatically provide the same reliability when the test inputs depart from the conditions represented in the calibration data. In practical applications, therefore, reliable flexibility bidding requires consideration of both predictive uncertainty and uncertainty in the inputs used to generate the predictions. 

\subsection{Limitations and future research}
\label{ssec: discussion limitations} 
Several limitations should be considered when interpreting the results. First, the study uses synthetic data generated from an industry-grade \gls{hems}, and the simulated \gls{hems} is deterministic. The resulting flexibility response therefore represents the behaviour of the implemented scheduling algorithm under the assumed techno-economic conditions rather than the heterogeneous behaviour of real prosumers. Evaluation using real-world prosumer data and more diverse \gls{hems} strategies is therefore an important direction for future work.
 
Second, the revenue-sharing incentive design relies on market price forecasts, while the discretisation of incentive factors and their ex-ante evaluation introduces a deliberate trade-off between computational tractability and solution optimality. The resulting solution is inherently sub-optimal relative to a formulation in which incentive prices are jointly optimised with flexibility predictions. Future work could investigate more adaptive incentive mechanisms and formulations that explicitly account for market price uncertainty.

The battery model implemented in the \gls{hems} does not explicitly account for degradation due to cycling or temperature-dependent battery efficiency. Incorporating these effects could provide a more realistic assessment of the economic and operational implications of flexibility provision. Furthermore, the centralised \gls{ml} training framework may raise concerns about data privacy and security, motivating the investigation of decentralised learning approaches such as federated learning.

Finally, the robustness analysis demonstrates the sensitivity of conformal calibration to departure from the calibration conditions. Future work should therefore investigate context-aware and adaptive conformal methods that can account for changing input uncertainty and distributional shifts, as well as more realistic temporally dependent forecast-error processes. Strategies for reallocating residual flexibility from conservative day-ahead bids to markets operating closer to real time also warrant further investigation.
%------------------------------------
%------------------------------------
\section{Conclusions and future scope}
\label{sec: conclusions}
This study investigated uncertainty-aware flexibility forecasting for aggregators participating in the Danish \gls{mfrr} capacity market under regulatory reliability constraints. Established \gls{uq} approaches, including \gls{mcd}, \gls{mcd}-\gls{cp}, and \gls{cqr}, were applied and systematically evaluated within a multivariate, multi-time-step aggregator bidding problem. Using a large-scale synthetic dataset generated from an industry-grade \gls{hems}, the approaches were assessed in terms of prediction interval reliability, economic performance, and robustness to perturbations in the input data.

The results show that conformal calibration can substantially improve the reliability of flexibility bids under the nominal conditions considered. Compared with uncalibrated \gls{mcd} predictions, the conformal approaches provide substantially higher joint coverage and reduce overbidding risk, with the more conservative approaches satisfying the P90 reliability requirement. At the same time, the results highlight a trade-off between reliability, interval conservatism, and economic value. \gls{cqr} and \gls{mcp} provide comparatively narrow intervals and retain higher economic value, whereas the more conservative \gls{mmcp}, \gls{pcp}, and \gls{ccp} approaches achieve lower overbidding at the expense of reduced day-ahead flexibility and profit. The incentive sensitivity analysis further indicates relatively weak flexibility supply elasticity of approximately 3--5\% for the simulated \gls{hems}, with low incentive factors generally yielding the highest aggregator profit under the conditions considered. Importantly, increasing perturbations to the prediction inputs progressively increased overbidding and reduced empirical joint coverage and economic performance, demonstrating the sensitivity of nominal conformal calibration to departures from the calibration conditions.

Overall, the study demonstrates the practical value of applying established \gls{uq} approaches to uncertainty-aware aggregator bidding, while highlighting the need to balance reliability, economic value, and robustness to input uncertainty. Further validation using real-world and heterogeneous prosumer and \gls{hems} data, together with adaptive approaches that account for changing input uncertainty and distributional shifts, is needed to assess the generalisability of these findings and support deployment in practical aggregator operations.

%------------------------------------
%------------------------------------
\section*{CRediT authorship contribution statement}
\noindent\textbf{Yogesh Pipada Sunil Kumar:} Conceptualisation, Methodology, Software, Formal analysis, Investigation, Writing - original draft, Writing - review and editing. \textbf{S. Ali Pourmousavi:} Conceptualisation, Supervision, Project administration, Funding acquisition, Writing - review and editing. \textbf{Jon A.R. Liisberg:} Supervision, Writing - review and editing. \textbf{Julian Lesmos-Vinasco:} Supervision, Writing - review and editing.
\section*{Acknowledgement}
\noindent This project is supported by the Australian Government Research Training Program (RTP) through the University of Adelaide, and a supplementary scholarship provided by Watts A/S, Denmark. During the preparation of this work, the authors used ChatGPT~\cite{chatgpt} in order to improve readability and language. After using this tool, the authors reviewed and edited the content as needed and take full responsibility for the content of the publication.
%----------------------------
%-----------------------------
\bibliography{bibliography_revised}
\bibliographystyle{elsarticle-num}
%-----------------------------
%-----------------------------
\appendix
\section{Data collection and input scenario sampling strategy}
\label{appendix 1}
This section details the data collection process and input scenario sampling strategy adopted in this study as part of the synthetic data generation framework illustrated in Fig.~\ref{fig: data gen block diagram}.

To construct the dataset for training, calibration and validation, we simulated the operation of the \gls{hems} for $\lvert\mathcal{I}\rvert$ residential prosumers located in the DK1 region of Denmark. The size of the solar \gls{pv} system of each prosumer was randomly sampled according to the distribution reported in the International Energy Agency’s national survey reports from 2012–2018~\cite{iea_denmark_2016, iea_denmark_2017, iea_denmark_2018}. The battery capacity for each household was estimated using a simplified sizing rule proposed in~\cite{battery_sizing, battery_sizing_2}, which sets the storage capacity to approximately 1.5 times the size of \gls{pv} for Denmark. To introduce variability, the size multiplier was randomly selected from the range \([1.2, 1.7]\). To ensure that the resulting capacities are practically deployable, each estimated battery size was matched to the nearest available model of the BYD residential product line~\cite{byd_batterybox_hv_au_v1_2}, adopting the corresponding nameplate capacity $b^{\text{cap}, i}$, the rated discharge power $b^{\text{dis}, i}$ and the round-trip efficiency of the product datasheet for each individual prosumer $i \in \mathcal{I}$.

The disaggregated load and solar generation data for each prosumer $i \in \mathcal{I}$ were required to simulate the \gls{hems} algorithm. The load data, $\boldsymbol{x^{\text{l}}, i}$, were obtained from the synthetic dataset developed by the authors in~\cite{ray_dataset}, containing 100,000 daily consumption profiles with hourly resolution, categorised into workdays and non-workdays. Solar generation data was obtained from the Renewables.ninja platform~\cite{renewable_ninja_3}, based on the methodology in~\cite{renewable_ninja_1, renewable_ninja_2}, which generates simulated hourly solar generation data for various \gls{pv} configurations and regions. Data were generated for 5,100 daily solar profiles across seven distinct panel tilt angles ($25^{\circ}, 36^{\circ}, 40^{\circ}, 42^{\circ}, 47^{\circ}, 58^{\circ}, 69^{\circ}$) for a 1~kW \gls{pv} system. Individual prosumer generation profiles, $\boldsymbol{x^{\text{s}}, i}$, were then derived by scaling these 1~kW profiles according to each prosumer’s assigned \gls{pv} capacity. Since the load and solar profiles were not temporally aligned, randomly paired daily load–solar profile combinations were created for each prosumer.

Electricity market data for Denmark were obtained using the Energinet Data Service API~\cite{energidataservice_api_guide}. The prosumer electricity sale price profiles, $\boldsymbol{\lambda^{\text{S}}}$, were derived from \gls{dam} daily price data for the DK1 bidding zone, covering the period from January~2023 to January~2025. The corresponding purchase price profiles, $\boldsymbol{\lambda^{\text{P}}}$, were calculated by adding applicable grid tariffs and a 25\% value-added tax (VAT) to the sale prices. The grid tariffs for the simulation period were obtained from one of the Denmark's distribution system operators, Cerius~\cite{cerius_tariffs, cerius_historical_tariff}. Similarly, market price profiles for \gls{mfrr} up-regulation capacity, $\boldsymbol{\lambda^{\text{U}}}$, were extracted for the period August~2021 to January~2025. Binary activation signals for \gls{mfrr} up-regulation, $\boldsymbol{a^{\uparrow}}$, were obtained for January~2021 to January~2025. To enhance data diversity, additional activation samples were synthetically generated by modelling the tetrachoric correlation structure of the historical activation signal. Multiple scenarios of the up-regulation incentive, \(\boldsymbol{\lambda}^{\text{I,u}}\), were generated by scaling the historical \gls{mfrr} up-regulation capacity prices with random multipliers independently drawn from a uniform distribution in \([0, 1]\) at each time interval, thus producing a wide range of incentive pricing scenarios. This scaling approach is consistent with the logical constraint that the incentive offered to prosumers must not exceed the market clearing price.

Because the various market datasets were not temporally aligned with the load and solar profiles, we generated combinations of randomly sampled market scenarios. Each scenario included a distinct set of daily values for electricity prices, up-regulation prices, and activation signals. To ensure consistency and allow aggregation at the cluster level, each market scenario was applied uniformly across all $\lvert\mathcal{I}\rvert$ prosumers, each with unique load, generation, and battery characteristics. This approach ensures a diverse yet coherent input dataset suitable for obtaining the aggregated behaviour of the prosumers under representative market and operational variability.
%----------------------------------------------
%----------------------------------------------
\section{Prediction interval evaluation metrics}
\label{appendix 2}
To assess the quality of the prediction intervals generated by our probabilistic forecasting algorithm, we employ a set of established evaluation metrics. Let $\left(\Theta^{(j)}, y^{(j)}\right)$ denote the $j^{\text{th}}$ instance in the test dataset $\mathcal{D}_{\text{test}}$. Let $\xi^{(j)} = \Phi_{1-\alpha}\left(\Theta^{(j)}\right)$ represent the corresponding prediction interval at a coverage level of $1 - \alpha$. The element-wise upper and lower bounds of this interval are denoted by $\boldsymbol{u}^{(j)} = \sup \xi^{(j)}$ and $\boldsymbol{l}^{(j)} = \inf \xi^{(j)}$, respectively. The resulting bounding hyperrectangle $\mathcal{B}^{(j)}_{\xi}$ is defined accordingly, as shown in Eq.~\eqref{eq: ch5 infinmum}, and serves as the basis for all interval-based evaluations in this case study. 

This adjustment only affects the \gls{pcp}-based predictions, leading to more conservative bounds, and does not affect the other methods, since $\xi$ remains a bounding hyperrectangle in those cases. Note that the output dimensionality $d$ is equal to the number of time steps $T$ over which the \gls{hems} algorithm is solved. Additionally, $\mathbbm{1}(\cdot)$ denotes the indicator function. The prediction interval evaluation metrics are formally defined in Eqs.~\eqref{eq: ch5 picp marginal}--\eqref{eq: ch5 is joint}. To provide a comprehensive assessment of the prediction intervals, we report both marginal metrics (evaluated per time step) and joint metrics (evaluated across the entire prediction horizon).
\begin{align}
    \label{eq: ch5 picp marginal}
    {}\quad & PICP_{t} = \frac{1}{\left|\mathcal{D}_{test}\right|} \sum_{j=1}^{\left|\mathcal{D}_{test}\right|}\mathbbm{1}\left(l^{(j)}_{t} \leq y^{(j)}_{t} \leq u^{(j)}_{t}\right) & \quad \forall t \in \mathcal{T}\\     
    \label{eq: ch5 picp joint}
    {}\quad & PICP_{joint} = \frac{1}{\left|\mathcal{D}_{test}\right|} \sum_{j=1}^{\left|\mathcal{D}_{test}\right|}\mathbbm{1}\left(y^{(j)} \in \mathcal{B}^{(j)}_\xi\right) & {}\\
    \label{eq: ch5 mpiw marginal}
    {} \quad& MPIW_{t} = \frac{1}{\left|\mathcal{D}_{test}\right|} \sum_{j=1}^{\left|\mathcal{D}_{test}\right|}\left(u^{(j)}_{t} - l^{(j)}_{t}\right) & \quad \forall t \in \mathcal{T}\\ %
    \label{eq: ch5 mpiw joint}
    {} \quad& MPIW_{joint} = \frac{1}{T} \sum_{t \in \mathcal{T}}MPIW_{t} & {}\\ 
    \label{eq: ch5 lpenalty}
   {}\quad &  LPenalty_{t}(\alpha) = \frac{1}{\left|\mathcal{D}_{test}\right|} \sum_{j=1}^{\left|\mathcal{D}_{test}\right|}\frac{2}{\alpha}\left[l^{(j)}_{t} - y^{(j)}_{t}\right]\cdot\mathbbm{1}\left(y^{(j)}_{t}\leq l^{(j)}_{t}\right) &  \quad \forall t \in \mathcal{T}\\
   \label{eq: ch5 upenalty}
   {}\quad &  UPenalty_{t}(\alpha) = \frac{1}{\left|\mathcal{D}_{test}\right|} \sum_{j=1}^{\left|\mathcal{D}_{test}\right|}\frac{2}{\alpha}\left[y^{(j)}_{t} - u^{(j)}_{t}\right]\cdot\mathbbm{1}\left(u^{(j)}_{t}\leq y^{(j)}_{t}\right) & \quad \forall t \in \mathcal{T}\\
   \label{eq: ch5 is marginal}
    {}\quad & IS_{t}(\alpha) = \frac{1}{\left|\mathcal{D}_{test}\right|} \sum_{j=1}^{\left|\mathcal{D}_{test}\right|}\left(MPIW_{t} + LPenalty_{t} + UPenalty_{t}\right) &  \quad \forall t \in \mathcal{T}\\   
    \label{eq: ch5 is joint}
    {}\quad & IS_{joint}(\alpha) = \frac{1}{T} \sum_{t \in \mathcal{T}}IS_{t} & {}
\end{align}

\gls{picp} quantifies the proportion of ground-truth values that fall within the predicted intervals, as defined in Eqs.\eqref{eq: ch5 picp marginal} and \eqref{eq: ch5 picp joint}. It serves as an empirical estimate of the coverage probability, especially as the size of the test dataset $\mathcal{D}_{\text{test}}$ approaches infinity. While \gls{picp} evaluates the reliability of the intervals, it does not account for their width. This aspect is captured by \gls{mpiw}, which measures the average width of the prediction intervals and reflects their sharpness, as shown in Eqs.\eqref{eq: ch5 mpiw marginal} and \eqref{eq: ch5 mpiw joint}. Lower \gls{mpiw} indicates narrower intervals and therefore greater sharpness.   

However, \gls{mpiw} alone does not indicate whether the prediction intervals successfully contain ground-truth values. To overcome this limitation, the \gls{is} metric is introduced, as defined in Eqs.\eqref{eq: ch5 is marginal} and \eqref{eq: ch5 is joint}. This metric combines the average interval width with explicit penalties for violations, where the ground-truth falls below the lower bound or above the upper bound of the interval, as described in Eqs.\eqref{eq: ch5 lpenalty} and \eqref{eq: ch5 upenalty}, respectively. In doing so, \gls{is} provides a more comprehensive evaluation of interval quality. By analysing these metrics together, we gain a clearer understanding of how accurate, sharp, and reliable the predicted intervals are.

In addition to the traditional prediction interval metrics described above, we also evaluate the frequency of overbid, defined as the proportion of intervals per day in which the true value of up-regulation flexibility \( y^{(j)}_{t} \) falls below the lower bound \( l^{(j)}_{t} \) of the predicted interval. This metric is formalised in Eq.~\eqref{eq: ch5 frequency of overbid}, which quantifies how often the predicted lower bound \( l^{(j)}_{t} \) fails to capture the actual flexibility value, effectively representing instances where the aggregator might have overbid in the \gls{mfrr} capacity market. 
\begin{equation}
    \label{eq: ch5 frequency of overbid}
    \text{Frequency of overbid} = \frac{1}{T} \sum_{t\in\mathcal{T}} \mathbbm{1}\left(y^{(j)}_{t} < l^{(j)}_{t}\right),
\end{equation}

This concept was originally introduced in~\cite{jalal_p90} to assess compliance with the P90 requirement mandated by Energinet, which stipulates that at least 90\% of the time, the actual flexibility should lie above the bid level. Hence, the mean frequency of daily overbid should remain below 10\% for our probabilistic forecast model to qualify for market participation. This requirement ensures that the aggregator does not systematically overestimate its capacity, thereby supporting reliable and secure market operations.
%-----------------------------------------------
%-----------------------------------------------

\section{Input perturbation generation procedure}
\label{app: noise generation}

This appendix describes the procedure used to generate input perturbations
for the robustness analysis presented in Section~\ref{ssec: ch5 robustness
analysis}. Two evaluations use the same input perturbation procedure: a
coverage/reliability evaluation and an economic evaluation. The perturbations
are applied only to test-time inputs, while the calibration data and
conformal calibration thresholds remain unchanged.

\subsection{Perturbed input features and noise levels}
\label{app: noise features}

The model input for one day consists of $6T+1$ features, where $T=24$,
comprising six time-varying feature blocks and one static feature. The
perturbation treatment applied to each input group is summarised in
Table~\ref{tab:app noise groups}.

\begin{table}[htpb]
    \centering
    \caption{Input feature groups and corresponding perturbation treatment.}
    \label{tab:app noise groups}
    \begin{tabular}{@{}llll@{}}
        \toprule
        Block & Symbol & Quantity & Perturbation treatment \\
        \midrule
        Purchase price
            & $\lambda^{P}$
            & Retail purchase price forecast
            & Coupled additive \\
        Selling price
            & $\lambda^{S}$
            & Retail selling price forecast
            & Coupled additive \\
        Incentive price
            & $\lambda^{I,u}$
            & Aggregator incentive price
            & Not perturbed \\
        Activation signal
            & $a^{\uparrow}$
            & \gls{mfrr} up-regulation activation forecast
            & Copula partial resampling \\
        Load
            & $x_l$
            & Aggregated load forecast
            & Multiplicative log-normal \\
        Solar
            & $x_s$
            & Aggregated \gls{pv} forecast
            & Multiplicative log-normal \\
        Battery sizing
            & $b^{\mathrm{cap}}, b^{\mathrm{dis}}$
            & Static aggregated nameplate
            & Not perturbed \\
        \bottomrule
    \end{tabular}
\end{table}

The perturbation magnitude is controlled by a scalar parameter $s$. For the
aggregate analysis, the same value of $s$ is applied simultaneously to all
perturbed input groups. For continuous inputs, $s$ scales the standard
deviation of the perturbation, whereas for the activation signal it specifies
the fraction of hours that are resampled. The noise levels considered are
\begin{equation}
    s \in \{0.01,\;0.05,\;0.10,\;0.20,\;0.30,\;0.50\}.
    \label{eq:app noise levels}
\end{equation}

Perturbations are generated in the min--max normalised input space used by
the prediction model, except for the coupled retail-price perturbation,
which is generated in physical units and subsequently re-normalised.
Independent pseudo-random streams are used for different input sources,
allowing the sources to be perturbed individually for the source-wise
analysis while retaining the same source-specific perturbations used in the
aggregate analysis.

\subsection{Load and solar perturbations}
\label{app: noise continuous}

Load and solar forecasts are non-negative and are therefore perturbed using
a multiplicative log-normal model. For each sample and hour $t$,
\begin{equation}
    \tilde{x}_{t}
    =
    x_{t}\exp(\varepsilon_t),
    \qquad
    \varepsilon_t \sim \mathcal{N}(0,s^2),
    \label{eq:app lognormal noise}
\end{equation}
where $s$ is the standard deviation of the logarithmic perturbation.
Independent perturbations are generated for each sample and hourly entry.
Because $\exp(\varepsilon_t)$ is strictly positive, the resulting load and
solar values remain non-negative.

\subsection{Price perturbations}
\label{app: noise prices}

The retail purchase and selling prices are generated from a common
day-ahead spot price. Their relationship can be expressed as
\begin{align}
    \lambda^{S} &= \mathrm{spot},\\
    \lambda^{P}
    &=
    \left(
        \mathrm{spot}
        +\tau_{\mathrm{tax}}
        +\tau_{\mathrm{var}}
        +\tau_{\mathrm{tariff}}
    \right)(1+\mathrm{VAT}),
\end{align}
which, for $\mathrm{VAT}=0.25$, gives
\begin{equation}
    \lambda^{P}
    =
    1.25\lambda^{S}+b_{\mathrm{fixed}},
    \label{eq:app price affine}
\end{equation}
where
\begin{equation}
    b_{\mathrm{fixed}}
    =
    1.25
    \left(
        \tau_{\mathrm{tax}}
        +\tau_{\mathrm{var}}
        +\tau_{\mathrm{tariff}}
    \right).
\end{equation}

To preserve this relationship during perturbation, a single shared
spot-price shock is applied to both retail prices in physical units. For
each hour $t$,
\begin{align}
    \tilde{\lambda}^{S}_{t}
        &= \lambda^{S}_{t}+\epsilon_t,\\
    \tilde{\lambda}^{P}_{t}
        &= \lambda^{P}_{t}+1.25\epsilon_t,\\
    \epsilon_t
        &\sim
        \mathcal{N}
        \left(
            0,
            s^2\sigma_{S,t}^{2}
        \right),
    \label{eq:app coupled price noise}
\end{align}
where $\sigma_{S,t}$ is the physical-unit standard deviation of the
selling-price column. The factor $1.25$ corresponds to the VAT multiplier
in Eq.~\eqref{eq:app price affine}, while the fixed component remains
unchanged. The perturbed prices are subsequently re-normalised before being
provided to the prediction model.

Additive rather than multiplicative perturbations are used for prices
because prices can take negative values or cross zero. Any additional
price-type input columns that are not retail purchase or selling prices are
perturbed independently according to
\begin{equation}
    \tilde{x}_{t}
    =
    x_t+\eta_t,
    \qquad
    \eta_t
    \sim
    \mathcal{N}
    \left(
        0,
        s^2\sigma_t^2
    \right),
    \label{eq:app other price noise}
\end{equation}
where $\sigma_t$ is the empirical standard deviation of the corresponding
input column.

The incentive-price input $\lambda^{I,u}$ is not perturbed. In the economic
evaluation, it represents the committed incentive cost derived from the
up-regulation market price and is therefore held fixed during the bidding
and settlement process. The battery-sizing features are also held fixed
because they represent static nameplate characteristics.

\subsection{Activation-signal perturbations}
\label{app: noise activation}

The \gls{mfrr} up-regulation activation signal
$a^{\uparrow}\in\{0,1\}^{T}$ is binary and exhibits temporal correlation
across the day. Therefore, independent bit-flips are not used. Instead,
perturbations are generated using partial resampling from a fitted
Gaussian-copula model.

A tetrachoric correlation matrix
$\boldsymbol{\Sigma}\in\mathbb{R}^{T\times T}$ and a vector of hourly
activation probabilities
$\boldsymbol{p}\in[0,1]^T$ are estimated from the historical activation
data used in constructing the synthetic dataset. A synthetic activation
profile is generated according to
\begin{align}
    \boldsymbol{z}
        &\sim
        \mathcal{N}(\boldsymbol{0},\boldsymbol{\Sigma}),\\
    \boldsymbol{u}
        &=
        \Phi_{\mathrm{cdf}}(\boldsymbol{z}),\\
    \tilde{a}^{\uparrow}_{t}
        &=
        \mathbb{1}(u_t<p_t),
    \label{eq:app activation copula}
\end{align}
where $\Phi_{\mathrm{cdf}}$ denotes the standard normal cumulative
distribution function applied element-wise. This construction preserves
the estimated hourly marginal activation probabilities and temporal
correlation structure.

For a given perturbation level $s$, a fresh copula activation profile is
generated and a fraction of the hours in each day is replaced by the
corresponding hours from the generated profile. Specifically,
$\lceil sT\rceil$ hours are selected randomly for replacement, while the
remaining hours retain their original values. Thus, increasing $s$
corresponds to an increasing fraction of the daily activation profile being
resampled. The copula generation and hour selection are seeded separately
for each repetition.

\subsection{Coverage and reliability evaluation}
\label{app: noise coverage}

For the coverage/reliability analysis, perturbations are applied only to the
test inputs. The calibration set $\mathcal{D}_{\mathrm{cal}}$ remains
unperturbed, and the conformal calibration thresholds $\hat{q}$ are
computed once using the clean calibration data and subsequently reused at
all perturbation levels. Neither the prediction models nor the conformal
calibration are retrained or recalibrated under the perturbed conditions.
All prediction intervals are evaluated against the corresponding
unperturbed ground-truth flexibility ${y}$.

Two configurations are considered. In the aggregate configuration, all
perturbed input sources are applied simultaneously at the same perturbation
level $s$. In the source-wise configuration, load, solar, price, and
activation inputs are perturbed individually, with the same source-specific
random streams used for the corresponding aggregate perturbations.

For the \gls{mcd}-based methods, $S=1000$ stochastic forward passes are
used to construct the predictive samples at each configuration. For
\gls{cqr}, a single deterministic forward pass is used. Five independent random realizations were evaluated at each perturbation level to reduce sensitivity to individual random draws and provide more robust estimates of the coverage and reliability metrics.

\subsection{Economic evaluation}
\label{app: noise economic}

The economic evaluation uses the aggregate perturbation configuration only.
Input perturbations are applied under the same test-only regime described
above, while the resulting bids are settled against the clean flexibility.
In addition to perturbations in the prediction inputs, uncertainty in the
up-regulation settlement price is introduced separately.

The up-regulation market price is represented using two values:
$\lambda^{U,\mathrm{pre}}$, which is used when the bid is formed, and
$\lambda^{U,\mathrm{post}}$, which determines realised settlement revenue.
The bid-time value is held fixed and determines the committed incentive
cost $\beta\lambda^{U,\mathrm{pre}}$. The settlement value is generated as
\begin{equation}
    \lambda^{U,\mathrm{post}}_t
    =
    \lambda^{U,\mathrm{pre}}_t+\xi_t,
    \qquad
    \xi_t
    \sim
    \mathcal{N}
    \left(
        0,
        s^2\sigma_t^2
    \right),
    \label{eq:app settlement noise}
\end{equation}
where $\xi_t$ is generated independently of the input-forecast
perturbations. At $s=0$, the settlement and bid-time prices are identical.

The input perturbation is scenario-keyed. For each scenario, a single
input-noise realisation is generated and applied identically across all
revenue-sharing values $\beta$ evaluated for that scenario. Different
noise realisations are used across scenarios and repetitions. The economic
results reported in Section~\ref{ssec: ch5 robustness analysis} uses an average of ten independent random realisations at each perturbation level. This reduces sensitivity to individual random draws and provides more robust estimates of the economic performance metrics.

\subsection{Summary of perturbation models}
\label{app: noise summary}

Table~\ref{tab:app noise models} summarises the perturbation distributions
and their level parameters.

\begin{table}[htpb]
    \centering
    \caption{Summary of perturbation models used in the robustness analysis.}
    \label{tab:app noise models}
    \begin{tabular}{@{}llll@{}}
        \toprule
        Input group & Perturbation & Distribution & Level parameter \\
        \midrule
        Load, solar
            & Multiplicative $x\exp(\varepsilon)$
            & $\varepsilon\sim\mathcal{N}(0,s^2)$
            & $s$ = log-SD \\
        Retail prices
            & Shared additive shock
            & $\epsilon\sim\mathcal{N}(0,s^2\sigma_S^2)$
            & $s\sigma_S$ \\
        Other price inputs
            & Additive $x+\eta$
            & $\eta\sim\mathcal{N}(0,s^2\sigma^2)$
            & $s\sigma$ \\
        Activation
            & Copula partial resampling
            & Gaussian copula $(\boldsymbol{\Sigma},\boldsymbol{p})$
            & $s$ = fraction of hours \\
        Settlement price
            & Additive $\lambda^{U,\mathrm{pre}}+\xi$
            & $\xi\sim\mathcal{N}(0,s^2\sigma^2)$
            & $s\sigma$ \\
        \bottomrule
    \end{tabular}
\end{table}

Across the experiments, the perturbation levels are
$s\in\{0.01,0.05,0.10,0.20,0.30,0.50\}$. Different input sources use
independent pseudo-random streams, while the retail purchase and selling
prices share a common spot-price shock. Continuous perturbations are
generated independently across samples and hours. The activation perturbation
retains within-day temporal dependence through the fitted correlation
structure. After perturbation, the normalised model inputs are clipped to
the interval $[0,1]$.

\end{document}